\documentclass{article}

\usepackage{microtype}
\usepackage{graphicx}
\usepackage{subfigure}
\usepackage{booktabs} 

\usepackage{hyperref}

\usepackage[accepted]{icml2022-arxiv}

\usepackage{amsmath}
\usepackage{amssymb}
\usepackage{mathtools}
\usepackage{amsthm}

\usepackage[capitalize,noabbrev]{cleveref}

\usepackage{amssymb}
\usepackage{amsmath}
\usepackage{amsthm}
\usepackage{thmtools, thm-restate}
\usepackage{enumerate}
\usepackage{enumitem}
\usepackage{notation}

\icmltitlerunning{Differentially Private Maximal Information Coefficients}

\begin{document}
\twocolumn[
\icmltitle{Differentially Private Maximal Information Coefficients}




\begin{icmlauthorlist}
  \icmlauthor{John Lazarsfeld}{yale}
  \icmlauthor{Aaron Johnson}{navy}
  \icmlauthor{Emmanuel Adeniran}{yale}
\end{icmlauthorlist}

\icmlaffiliation{yale}{Department of Computer Science, Yale University}
\icmlaffiliation{navy}{U.S. Naval Research Laboratory}

\icmlcorrespondingauthor{John Lazarsfeld}{john.lazarsfeld@yale.edu}

\icmlkeywords{Differential Privacy}

\vskip 0.3in
]





\printAffiliationsAndNotice{}  

\begin{abstract}
The Maximal Information Coefficient (MIC) is a powerful statistic to identify
dependencies between variables. However, it may be applied to sensitive data,
and publishing it could leak private information. As a solution, we present
algorithms to approximate MIC in a way that provides differential privacy. We
show that the natural application of the classic Laplace mechanism yields
insufficient accuracy. We therefore introduce the $\micr$ statistic, which is a
new MIC approximation that is more compatible with differential privacy. We
prove $\micr$ is a consistent estimator for MIC, and we provide two
differentially private versions of it. We perform experiments on a variety of
real and synthetic datasets. The results show that the private $\micr$
statistics significantly outperform direct application of the Laplace mechanism. Moreover, experiments on real-world datasets show accuracy that is
usable when the sample size is at least moderately large.
\end{abstract}


\section{Introduction}

The Maximal Information Coefficient (\mic{}) is a powerful and relatively new
tool to detect correlations in data~\cite{reshef2011,reshef16-mice}. \mic{} uses
mutual information to detect general dependencies between numeric attributes, in
contrast to a more common statistic such as Pearson’s correlation coefficient,
which is only designed to detect linear relationships. \mic{} is thus
particularly suited to identify novel relationships in complex data, as in
that setting it is unknown which properties might be related and how.

However, many datasets that are valuable for such data mining
(such as medical or economic data) contain sensitive personal information.
Moreover, even
publishing just the statistics that result from a correlation analysis can
reveal private details of the individuals comprising the
data~\cite{homer2008resolving,wang2009learning}. A scientist or government
analyst must therefore consider not only the effectiveness of their statistical
techniques, but they must also take into account if the resulting statistics can
be published in a way that protects privacy.

Differential privacy (DP)~\cite{dwork-roth} has emerged as the leading method
for privacy-preserving data publishing. Major companies and institutions use
differential privacy to publish statistics about their sensitive
data~\cite{cormode2018privacy,dwork2019differential}. \mic{} appears at first to
be well-suited to being published in a differentially private way. At a high
level, for a pair of numeric variables, it measures the maximum mutual
information over possible \emph{grids} partitioning their joint range. For any
given grid, the effect of changing just one data point, which is the the main
criterion for differential privacy, changes the distribution of points very
little. This fact suggests that a differentially private \mic{} could be
designed with high accuracy.

However, \mic{} is difficult to compute, and instead it is suggested to
approximate it using the \mice{} statistic~\cite{reshef16-mice}. \mice{} is
efficiently computable because it restricts its optimization to subgrids of a
master \emph{mass equipartition} of the dataset, that is, a grid in which each
row and column contains the same number of data points. This master grid depends
on the data, though, and we obtain a bound on the change in \mice{} from
altering one data point that is significantly higher than would be expected for
\mic{}.

We therefore investigate a new method to approximate \mic{} that is less
sensitive to small changes in the dataset. We propose the \micr{} statistic,
which optimizes over subgrids of a master \emph{range equipartition}, that is, a
grid in which the rows and columns divide the range equally. \micr{} is
efficiently computable, and we obtain a bound on its sensitivity to input
perturbations that is lower than our bound for \mice{}, both asymptotically and
concretely. We prove that \micr{} converges in probability to \mic{} (or, more
properly, to the analogous statistic defined over distributions).

We then present two differentially private versions of \micr{}, representing
fundamentally different approaches to adding DP noise. \micrlap{} uses the
classic Laplace mechanism, which adds random noise to the function
\emph{output} (in our case, the non-private \micr{} value). In contrast, \micrgeom{} perturbs
the \emph{input}, accomplished by adding a random count sampled from the
geometric distribution to each grid cell, before then computing \micr{}. We
prove that the error added to \micr{} by each of these mechanisms goes to zero
as the size of the dataset increases.

Finally, we implement and experimentally analyze the two \micr{} mechanisms and
the naive application of the Laplace mechanism to \mice{} (\micelap{}). Our
experiments use the synthetic and real datasets used to evaluate the original
\mic{} statistics. The results show that the \micr{} mechanisms both
significantly outperform \micelap{}. Comparing the \micr{} mechanisms, we
observe that \micrlap{} has lower bias but higher variance than \micrgeom{}.
Moreover, the experiments on real-world datasets show usable accuracy for a
reasonable level of privacy when the sample size is at least moderately large.
For example, with $\epsilon=1$ we obtain average errors as low as 0.068 and
0.016 on datasets with 337 and 4381 datapoints, respectively, where MIC itself
can range from 0 to 1 (low to high correlation).





\section{Preliminaries}
\label{sec:prelims}

We begin by introducing notation and concepts
that are used to define $\mic$ and its related approximations.

\subsection{Datasets, Grids, and Distributions}

We define a \textit{dataset} $D$ as a sequence of $n$ points in $\R^2$.
To distinguish between $x$ and $y$ coordinates,
we write $D = (D_x, D_y)$, where $D_x$, $D_y$ are
sequences of $n$ points in $\R$.

For integers $k, \ell \ge 2$, a
$k \times \ell$ \textit{grid} $G = (P, Q)$ on $\R^2$
is comprised of a size-$k$ partition
$P = \{P_1, \dots, P_k\}$ of the $y$-axis,
and a size-$\ell$ partition $Q =\{Q_1, \dots, Q_\ell\}$
of the $x$-axis.
We say that a $k \times \ell$ grid $G$ has
$k\ell$ total \textit{cells},
and we let $\G(k, \ell)$ denote the set of all
grids with $k\ell$ total cells.

For $k \le j$, a single-axis partition $P = \{P_1, \dots, P_k\}$,
is a \textit{subpartition} of $C = \{C_1, \dots, C_j\}$
(denoted by $P \subseteq C$) if every $P_i$ is the
union of adjacent intervals in $C$.
For example, let $P$=\{[0, 0.2), [0.2, 0.5), [0.5, 0.8), [0.8, 1]\}
be a size-4 partition of the interval [0, 1].
Then $Q$=\{[0, 0.5), [0.5, 1]\} is a size-2 subpartition
of $P$ because each element of $Q$ is the union of two elements
of $P$ that are adjacent on $[0,1]$.

We call $P$ a size-$k$ \textit{range equipartition}
of an interval $I = [x_0, x_1]$ when $P$ is
a size-$k$ partition of $I$ and all parts of $P$
are intervals of length $(x_1-x_0)/k$,
We call a grid $G = (P, Q)$ a \textit{subgrid}
of $\Gamma = (C_x, C_y)$  (denoted by $G \subseteq \Gamma$) if
$P$ and $Q$ are subpartitions of $C_x$ and $C_y$ respectively.

For a $k \times \ell$ grid $G = (P, Q)$, and for a point
$d = (d_x, d_y) \in \R^2$, we define the point-mapping
function $\phi$, where $\phi(d, G) = (i, j)$
iff $d_x \in Q_j$ and $d_y \in P_i$.
Then for a dataset $D$ of $n$ points and a $k \times \ell$
grid $G$, we define $\A_{D, G} \in \mathbb{Z}^{k \times \ell}$
as the \textit{count matrix} for $D$ and $G$.
For all $(i,j) \in [k] \times [\ell]$, $\A_{D, G}$ has
entries
%
$$
a(i, j) =
\left|
  \{d \in D \;:\; \phi(d, G) = (i, j) \}\right|,
$$
meaning $\sum_{i,  j} a(i, j) = n$.
For a $k \times \ell$ grid $G = (P, Q)$,
we call $P$ (resp. $Q$) a \textit{mass equipartition} if
all row sums (resp. column sums) of $\A_{D, G}$ are equal.

\begin{figure}
  \centering
  \includegraphics[width=0.45\textwidth]{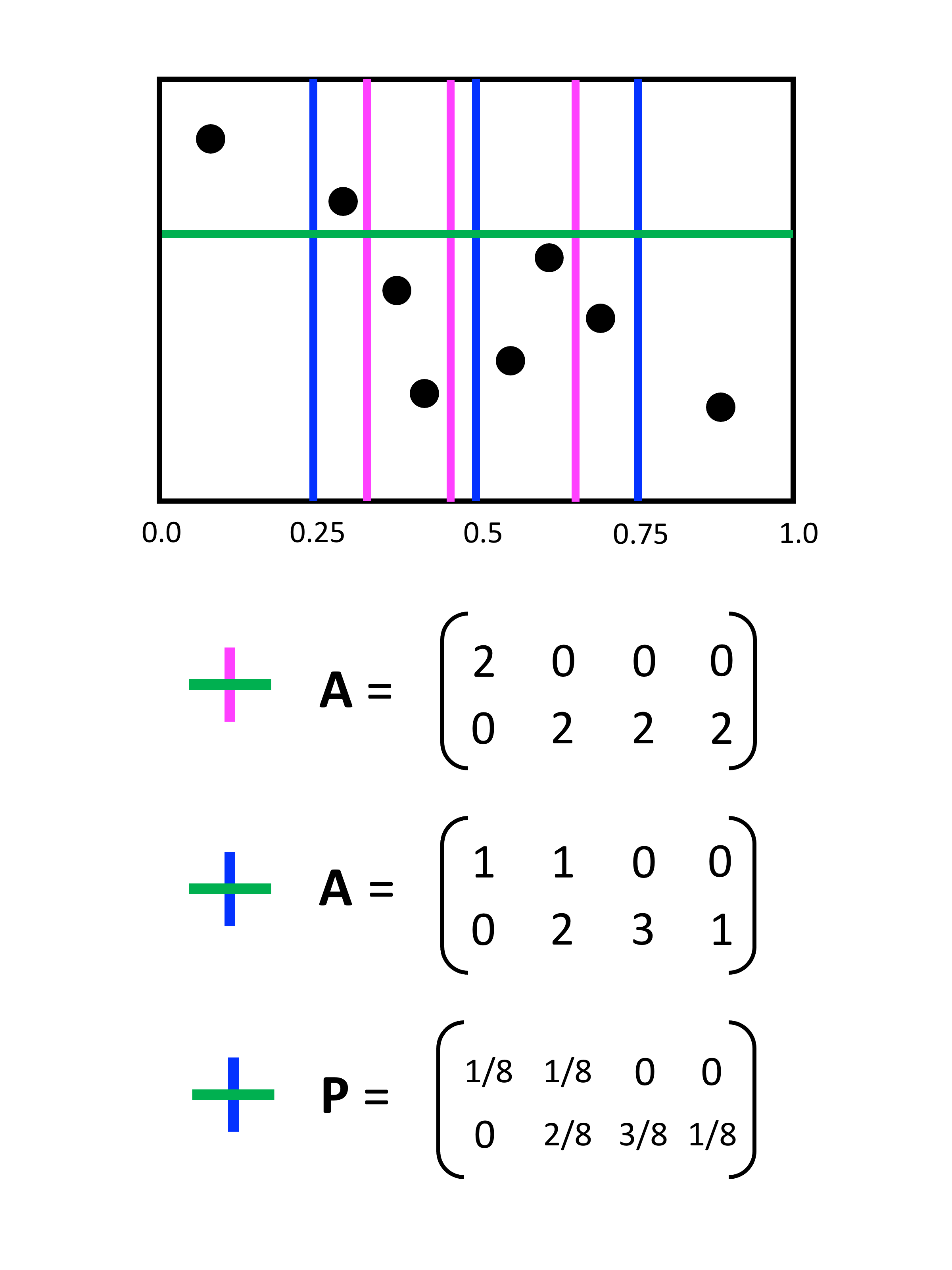}
  \vspace*{-1em}
  \caption{
    An example dataset of $n$=8 points (black dots).
    The pink vertical lines represent a
    size-4 \textit{mass} equipartition,
    and the blue vertical lines represent a
    size-4 \textit{range} equipartition of the
    interval $[0, 1]$.
    The first matrix shows the count matrix
    for the grid comprised of the pink column partition
    and green row partition.
    The second matrix shows the count matrix for the
    grid comprised of the blue column partition and
    green row partition.
    Notice how the count matrix can vary depending
    on whether the column partition
    is a mass (pink) or range (blue) equipartition.
    The final matrix is the normalized count matrix
    for the blue/green grid.
  }
  \label{fig:grid-ex}
\end{figure}

If $G$ is a subgrid of a grid $\Gamma$,
observe that each entry of $\A_{D, G}$ can be ``generated''
by summing adjacent entries in $\A_{D, \Gamma}$,
and we write $\psi(\A_{D, \Gamma}, \Gamma, G) = \A_{D, G}$
to denote the matrix whose entries are generated via this process.

Given $\A_{D, G}$, the matrix $\P_{D, G} \in \R^{k \times \ell}$
with entries
$p(i, j)$ is the \textit{normalized count matrix}
for $D$ and $G$, where
\begin{equation*}
  \P_{D, G} = (1/n) \cdot \A_{D, G}.
\end{equation*}
We use $p(i, *) = \sum_{j} p(i, j)$ and
$p(*, j) = \sum_{i} p(i, j)$ to denote the row
and column sums of $\P_{D, G}$ respectively.

Figure~\ref{fig:grid-ex} shows an example dataset
to illustrate the differences
between mass and range-based equipartitions
and their resulting count matrices.

For a fixed $D$ and $G$ of size $k\times\ell$, we write $D|_G$
(read as \textit{$D$ partitioned by $G$}) to denote
the joint distribution over the space $[k] \times [\ell]$,
with a probability mass function (PMF) given by the entries of $\P_{D, G}$.
Then the discrete mutual information of $D|_G$
(equivalently, of $\P_{D, G}$) is computed by
\begin{equation*}
I(D|_G) = I(\P_{D, G}) =
\sum_{i, j} p(i, j) \log_2 \frac{p(i ,j)}{p(i, *)p(*, j)}.
\end{equation*}
We then define the normalized $I^\star(D|_G)$ by
\begin{equation*}
  I^\star(D|_G) := \frac{I(D|_G)}{\log_2 \min\{k, \ell\}},
\end{equation*}
which ensures $I^\star(\cdot)$ is always in the range $[0, 1]$.

For a jointly distributed pair of random variables
$\Pi = (X, Y)$ and a $k \times \ell$ grid $G$,
we define $\Pi|_G$ and $\P_{\Pi, G}$ similarly,
where $\P_{\Pi, G}$ has entries
$p(i,j) = \Pr(X \in Q_j, Y \in P_i)$.
So $\Pi|_G$ is a discrete probability distribution
over $[k] \times [\ell]$,
and we compute $I^\star(\Pi|_G)$ as above.


\subsection{The MIC, MIC*, and MICe Statistics}

Given a dataset $D$, the $\mic$ statistic measures correlation
by finding the grid $G$ where $I^\star(D|_G)$ is maximized.
If the variables represented by $D$ have some relationship,
MIC will identify the grid containing the intervals where
each variable gives the most information about the other.

For any $D$ with $n$ points, it is easy to see that (assuming
no points with the same $x$ or $y$ value) there exists
an $n\times n$ grid $G$ that separates each point into
its own cell, which means $I^\star(D|_G) = 1$.
On the other hand, grids with too few total cells may
not be able to capture more complex relationships.
To negotiate this tradeoff, the $\mic$ statistic requires a
\textit{maximum grid size} parameter $B := B(n)$, where only
grids with at most $B(n)$ cells are considered,
but where $B(n)$ is expected to grow with $n$.
Defined formally:

\begin{definition}[$\mic$ statistic~\cite{reshef2011}]
  \label{def:mic}
  For a dataset $D$ of size $n$ and $B := B(n)$,
  let $\M^{\G}_D$ denote its
  \textit{characteristic matrix}
  with $(k, \ell)$ entries
  $\max_{G \in \G(k, \ell)} I^\star(D|_G)$.
  Then
  $\mic(D, B)
  = \max_{k,\ell:\; k\ell\le B(n)} \left(\M^{\G}_D\right)_{k, \ell}
  $\;.
\end{definition}

When the dataset $D$ can be modeled as a sample of points
drawn from a joint distribution $\Pi$,
then $\mic$ can be viewed as an \textit{estimator}
of the analogous $\micstar$ statistic defined
for the distributional setting:
\begin{definition}[$\micstar$ statistic~\cite{reshef16-mice}]
  For a jointly-distributed pair of random variables
  $\Pi = (X, Y)$, let $\M^{\G}_\Pi$
  denote its characteristic matrix with $(k, \ell)$ entries
  $\max_{G \in \G(k, \ell)} I^\star(\Pi|_G)$.
  Then $\micstar(\Pi) = \sup \M^\G_\Pi$.
\end{definition}
When $B(n) = O(n^\alpha)$ for some $\alpha \in (0, 0.5)$,
$\mic$ is also a \textit{consistent} estimator
of $\micstar$, meaning that
for a dataset $D_n$ of $n$ points drawn i.i.d. from $\Pi$,
$\mic(D_n, B)$ converges in probability
to $\micstar(\Pi)$ as $n\to \infty$
\cite{reshef16-mice, arxivnote}.
%
However, for every $k$, $\ell$ with $k\ell\le B(n)$,
$\mic$ is a maximization over the
set $\G(k, \ell)$ of \textit{all} grids with $k\ell$ cells,
which for datasets of size $n$ makes computing $\mic$ infeasible
in practice.

To this end, \citet{reshef16-mice}
introduced an efficiently-computable \textit{approximation}
of $\mic$ called $\mice$.
This newer statistic  approximates $\mic$ by
defining its characteristic matrix entries
as maximizations over \textit{dataset-dependent subsets}
of $\G(k, \ell)$.
Specifically, for a dataset $D = (D_x,  D_y)$, a constant $c > 0$,
and for all $2 \le k \le \ell$, the set
$\calE(D, c, k, \ell)$ contains all grids $G = (P, Q)$
where $Q$ is a size-$\ell$ mass equipartition of $D_x$
and $P$ is a size-$k$ subpartition of a size-$c\ell$
mass equipartition of $D_y$. The set is defined symmetrically
for $k > \ell$. Then $\mice$ is defined
similarly to $\mic$ but replaces the maximizations over
$\G(k, \ell)$ with $\calE(D, c, k, \ell)$:

\vspace*{0.5em}
\begin{definition}[$\mice$ statistic, \citet{reshef16-mice}]
  \label{def:mice}
  For any dataset $D$ of $n$ points, constant $c > 0$, and $B := B(n)$,
  let $\M^{\calE(D, c)}_D$ be the \textit{equicharacteristic} matrix
  with ($k, \ell$) entries
  $\max_{G \in \calE(D, c, k, \ell)} I^\star(D|_G)$.
  Then
  $$\mice(D, B, c) = \max\limits_{k,\ell:\; k\ell \le B(n)}
  \bigl(\M^{\calE(D, c)}_D\bigr)_{k, \ell}.
  $$
\end{definition}

Each ($k, \ell$) entry of the equicharacteristic matrix
can be reduced to a maximization defined only over
subpartitions on a \textit{single} axis.
This maximization can be computed efficiently
using the dynamic-programming algorithm $\optimizeaxis$
of \citet{reshef2011}.
%
Then in total, $\mice(D, B, c)$ can be computed
in $O(c^2 B^4)$ time for any $B$ and $c$,
and this becomes $O(c^2 n^{4\alpha})$
when $B(n) = O(n^{\alpha})$.
Here, the constant $c$ can be viewed as an
approximation parameter, where a larger value leads to a
better approximation of $\mic$,
but at the expense of slower computation\footnote{%
  Note that \citet{reshef16-mice} originally defined MICe
  with a more complicated dependence on the parameter $B$ to
  constrain the maximization space.
  The result is faster computation but with
  lower accuracy for a given $n$,
  though in practice
  the two variants are similar.
  We discuss this further in
  Appendix~\ref{sec:app:real-exps:imp}.
  }.
More details of computing $\mice$ are given
in Appendix~\ref{sec:app:mice-alg}.

In addition to its efficiency, $\mice$ is
still a consistent estimator of $\micstar$
when $B(n) = O(n^{\alpha})$ for $\alpha \in (0, 0.5)$
\cite{reshef16-mice, arxivnote}.
In practice, \citet{reshef2018empirical, reshef16-mice}
suggested using $\alpha$=0.6 and $c$=15
after evaluating the statistic on synthetic
and real data, and so we treat these as
its default settings.
Given its computational efficiency \textit{and}
consistency, we consider $\mice$ as a starting
point for developing private approximations
of $\mic$, but we first briefly recall
a few definitions and tools related to differential privacy.

\subsection{Differential Privacy}

For two datasets $D = (d_1, \dots, d_n)$ and
$D' = (d'_1, \dots, d'_m)$,
we say $D$ and $D'$ are \textit{neighboring}
(denoted $D \sim D'$)
if $n=m$ and there exists at most one index $i$ where $d_i \neq d'_i$.
Differential privacy ensures that the output
distributions of a (randomized) algorithm
are similar when run on $D \sim D'$:
\begin{definition}[\citet{dwork-roth}]
  For $\eps > 0$, a randomized algorithm
  $\mathcal{A}: \mathbb{R}^{2\times n} \to \R$
  is $\eps$-differentially private ($\eps$-DP) if, for every $D \sim D'$
  and every $I \subseteq \R$,
  $
  \Pr(\mathcal{A}(D) \in I) \le \exp(\eps) \Pr(\mathcal{A}(D') \in I) .
  $
\end{definition}
In our setting, using an $\eps$-DP mechanism to estimate $\mic$
implies that the output leaks little information about any
$d_i\in D$ (see \S~\ref{sec:discussion} for more on the privacy semantics).

One common approach for designing $\eps$-DP mechanisms is
to add random noise to the output of a non-private function.
When doing so, an important consideration is the \textit{sensitivity}
of the function, which is the maximum possible change
in function value over neighboring $D \sim D'$:

\begin{definition}[Sensitivity]
  The ($\ell_1$) \textit{sensitivity} of a function
  $f: \R^{2\times n} \to \R$ is
  $\max_{D \sim D'} | f(D) - f(D') |$.
\end{definition}

For a function $f$ with sensitivity $\Delta$,
the Laplace mechanism of \citet{dworklap}
adds noise from a zero-mean Laplace distribution with
parameter $\Delta/\eps$. The result is $\eps$-DP.

For convenience, we will write $\lap(b)$ to denote
a random variable with a zero-mean Laplace distribution
with parameter $b$,
which has density $f(x)=e^{-|x|/b}/(2b)$.
Also, for $x \in \R$, $[x]_{0, 1}$ denotes that $x$
is trunctated to the range $[0, 1]$.






\section{MICe-Lap Mechanism}
\label{sec:mice-lap}

We begin by considering the compatability of $\mice$
with the standard Laplace mechanism.
Doing so requires obtaining a bound on the sensitivitiy
of $\mice$, which for datasets of size $n$, $B(n)$,
and $c > 0$ we denote by $\Delta_n(\mice, B, c)$.
Our first result gives an upper bound
on this quantity.
\begin{restatable}{thm}{micesens}
  \label{thm:mice-sens}
  For any $B := B(n)$, $c > 0$, and $n \ge 6$,
  $
  \Delta_n(\mice, B, c)
  \;\le\;
  B \cdot ((2 \log_2 n)/n + 4.8/n).
  $
\end{restatable}
Using the suggested setting $B(n) = n^{\alpha}$
\cite{reshef2011, reshef16-mice},
$\Delta_n(\mice, B, c) = O( (\log_2 n)/n^{1-\alpha})$,
which is asymptotically larger than the  $O(\log_2 n /n)$
sensitivity of $\mic$ for $\alpha \in (0, 1)$
(this $\mic$ sensitivity bound can be obtained from the proof of
Theorem~\ref{thm:micr-sens}).
Theorem~\ref{thm:mice-sens} is proved in Appendix~\ref{sec:app:mice-sens:ub},
and the main intuition is that because $\mice$
maximizes over a dataset-dependent set of grids,
for $D\sim D'$, a worst-case change in $I^\star$ can occur when
the grids for $D$ have no cells in common with those of $D'$,
which changes $I^{\star}$ a bit for each of the at most $B$ cells.
%

This sensitivity bound then yields
the $\eps$-DP $\micrlap$ mechanism by adding Laplace noise and
truncating:
\begin{mech}[$\micelap$]
  For any dataset $D$ of size $n \ge 6$,
  $B := B(n)$, $c > 0$, and $\epsilon > 0$,
  let $\Delta := B\cdot((2 \log_2 n/n) + 4.8/n)$.
  Then
  $\micelap(D, B, c, \epsilon)
  := \left[\mice(D, B, c) + \lap(\Delta/\epsilon)\right]_{0, 1}$.
\end{mech}
\begin{restatable}{thm}{micelapdp} \label{thm:micelap-dp}
  $\micelap(\cdot, B, c, \epsilon)$ is $\epsilon$-DP.
\end{restatable} 
Theorem~\ref{thm:micelap-dp} is proved in Appendix~\ref{sec:app:mice-sens}.
Given that the Laplace sampling
can be done in constant time, the runtime
of computing $\micelap$ is asymptotically equivalent
to that of $\mice$ because only output noise is added.
%
%

Additionally, due to the $[0, 1]$ truncation,
the standard deviation of the mechanism is bounded
by that of $\lap(\Delta/\eps)$.
So for $B = n^{\alpha}$, the mechanism's standard deviation
is at most
$
  (\sqrt{2}/\eps) \cdot
  ((2 \log_2 n /(n^{1-\alpha})) + (4.8/n^{1-\alpha})) .
$
While for a fixed $\epsilon$ this value decreases with $n$,
using the suggested $\alpha$=0.6,
this quantity is 1.38 when $\eps$=1 and $n$=5000.
Given that the \mice{} value is in $[0,1]$,
the outputs of $\micelap$
have intolerably high error, even for
moderately large values of $\eps$ and $n$.
This result motivates designing an alternative
$\mic$ approximation with lower-error private variants.





\section{MICr Mechanisms}
\label{sec:micr}

\subsection{Non-private MICr Statistic}

Despite being efficiently computable and a consistent
estimator, the dataset dependence and
resulting high sensitivity of $\mice$ precludes
a straightforward, low-error private variant.
To that end, we introduce
the $\micr$ approximation for $\mic$.
It remains both efficiently computable
and a consistent estimator of $\micstar$, but it
optimizes over dataset-\textit{independent}
sets of grids, yielding a lower sensitivity. Indeed,
its sensitivity matches that of $\mic$,
making it more compatible with designing
differentially private variants.

We begin by describing the (non-private) $\micr$ statistic
before introducing two differentially private variants.
The key difference between $\mice$ and $\micr$ is
that the former maximizes over mass equipartitions
while the latter maximizes over \textit{range}
equipartitions.
Therefore, in addition to a maximum grid size parameter $B$
and a finite $c > 0$, $\micr$ also
takes as a parameter a range $L \subset \R^2$ of the
form $L = [x_0, x_1] \times [y_0, y_1]$.
When computing $\micr$, we require that the
coordinates of all points of $D$ lie within $L$.
This requirement can be satisfied by most types of data,
for example those with a defined range or using
conservative maxima and minima estimates.
$L$ must be dataset-independent.

For any $D$ restricted to range $L = L_x \times L_y$ and $c > 0$,
the dataset-independent sets of grids used to define
the entries of the analogous equicharacteristic matrix
for $\micr$ are defined as follows.
First, for an interval $I \subset \R$, let
$R_{I, \ell}$ denote the size-$\ell$ range equipartition
of $I$, and let $\calP(I, k, [j])$ denote the
set of all size-$k$ subpartitions of a
size-$j$ range equipartition of $I$.
Then for $k \le \ell$, we define
$\calR(L, c, k, \ell)$ as the set of all grids $G = (P, Q)$
where $Q = R_{L_x, \ell}$ and $P \in \calP(L_y, k, [c\ell])$.
When $k \le \ell$, we call the grid
$\Gamma_{L, c, \ell} = (R_{L_y, c\ell}, R_{L_x,\ell})$
the \textit{master} range equipartition for $\calR(L, c, k, \ell)$.
When $k > \ell$ , the set $\calR(L, c, k, \ell)$
and the master grid $\Gamma_{L, c, k}$ are defined symmetrically.
The full definition of $\micr$ then follows similarly
to $\mice$:

\vspace*{0.8em}
\begin{definition}[$\micr$ statistic]
  \label{def:micr}
  For any dataset $D$ of $n$ points with range restricted to $L$,
  $c > 0$, and $B := B(n)$, let
  $\M^{\calR(L, c)}_D$ denote the range equicharacteristic matrix
  with $(k, \ell)$'th entry
  $\max_{G \in \calR(L, c, k, \ell)} I^\star(D|_G)$.
  Then
  \begin{equation*}
    \micr(D, L, B, c) = \max\limits_{k, \ell:\; k\ell\le B(n)}
    \big(\M^{\calR(L, c)}_D\big)_{k, \ell}.
  \end{equation*}
\end{definition}

As with $\mice$, the $\micr$ statistic can be
computed in $O(c^2 B^{4})$ time using the $\optimizeaxis$ routine,
and $c$ can again be viewed as an approximation parameter.
We also prove that $\micr$ is still a consistent
estimator of $\micstar$ when $B(n) = O(n^{\alpha})$
for $\alpha \in (0, 0.5)$.

\vspace*{0.8em}
\begin{thm}
  \label{thm:micrcons:informal}
  $\micr$ is a consistent estimator of $\micstar$.
\end{thm}

Details of the computation and runtime of $\micr$
appear in Appendix~\ref{sec:appendix:micr-overview},
and a more formal statement of the consistency result
is given by Theorem~\ref{thm:micr-cons} in Appendix~\ref{sec:appendix:micr-cons}.

In addition to consistency, we leverage the fact
that the sets $\calR(L, c, k,\ell)$
are dataset-independent to prove an upper bound
on the $\ell_1$ sensitivity of $\micr(\cdot, L, B, c)$,
which for datasets of size $n$ we denote by
$\Delta_n(\micr, L, B, c)$.

\begin{restatable}{thm}{micrsens}
  \label{thm:micr-sens}
  For any $L$, $c > 0$, $B := B(n)$, and $n \ge 4$,
  $\Delta_n(\micr, L, B, c)
  \le
  (4 \log_2 n)/n + 6/n$.
\end{restatable}

Compared to the sensitivity bound for $\mice$,
the bound here loses the multiplicative $B$ factor.
This is because the sets of grids considered by $\micr$ are
\textit{fixed} for all datasets of the same size, and
for any $G$ and $D \sim D'$, the count
matrices $\A_{D, G}$ and $\A_{D', G}$ can differ by at most 1
at exactly 2 entries.
The proof of the theorem is developed in
Appendix~\ref{sec:appendix:micr-sens}.

Equipped with the fact that $\micr$ is both a
consistent estimator of $\micstar$ \textit{and} has low sensivity,
$\micr$ is a more suitable base statistic for designing high-utility private variants.
To this end, because two classical approaches to designing differentially
private mechanisms (\textit{output perturbation}, as with the Laplace mechanism,
and \textit{input perturbation}, as with private histograms \cite{dwork-roth})
appear to have equal potential for achieving this task, we designed
two private variants, each one following a different approach.

\subsection{MICr-Lap Mechanism}

The first private variant is
analogous to $\micelap$, but using the smaller
$\micr$ sensitivity bound from Theorem~\ref{thm:micr-sens}.

\vspace*{1em}
\begin{mech}[$\micrlap$]
  \label{mech:micr-lap}
  For any dataset $D$ of size $n \ge 4$ restricted to $L$,
  any $B := B(n)$, any $c > 0$ and any $\epsilon > 0$, let
  $\Delta := (4 \log_2 n)/n + 6/n.$
  Then
  \begin{align*}
  &\micrlap(D, L, B, c, \epsilon) := \\
    \bigskip
    &\hspace*{6em}
      \left[\micr(D, L, B, c) + \lap(\Delta/\epsilon)\right]_{0, 1}.
  \end{align*}
\end{mech}

\begin{restatable}{thm}{micrlapdp}
  \label{thm:micrlap-dp}
  $\micrlap(\cdot, L, B, c, \epsilon)$ is $\epsilon$-DP.
\end{restatable}

Again using the standard deviation of the Laplace mechanism,
for any $B$, the standard deviation of $\micrlap$ is at most
$
(\sqrt{2}/\eps)\cdot
((4 \log_2 n)/n + 6/n) .
$
When $n$=5000 and $\eps$=1 this value is 0.02, which
means (as we show in Section~\ref{sec:exps}) that
the error of the private output is likely small enough
for the mechanism to be useable in practice.

Also, because for fixed $\eps$ the standard deviation
is decreasing with $n$, and using the consistency of
$\micr$, it is straightforward to show that
$\micrlap$ is still a consistent estimator of $\micstar$,
and we prove this in Theorem~\ref{thm:micrlap-cons}
in Appendix~\ref{sec:appendix:micr-lap}.



\subsection{MICr-Geom Mechanism}
\label{sec:micrgeom}

Our second private variant, $\micrgeom$, adds noise
during the computation of each entry in
the $\micrgeom$ range-equicharacteristic matrix.

Recall for subpartitions $G \subseteq \Gamma$ that
the count matrix $\A_{D, G}$ (and thus $\P_{D, G}$) can
be generated via the function $\psi(\A_{D, \Gamma}, \Gamma, G)$,
which doesn't depend directly on the coordinates of $D$.
Then the $k, \ell$ entry of the
range-equicharacteristic matrix for $\micrgeom$
is a maximization of $I^\star$ over grids $G \in \calR(L, c, k, \ell)$,
but where the count matrix for each grid is generated
via $\psi(\hatA, \Gamma, G)$, where $\hatA$ is
some \textit{noisy} approximation
of $\A_{D, \Gamma}$ and $\Gamma := \Gamma_{L, c, \ell}$
is the master range-equipartition for $\calR(L, c, k, \ell)$.

Specifically, as established earlier,
for any neighboring $D \sim D'$ of size $n$,
at most two corresponding entries of
$\A_{D, \Gamma}$ and $\A_{D', \Gamma}$ can differ,
and they differ by at most 1.
Thus we generate a noisy version of $\A_{D, \Gamma}$
by producing independent, noisy estimates for each of its entries
using the $\epsilon$-DP Truncated Geometric mechanism of \citet{ghosh12},
which applies to counts with sensitivity at most 1.
Intuitively, the $\truncgeom(\epsilon, n, f)$ is
a doubly-geometric distribution centered at $f$
with parameter $e^{-\epsilon}$, and with
truncation at $f$ and $n-f$.
We summarize the distribution here:
\begin{definition}[$\truncgeom$, \cite{ghosh12}]
  \label{def:truncgeom}
  For any $\epsilon > 0$,  $n$, and $0 \le f \le n$,
  let $\truncgeom(\epsilon, n, f)$ be a discrete distribution
  over $\{0, \dots, n\}$. Set $\rho :=  e^{-\epsilon}$. Then:
  \begin{itemize}[leftmargin=1em]\vspace*{-1em}
  \setlength\itemsep{-0.1em}
  \item[-]
    $\truncgeom(\epsilon, n, f) = 0$ w.p.
    $\rho^f/ (1 + \rho)$.
  \item[-]
    $\truncgeom(\epsilon, n, f) = i$ w.p.
    $((1-\rho)/(1+\rho)) \rho^{|f-i|}$
    for all $1 \le i \le n - 1$.
  \item[-]
    $\truncgeom(\epsilon, n, f) = n$ w.p.
    $\rho^{(n-f)}/ (1 + \rho)$.
  \end{itemize}
\end{definition}

We then define the $\micrgeom$ mechanism formally and give
its privacy guarantee as follows:

\begin{mech}[$\micrgeom$]
  \label{mech:micr-geom}
  Fix any dataset $D$ of size $n$ restricted to
  $L$, any $c > 0$, any $B := B(n)$, and any $\epsilon > 0$.
  \begin{itemize}[leftmargin=1.1em]\vspace*{-1em}
    \setlength\itemsep{-0.1em}
  \item[1.]
    For every $k, \ell \ge 2$, let $\Gamma$ denote the
    master range-equipartition grid for $\calR(L, c, k, \ell)$.
    Let $\hatA := \hatA^\eps_{D, \Gamma}$ be the \textit{noisy}
    count matrix whose $(i, j)$ entry is given by
    $\truncgeom(\eps/2, n, a(i, j))$
    (where $a(i, j)$ is the corresponding entry of
    $\A_{D, G}$).
    
    Let $\hatn$ be the sum of entries in $\hatA$,
    and let $\hatP = (1/\hatn) \cdot \hatA$.
  \item[2.]
    For every $G \in \calR(L, c, k, \ell)$,
    let $\hatP_G := \psi(\hatP, \Gamma, G)$.
  \item[3.]
    Let $\hatM^{\calR(L, c)}_{D, \epsilon}$ be the \textit{noisy}
    range-equicharacteristic matrix with $(k, \ell)$ entry
    $\max_{G \in \calR(L, c, k, \ell)} I^\star(\hatP_G)$. Then:
  \end{itemize}\vspace*{-1em}
  $$\micrgeom(D, L, B, c, \epsilon)
  = \max\limits_{k, \ell:\; k \ell \le B(n)} \big(\hatM^{\calR(L, c)}_{D, \epsilon}\big)_{k, \ell}.$$
\end{mech}

\begin{restatable}{thm}{micrgeomdp}
  \label{thm:micrgeom-dp}
  $\micrgeom(D, L, B, c, \epsilon)$ is $\epsilon$-DP.
\end{restatable}

With a linear-time additive preprocessing step,
samples from the $\truncgeom$ distribution can be done
in constant time, and thus the running time of computing
$\micrgeom$ is asympotically equivalent to $\micr$
and $\micrlap$. The details of both the
privacy statement and the runtime analysis are
given in Appendices~\ref{sec:app:micrgeom-privacy}
and~\ref{sec:app:micrgeom-comp} respectively.

Additionally, we prove that for a fixed $\eps$ and $c$,
the error introduced from using the noisy count matrices
$\A^{\eps}_{D, \Gamma}$ goes to $0$ as $n$ grows:

\begin{restatable}[Added error of $\micrgeom$]{thm}{micrgeomerr}
  \label{thm:micrgeom-err}
  Fix any $\alpha \in (0, 0.5)$, finite $c > 0$,
  $\epsilon > 0$, and dataset $D$ of size $n$.
  For sufficiently large $n$, there exists
  some $a > 0$ such that
  \vspace*{-0.6em}
  \begin{equation*} \vspace*{-0.6em}
    \left| \big(\widehat \M^{\calR(L, c)}_{D, \epsilon}\big)_{k, \ell}
    - \big(\M^{\calR(L, c)}_{D}\big)_{k, \ell} \right|
    = O((c/\epsilon)n^{-a})
  \end{equation*}
  for all $k \ell \le B(n) = O(n^{\alpha})$ simultaneously
  with probability at least $1 - O(n^{-2})$.
\end{restatable}

Intuitively, the dependence on $c$ in the error bound
is a result of having a master range-equiparitition size of
$c\ell^2$ (wlog when $k\le \ell$).
So choosing larger $c$ yields a better approximation of $\mic$
but results in a ``noisier'' $\hatA$ (and thus a slower
convergence to the non-private $\micr$). We explore this tradeoff
more experimentally in Section~\ref{sec:exps}.
Together with the fact that $\micr$ is
a consistent estimator of $\micstar$ (Theorem~\ref{thm:micr-cons}),
it is straightforward to prove that $\micrgeom$ is
also a consistent estimator of $\micstar$.
We prove this in Theorem~\ref{thm:micrgeom-cons}
in Appendix~\ref{sec:appendix:micr-geom-cons}.



\section{Experimental Evaluation}
\label{sec:exps}

To investigate their utility, we evaluated the three
private mechanisms on both synthetic and real datasets.
We used synthetic data (following the methodology of
\citet{reshef2011, reshef16-mice})
to help better understand the error of our mechanisms
at growing sample sizes over a \textit{variety} of
distributions designed to capture different types of relationships.
Using real data helps to additionally verify the utility
of our mechanisms in practice at specific, fixed sample sizes.

The code and data used to obtain our experimental results
can be accessed at \url{https://github.com/jlazarsfeld/dp-mic},
and more implementation details are given in
Appendix~\ref{sec:app:real-exps:imp}.

\subsection{Synthetic Data}
\label{sec:exps:synth}

We considered the family of 21 functional
relationships introduced by \citet{reshef2011}.
Similar to their later work \cite{reshef16-mice},
for every relationship we defined
9 joint distributions, each generated by placing $k$=100 independent
bivariate Gaussian distributions (with zero correlation
and identical variances)
centered at points evenly-spaced along the function graph.
The 9 distributions were parameterized by an $R^2$ value
in $\{0.1, 0.2, \dots, 0.9\}$ that determined the variances of
each Gaussian. 
The result is a set $\mathcal{Q}$ of 189 joint distributions
bounded in range by $[0, 1] \times [0, 1]$,
each representing a functional relationship with varying levels of noise.
For each $\Pi \in \mathcal{Q}$, we computed
an approximation of $\micstar(\Pi)$ using the (provably convergent)
method of \citet{reshef16-mice}.
For $n$ ranging from 25 to 10,000 and $\eps \in \{0.1, 1.0\}$,
we measured the accuracy of each mechanism (wrt $\micstar(\Pi)$)
on datasets of $n$ points sampled i.i.d. from $\Pi$.
The full details of this dataset generation process
are included in Appendix~\ref{sec:app:synth-exps}.

Given that $\micelap$ simply adds noise to $\mice$,
we set its parameters to $B(n) = n^{0.6}$ and $c=15$,
matching the suggested settings for $\mice$.
However, because $\micrlap$ and $\micrgeom$ use
a different base statistic, we evaluated
these mechanisms with varying $B$ and $c$ to better determine
optimal parameter settings for
a fixed $n$ and $\epsilon$.

\paragraph{Parameter Tuning for MICr-Geom, MICr-Lap:}
We considered sample sizes of
$n \in$ \{25, 250, 500, 1000, 5000, 10000\},
$\epsilon \in \{0.1, 1.0\}$, and
various values of $B$ between 4 and 150.
Although the consistency guarantees for
$\micrgeom$ and $\micrlap$ are phrased in terms
of $B := B(n) = O(n^\alpha)$, defining
$B$ in absolute terms helps us better determine
optimal values of $B$ for each mechansim.
For computational considerations, we fixed $c=5$ for
the $\micrlap$ mechanism (we found the
mechanism to be insensitive to larger values of $c$),
and for the $\micrgeom$ mechanism we considered
$c \in \{1, 2\}$ (under the consideration that
larger $c$ could worsen accuracy).

For each combination of $(n, \epsilon, B, c)$,
distribution $\Pi \in \calQ$, and mechanism,
we ran 50 iterations of the following process: (1)
construct a dataset $D$ by sampling $n$ points i.i.d.
from $\Pi$, and (2) run the private mechanism on $D$.
For each mechanism, $n$, and $\epsilon$,
we minimized an objective function
over the $(B, c)$ parameters that involved
a weighted sum of the mechanism's
average absolute error (wrt $\micstar$)
across all distributions in $\mathcal{Q}$.
The objective function was designed to
choose parameters that could ensure
parity in error across both low and high-correlation distributions,
and the exact description is given in Appendix~\ref{sec:app:synth-exps}.

The optimized $B$ and  $c$ parameters for $\micrlap$
and $\micrgeom$ are summarized in Table~\ref{tab:opt-BC}
in Appendix~\ref{sec:app:synth-exps}.
For both mechanisms and $\epsilon$,
the optimal $B$ values are generally increasing with $n$,
which aligns with the intuition that the mechanisms
converge toward $\micstar$ with larger $n$.
The optimal values ranged from $12$ to $150$
for $B$ and $1$ to $5$ for $c$.

\paragraph{Bias/Variance Evaluation:}
Using the parameters from Table~\ref{tab:opt-BC}
for $\micrlap$ and $\micrgeom$ and $(B(n), c)=(n^{0.6}, 15)$
for $\micelap$, for each mechanism we compared the bias
(average signed error wrt to $\micstar$)
and variance (of the 50 private iterations for a fixed distribution)
over \textit{all} 189 distributions in $\calQ$ as $n$ grows.
The results for $\eps$=1 are summarized in Figure~\ref{fig:synthetic-bv},
and analogous plots for $\eps$=0.1 are
given in Figure~\ref{fig:synthetic-bv:eps10} of Appendix~\ref{sec:app:synth-exps}.
In both subplots in Figure~\ref{fig:synthetic-bv},
for every value of $n$, we show boxplots of the bias (resp. variance)
for each mechanism over \textit{all} $\Pi \in \cal{Q}$.
Recall each box represents the interquartile range (IQR, 25'th to 75'th quantiles)
of the data, the whiskers appear above and below these quantiles by
an additional 1.5x of the IQR,
and outlier points beyond the whiskers are plotted individually.
\begin{figure}[t!]
  \vspace*{1em}
  \begin{center}
    \includegraphics[width=\linewidth]{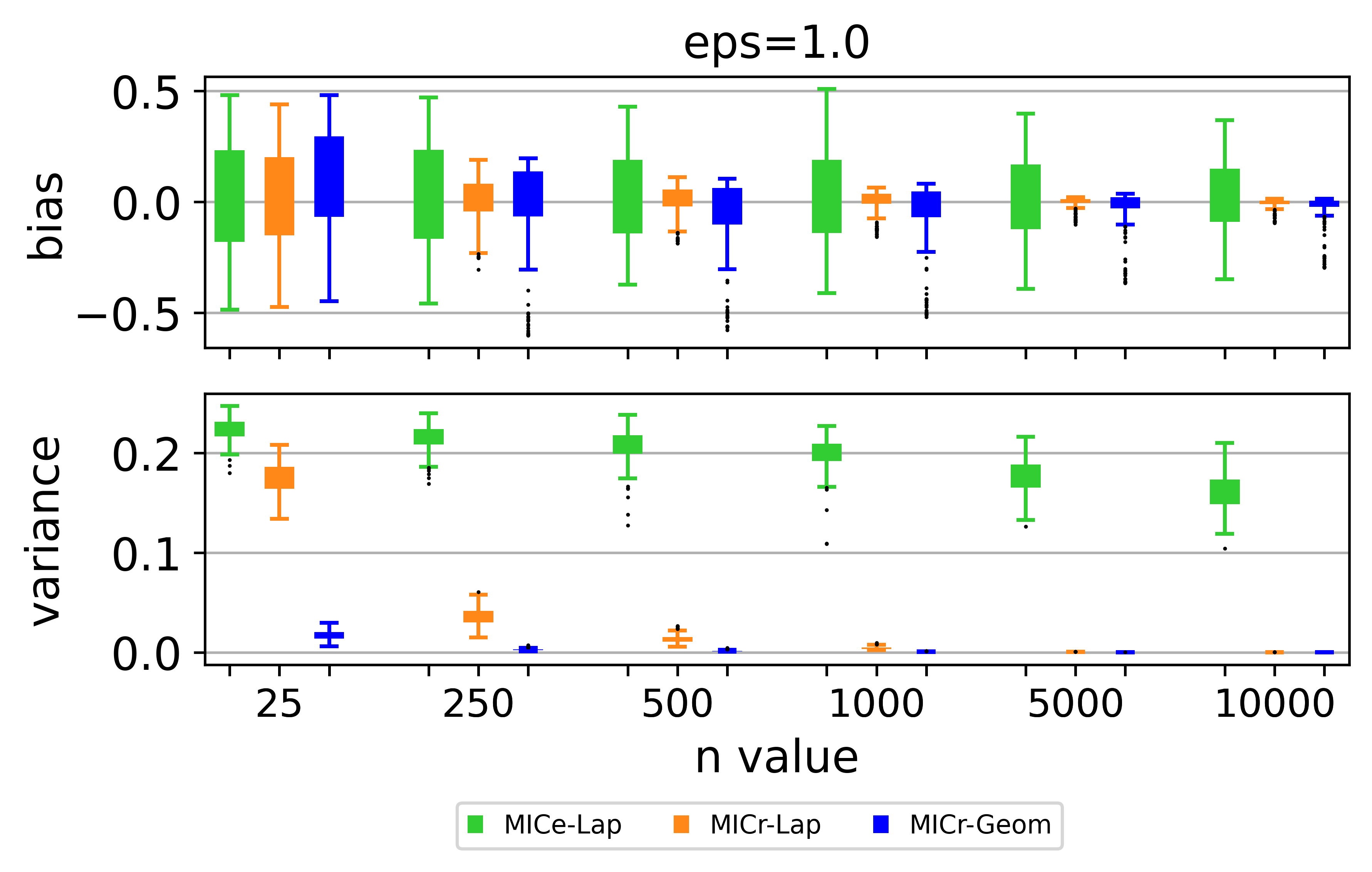}
  \end{center}\vspace*{-1.5em}
  \caption{\small{%
      Boxplots of the bias and variance
      of each mechanism (over 50 iterations)
      over all $\Pi \in \mathcal{Q}$ for $\eps$=1
      and varying $n$.
    }} \vspace*{-1em}
  \label{fig:synthetic-bv}
\end{figure}

For $\eps$=1, the IQR of the bias of each mechanism
generally drops with $n$, and this is most prominent
for $\micrlap$ and $\micrgeom$.
In particular, when n=5000, the median bias of
$\micrlap$ is 0.01 with min and max biases of
-0.1 and 0.02, which we consider tolerably low.
On the other hand, while the IQR of
the bias of $\micrgeom$ at $n$=5000 is also small,
the outlier points indicate that the mechanism has
large \textit{negative} bias on a subset of distributions in $\mathcal{Q}$.
Specifically, while the median and max bias of $\micrgeom$
at $n$=5000 are 0.01 and and 0.04 respectively,
the min bias is -0.37.
Because the $\micrgeom$ mechanism generates noisy counts
for \textit{every} cell of a master grid,
we found this negative bias to occur on datasets
for which the non-noisy master count matrix
contains a large submatrix with mostly zero entries,
and this corresponds to datasets restricted to $L$
that have large sub-regions with no points.
When $\eps$=0.1, Figure~\ref{fig:synthetic-bv:eps10} (Appendix~\ref{sec:app:synth-exps})
shows similar trends for the bias of each mechanism,
but where decreases with $n$ are slower.

Additionally, the variances of $\micrlap$ and $\micrgeom$
are significantly smaller than for $\micelap$.
For $\micrgeom$ this variance is particularly low, even for small $n$.
For example, for $n$=250 and $\eps$=1, the
median variance of $\micrgeom$ is 0.003 and these medians
decrease for larger $n$.
In general, we observe (especially for smaller $n$)
a bias/variance tradeoff between $\micrlap$ (less bias, more variance)
and $\micrgeom$.

\subsection{Real Data}

We also evaluated the utility of the three private mechanisms on
two sets of data used in the experiments of \citet{reshef2011}:
the Spellman data and the Baseball data.
Note that these datasets do not necessarily contain sensitive
information, and these sources were chosen mainly because of their
previous use by \citet{reshef2011}.
Because both sets contain multiple columns,
we constructed for each source a \textit{collection} of datasets
corresponding to different pairs of columns,
and we first describe this process in more detail.


\paragraph{Dataset Description:}
The Spellman data \cite{spellman, reshef2011} contains gene expression
measurements for $4381$ genes in the yeast organism,
where each gene has a timeseries (at common, fixed time points)
of $n$=23 measurments. Modeling the methodology of \citet{reshef2011},
for each gene, we consider the dataset $D = ([23], T_i)$,
where $T_i$ is the timeseries for the $i$'th gene.
For each $i$, we let $l_i = |\max T_i - \min T_i|$,
$y_0 = (\min T_i) - l_i/100$, $y_1 = (\max T_i) + l_i / 100$,
and we use the range bounds $L_i = [0, 24] \times [y_0, y_1]$.
This results in a collection of $m = 4381$ datasets $D$,
each of size $n=23$, and we refer to this as the Spellman23 collection.

Because the size of each dataset in Spellman23 is small (n=23),
and given that we expect the error of our mechanisms to
decrease with larger $n$, we also constructed a collection of
higher-dimensional datasets from the Spellman source as follows:
for each fixed time index $t \in [23]$, let $C_t$ denote
the set of measurements for all $4381$ genes at time index $t$.
Then for each unique pair $t, v \in [23]$,
we constructed the dataset $D = (C_{t}, C_{v})$.
We set global range bounds by considering the
maximum and minimum values across all $T_i$,
denoted by $\ell_0$ and $\ell_1$ respectively,
and by setting $x_0 = \ell_0 - |\ell_1 - \ell_0|/ 100$,
$x_1 = \ell_1 + |\ell_1 - \ell_0|/100$ and $L = [x_0, x_1] \times [x_0, x_1]$
for all $i$.
The result is a collection of $m = 253$ datasets,
each of size $n=4381$. We refer to this as the Spellman4381 collection.

Finally, the Baseball dataset \cite{reshef2011, baseball2}
contains values of $129$ performance-related statistics for
$337$ players from the 2008 MLB season. Again following \citet{reshef2011},
for the $i$'th statistic, we let $C_i$ denote the set of corresponding
values across all $337$ players, and for each unique pair $i, j \in [129]$,
we constructed the dataset $D = (C_i, C_j)$.
We generated range bounds
$[x_0, x_1]$ for $C_i$ and $[y_0, y_1]$ for $C_j$
using the same process as for each $T_i$ in Spellman23.
We then set $L = [x_0, x_1] \times [y_0, y_1]$.
The result is a collection of $m = 8256$ datasets of size
$n$=337, which we refer to as the Baseball collection.

For the purposes of these experiments, the range bounds $L$ for
each collection of datasets were constructed manually.
However, we assume in general that practitioners with
domain-specific knowledge can set reasonable bounds for $L$
\textit{without} observing a particular dataset.
For example, a practitioner with baseball acumen could
set the range for the statistic ``Number of Games Played''
to $[-1, 183]$, since there are 182 games in an MLB season.

\begin{table}[tb!]
  \centering
  \small
  \vspace*{1em}
\begin{tabular}{r | rrr}
  & $\micelap$
  & $\micrlap$
  & \micrgeom \\
  \hline
  Spellman23
  & 0.18 \;(0.23)
  & 0.14 \;(0.17)
  & 0.24 \;(0.01) \\
  Baseball
  & 0.30 \;(0.21)
  & 0.02 \;(0.02)
  & 0.06 \;(9e-4) \\
  Spellman4381
  & 0.26 \;(0.17)
  & -0.01 \;(4e-4)
  & 0.02 \;(1e-4)
\end{tabular}
\caption{The median bias (average signed error wrt $\mice$ over 100 runs)
  and median variance (over 100 iterations) of each private mechanism
  across all datasets of each collection for $\eps$=1.
} 
\label{tab:real-bv}
\end{table}

\begin{figure}[tb!]
  \vspace*{1em}
  \begin{center}
    \includegraphics[width=0.95\linewidth]{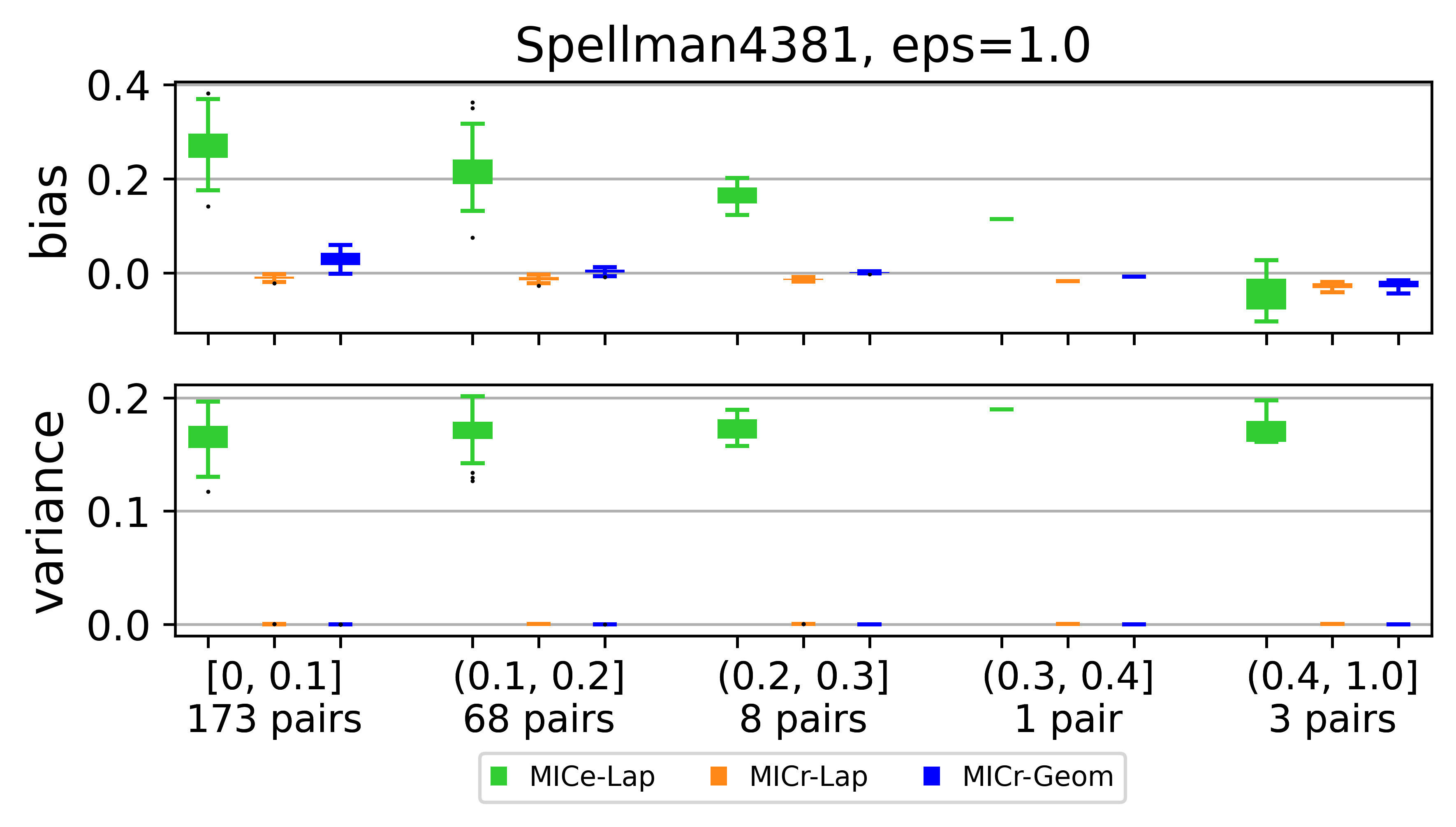}
    \includegraphics[width=0.95\linewidth]{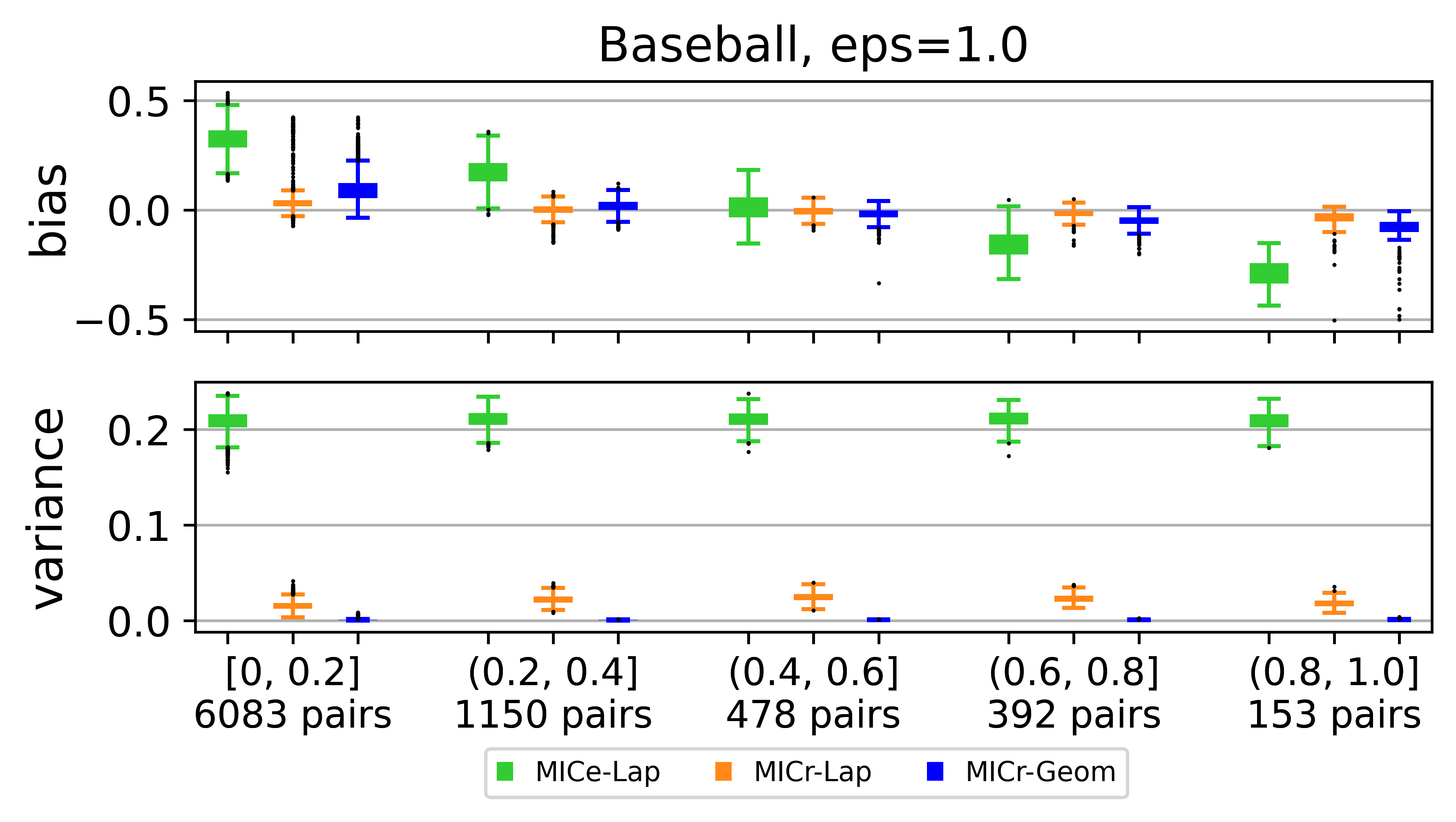}
  \end{center}\vspace*{-1em}
  \caption{{\small%
      Bias and variance boxplots for each mechanism
      over datasets (pairs) in the Spellman4381 collection (top)
      and Baseball collection (bottom) binned by
      non-private $\mice$ score for $\eps$=1.
    }}
  \label{fig:real-bw}
\end{figure}

\paragraph{Evaluation of Private Error:}
For each dataset in the three collections
described above, we measured the error of all private mechanisms
with respect to the dataset's non-private $\mice$ score
using parameters $B(n) = n^{0.6}$ and $c=15$.
Although $\micr$ shares similar theoretical properties
as $\mice$, we primarily
view $\micr$ as a conduit for private approximations of $\mic$.
Thus measuring private error with respect to
the exisiting $\mice$ statistic allows for a more
realistic evaluation of our mechanisms' utility.

For every dataset, we ran 100 computations
of each private mechanism for $\eps$=1 and $\eps$=0.1
(we mainly describe here the results for $\epsilon$=1
but the corresponding tables and figures for $\eps$=0.1
can be found in Appendix~\ref{sec:app:real-exps}).
For $\micrgeom$ and $\micrlap$, we set the $B$ and $c$
parameters by linearly interpolating values
from Table~\ref{tab:opt-BC}.
For $\micelap$, we set $B(n) = n^{0.6}$ and
$c = 15$ (matching the settings from the
synthetic data evaluation).

For each collection of datasets, Table~\ref{tab:real-bv}
lists the median bias (average signed error wrt $\mice$)
and median variance (over 100 iterations) of each private mechanism
over \textit{all} datasets in the collection.
For both $\micrlap$ and $\micrgeom$, the median bias
drops significantly for the Baseball and
Spellman4381 datasets compared to Spellman23.
Similar to our evaluation on synthetic data, this
supports the intutition that the error incurred
due to the privatization of $\micr$ (in both mechansims)
decreases with $n$.
Additionally, although $\micelap$ has relatively small
median bias for Spellman23,
its median average \textit{unsigned} error was the largest
(0.46 compared to 0.39 and 0.25 for $\micrlap$ and $\micrgeom$,
respectively).
While the median average unsigned errors for $\micrlap$ and $\micrgeom$
on Spellman4381 (0.016 and 0.019) and on Baseball (0.097 and 0.068)
are low, Table~\ref{tab:real-bv} shows these mechanisms still incur
large error on Spellman23, which indicates that none of
the private mechanisms are accurate enough for datasets
as small as $n$=23.


On the other hand, to better understand the
practicality of the mechansims in the higher-dimensional regime,
we further analyze their error on datasets in
the Baseball and Spellman4381 collections,
with a particular focus on how error varies
for low and high-correlation datasets
(using low and high $\mice$ as a proxy).
To this end, each dataset of the Spellman4381 collection
is \textit{binned} according to its non-private $\mice$ score
using interval endpoints $[0, 0.1, 0.2, 0.3, 0.4, 1]$.
Similarly, each dataset in the Baseball collection is binned
by its $\mice$ score using interval endpoints
$[0, 0.2, 0.4, 0.6, 0.8, 1]$.
For both collections, we analyzed the bias and variance of each
mechanism across all pairs in a \textit{fixed bin}.
Figure~\ref{fig:real-bw} displays bias and variance boxplots
for each mechanism using $\eps$=1 for both collections,
and similar plots for $\eps$=0.1 are
provided in Appendix~\ref{sec:app:real-exps}.

For both collections, the IQRs for bias
for both $\micrlap$ and $\micrgeom$
in every bin are smaller and with medians closer to zero than
those of $\micelap$.
In addition, in both collections the variances for $\micrlap$ and $\micrgeom$
are significantly smaller across all bins compared to $\micelap$,
and in general we see a bias/variance tradeoff
between $\micrlap$ and $\micrgeom$ similar to the synthetic data.

For $\micrgeom$, we also notice
on both collections that for datasets with lowest $\mice$ scores,
the mechanism tends to have a slightly more \textit{positive} bias
(medians of 0.03 and 0.09 for Spellman4381 and Baseball) compared to
$\micrlap$ (medians of -0.01 and 0.03).
This positive bias of $\micrgeom$ likely occurs because generating
private counts using $\truncgeom$ on datasets with
low $\mice$ scores (which have more ``uniform'' scatterplots),
can \textit{reduce} some of the noise in the dataset.
In contrast, the slight \textit{negative} bias of $\micrgeom$
and $\micrlap$ on datasets with the largest $\mice$ values
(in both collections, but particularly for Baseball)
is likely due to underlying approximation differences between
(non-private) $\mice$ and $\micr$.

Also, while some outlier datasets with magnitudes of bias as
large as 0.5 exist in the Baseball collection
for both $\micrlap$ and $\micrgeom$, they are most prevalent for
datasets with the smallest and largest $\mice$ scores
(left and right-most bins).
This means it is less likely that the output of
either mechanism on one of these datasets would
suggest a highly-correlated pair (say, output above 0.5)
when the true $\mice$ score is low (or vice versa).
Moreover, for the Spellman4381 collection,
these slight positive and negative biases are less apparent,
and the min and max biases of -0.04 and -0.002 for $\micrlap$
and -0.04 and 0.06 for $\micrgeom$ across \textit{all} bins
indicates that for $n$ as small as 4381, the error of
these two mechanisms is sufficiently small to be useable in practice.


\section{Related Work}

$\mic$ is one of several recently proposed statistics
for detecting non-linear dependencies between variables
in the non-private setting.
Other works that propose measures like the Randomized Dependence Coefficient
\cite{rdc} and the $\xi_n$ coefficient \cite{chatterjeePeerReviewed}
compare results directly with $\mic$ and claim slightly improved
computational properties and more intuitive dependence measurements for certain
classes of distributions.
However, $\mic$ and $\mice$ have been
more thoroughly investigated
\cite{reshef16-mice, reshef2018empirical, reshef20theory}
and analyzed \cite{kinney2014, reshef2014cleaning}
and has found wide use in practice, particularly
in bioinformatics settings \cite{albanese, cao2021}. 

While no private variants of competing measures
of dependence have been introduced, much work
has been devoted to differentially private estimation
of other statistical properties,
for example mean estimation \cite{dp-mean},
hypothesis selection \cite{dp-hypothesis},
quantiles \cite{dpquantiles}, and false discovery
rate control \cite{paprika}. 

We contribute to both lines of work by
introducing low-error, private variants for measuring
non-linear dependencies.






\section{Discussion and Future Work} \label{sec:discussion}

In practice, publishing MIC values privately requires a broad consideration of
the analysis process. If multiple statistics are published, then each must be
allocated some fraction of an overall privacy budget. Moreover, choosing which
statistics to publish based on the data may leak information, and thus the
selection process itself should be differentially private. To optimize accuracy,
we recommend that the parameters $B$ and $c$ be optimized for the size $n$ of
the dataset of interest and the desired privacy level $\epsilon$, though linear
interpolation across precomputed values can be used for speed. The privacy
semantics of the results depend on what each data point represents. For example,
we obtain user-level privacy if distinct individuals contribute each point
and event-level privacy if points represent different events.

There are several promising directions to improve and apply this work. We have
not yet considered how to use the local sensitivity~\cite{nissim2007smooth} of a
given dataset to improve the accuracy of a DP MIC statistic. Also, it may be possible to
remove the requirement for \micr{} that the range of the data be known in advance by
estimating it in a differentially private way. Moreover, we note that the noisy
master grid of \micrgeom{} is already differentially private, and so
more-accurate methods to estimate \mic{} from this object
seem possible, potentially by
exploiting the regular pattern of bias we observed experimentally
(where bias decreases with non-private \mice{}).

\section*{Acknowledgments}
This work was supported by the Office of Naval Research.
The authors would like to thank Joan Feigenbaum for
advising and supporting the early stages of this work,
Argyris Oikonomou and Grigoris Velegkas for
their contributions in an initial analysis
of the sensitivity of MIC, and Yakir Reshef for
helpful feedback on the consistency of MICe.

\newpage
\bibliography{dpmic-refs}
\bibliographystyle{icml2022}

\appendix
\onecolumn

\vspace*{-2em}
\begin{center}
  \Large{\textbf{Appendices}}
\end{center}


\section{MICe and MICe-Lap Details}
\label{sec:app:mice-sens}

This section provides more details on the $\mice$
and $\micelap$ statistics and develops
the proofs of Theorem~\ref{thm:mice-sens} (MICe sensitivity)
and Theorem~\ref{thm:micelap-dp} (MICe-Lap privacy). 

\subsection{Computing MICe}
\label{sec:app:mice-alg}

We first give more details on computing MICe
(which also provides the foundation for computing MICr).

For any $c > 0$ and $B := B(n)$, $\mice$ can be computed efficiently
using the $\optimizeaxis$ routine of \citet{reshef2011}:
for a fixed dataset $D$ of size $n$ and a fixed (wlog) $k \le \ell$,
$\optimizeaxis$ takes as
input a fixed column partition $\Gamma_x$ of size $\ell$,
a fixed master row partition $\Gamma_y$ of size $\hat k$, and
simultaneously outputs the size-$k$ subgrid $P_k \subseteq \Gamma_y$
(where $k \le \hat k$)
that maximizes $I^\star(D|_{(P_k, \Gamma_x)})$ for all
$k \le \min\{\ell, B(n)/\ell\}$
in $O((\hat k)^2 k \ell)$ time. 

We refer the reader to Appendix 3 of \citet{reshef2011}
for exact implementation details of $\optimizeaxis$,
which uses a dynamic-programming-based approach to perform
the maximization.
Using this subroutine, the
runtime of computing $\mice$ from Definition~\ref{def:mice}
can be stated as follows:

\vspace*{0.8em}
\begin{thm}[\citet{reshef16-mice}]
  \label{thm:mice-runtime}
  For any dataset $D$ of size $n$,  $B := B(n)$, and $c > 0$, 
  $\mice(D, B, c)$ can be computed using $\optimizeaxis$
  in $O(c^2 B^4)$ time. 
\end{thm}

\vspace*{-1em}
\begin{proof}
  As stated in Definition~\ref{def:mice}, for a fixed $\ell$,
  for every $k \le \ell$, the $(k, \ell)$'th entry of the equicharacteristic matrix
  uses a master row mass equipartition of size $\hat k = c \cdot \ell$, 
  for some fixed $c > 0$.
  Then for each $\ell = 2, \dots, B/2$, the $\optimizeaxis$ routine
  runs in $O(c^2 \ell^2 \cdot k \ell) = O(c^2 \ell^2 B)$ time,
  since $k\ell \le B$. 
  In sum, running this routine for each $\ell = 2, \dots B/2$
  (to compute all equicharacteristic matrix
  entries where $\ell \ge k$) takes $O(c^2 B^4)$ time.
  By symmetry, computing all equicharacteristic matrix entries where
  $k > \ell$ also takes $O(c^2 B^4)$ time, and thus the
  theorem statement follows.
\end{proof}

\subsection{MICe Sensitivity Upper Bound}
\label{sec:app:mice-sens:ub}

We now develop the proof of Theorem~\ref{thm:mice-sens},
which bounds the $\ell_1$-sensitivity of the $\mice$ statistic.

First, recall from Section~\ref{sec:prelims} that
for a fixed $k \le \ell$ and for a fixed $D$ of $n$ points,
$\mathcal{E}(D, c, k, [\ell])$ is the set of all grids $G$
with $k$ rows and $\ell$ columns where the corresponding
distribution $D|_G$ has mass-equipartitioned columns.
Equivalently, for $G \in \mathcal{E}(D, c, k, [\ell])$,
the count matrix $\A_{D, G}$ is of size $k \times \ell$,
and its column sums are all equal\footnote{%
  In the case that $n/\ell$ is not whole, let the first
  $\ell-1$ columns of $\A_{D, G}$ contain
  $\left\lceil (n/\ell)\right\rceil$
  points each, and let $\ell$'th column contain the remainder
  of the points.
  Additionally, we assume that the coordinates of
  every point in $D$ are unique, since we can perturb
  the position of each point by a small amount without
  affecting the output of the $\mice$ statistic.
}.
For $k > \ell$, the set $\mathcal{E}(D, c, [k], \ell)$
is analogoulsy defined as the set of all $k \times \ell$
grids $G$ where $D|_G$ has equipartitioned rows.

Recall also from Section~\ref{sec:prelims} that for a dataset
$D$ of $n$ points, a maximum grid size parameter $B := B(n)$,
and a constant $c > 0$, we define
\begin{equation}
  \mice(D, B, c)
  \;=\;
  \max_{k\cdot\ell \le B} \; \left(\M^{\calE(D, c)}_D \right)_{k, \ell} \;, 
  \label{eq:mice-sens:intro:def1}
\end{equation}
where $\M^{\calE(D, c)}_D$ is the equicharacteristic matrix of $D$.

Then for any $(k, \ell)$ pair, the $(k, \ell)$ entry of $\M^{\calE(D, c)}_D$ is 
computed by
\begin{align}
  \left(\M^{\calE(D, c)}_D \right)_{k, \ell}
  \;=\;
  \begin{cases}
    \begin{split}
      &\max_{G \in \mathcal{E}(D, c,  k, [\ell])}
      \; \frac{I(D|_G)}{\log_2 k}
      \;\;\;\; \text{if $k \le \ell$}  \\
      \\
      &\max_{G \in \mathcal{E}(D, c, [k], \ell)}
      \; \frac{I(D|_G)}{\log_2 \ell}
      \;\;\;\; \text{if $k > \ell$} 
      \;\;\; .
    \end{split}
    \label{eq:mice-sens:intro:def2}
  \end{cases}
\end{align}


Fix $k \le \ell$ (which we assume without loss of generality,
since the analogous arguments when $k > \ell$ follow by symmetry). 
Because the set $\mathcal{E}(D,c, k, [\ell])$ depends on $D$, it follows
that $\mathcal{E}(D, c, k, [\ell]) \neq \mathcal{E}(D',c, k, [\ell])$
for datasets $D \neq D'$. 
This means that in \eqref{eq:mice-sens:intro:def2},
the values $(\M^{\calE(D, c)}_D)_{k, \ell}$ and
$(\M^{\calE(D', c)}_{D'})_{k, \ell}$ are defined as maximizations
over nonequal constraint sets. 

To derive an upper bound on the sensitivity of $\mice(\cdot, B, c)$,
it is helpful to define these two maximizations using a \textit{common}
constraint set --- one that depends on $n$, but not on $D$ or $D'$.
To this end, we will proceed by developing a new, equivalent formulation
for entries in $\M^{\calE(D, c)}_D$, where the constraint set depends only on $n$.

\subsubsection{Alternate Formulation of Equicharacteristic Matrix}

For a $k \times \ell$ grid $G$ and a fixed dataset $D$ of size $n$,
recall that $D|_G$ is a joint probability distribution
over the discrete space $\{1, \dots, k\} \times \{1,\dots, \ell\}$.
So the distribution can be represented by 
the count matrix $\P_{D, G} \in \R^{k \times \ell}$, where
\begin{equation*}
  \P_{D, G} \;=\; \frac{1}{n} \cdot \A_{D, G} \; .
\end{equation*}
Since by definition $\sum_{i, j} \; (\A_{D,G})_{i,j} = n$, we must have
$\sum_{i, j} \; (\P_{D, G})_{i,  j} = 1$.
For readability, when $D$ and $G$ are fixed and clear from context,
we will drop the subscripts and write $\A$ and $\P$. 

The row and column marginal distributions of $D|_G$ can also
be represented in vector form.
First, letting $(\a_{D, G})_{i, *}$ and $(\a_{D, G})_{*, j}$ denote
the $i$'th row vector and $j$'th column of $\A_{D, G}$ respectively,
we define $\r_{D, G} \in \mathbb{Z}^k$ and $\c_{D, G} \in \mathbb{Z}^\ell$ as
\begin{align*}
  &\r_{D, G}
    \;=\;
    \left(
    \| (\a_{D,G})_{1, *} \|_1,
    \dots,
    \| (\a_{D, G})_{k, *} \|_1
    \right) \\
  &\c_{D, G}
    \;=\;
    \left(
    \| (\a_{D, G})_{*, 1} \|_1,
    \dots,
    \| (\a_{D, G})_{*, \ell} \|_1
    \right)  \;,
\end{align*}
where $\| \r_{D, G}\|_1 = \| \c_{D, G} \|_1 = n$.
Then the row and column marginal distributions of $D|_G$
are given by $(\r_{D, G}/n)$ and $(\c_{D, G}/n)$ respectively. 

Now observe that by definition,
the mutual information $I(D|_G)$
depends only on the joint distribution $D|_G$, and not on
the individual coordinates of points in $D$.
Additionally, because we can represent $D|_G$
using the matrix $\P = \tfrac{1}{n} \cdot \A$,
the joint distribution $D|_G$ depends only on
the entries of $\A$, and not on the absolute
positions of the column and row dividers of $G$. 

It follows trivially that for a fixed dataset
$D$ of size $n$ and two different $k\times \ell$
grids $G$ and $G'$, if $\A_{D, G} = \A_{D, G'}$, then
$\r_{D, G} = \mathbf{r'}_{D, G}$, 
$\c_{D, G} = \mathbf{c'}_{D, G}$,
and  $\P_{D, G} = \P_{D, G'}$.
Moreover, for a fixed $D$,
if $\r_{D, G} = \r_{D, G'}$
and $\c_{D, G} = \c_{D, G'}$ then
$\A_{D, G} = \A_{D, G'}$ and 
$\P_{D, G} = \P_{D, G'}$
(i.e., if $D|_G$ and $D|_{G'}$ have the same marginal row
and column distributions, then $D|_G = D|_{G'}$).

Together, these two observations imply that for a fixed $D$ and any
two grids $G$ and $G'$, the two distributions $D|_G$ and
$D|_{G'}$ are equal if and only if $\r_{D, G} = \mathbf{r'}_{D, G}$
and $\c_{D, G} = \mathbf{c'}_{D, G}$.
We summarize these observations and equivalences with the following
lemma.

\begin{restatable}{lem}{micesensframework}
  \label{lem:mice-sens:framework}
  Fix a dataset $D$ of $n$ points, and fix any $k$ and $\ell$.
  For any grids $G$ and $G'$ of size $k \times \ell$, the following
  statements are all equivalent:
  \begin{align*}
    D|_G \;&=\; D|_{G'}
           \tag{i}
           \label{eq:framework:i} \\
    \A_{D, G} \;&=\; \A_{D, G'}
                  \tag{ii}
                  \label{eq:framework:ii} \\
    \P_{D, G} \;&=\; \P_{D, G'}
                  \tag{iii}
                  \label{eq:framework:iii} \\
    \r_{D, G} = \r_{D, G'}
    \;\;&\text{and}\;\;
    \c_{D, G} = \c_{D, G'} \;.
    \tag{iv}
    \label{eq:framework:iv}
  \end{align*}
\end{restatable}

\begin{proof}
  Statements \eqref{eq:framework:i} and \eqref{eq:framework:ii}
  are equivalent by the definition of a count matrix $\A$, and
  \eqref{eq:framework:iii} is equivalent to \eqref{eq:framework:ii}
  by the definition of the $\P$.
  Statement \eqref{eq:framework:ii} also implies \eqref{eq:framework:iv}
  by the definition of the vectors $\r$ and $\c$.

  To complete the proof, we show that \eqref{eq:framework:iv} implies
  \eqref{eq:framework:ii}.
  For this, notice that for a fixed $D$, the entries of $\A_{D, G}$
  can be computed given the vectors $\r_{D, G}$ and $\c_{D, G}$.
  For example, the entry $(\A_{D, G})_{1, 1}$ is the number of points in
  $D$ whose $y$-coordinate has descending rank at most $\r_{D, G}(1)$,
  and whose $x$-coordinate has ascending rank at most $\c_{D, G}(1)$;
  the entry $(\A_{D, G})_{1, 2}$ is the number of points in $D$
  whose $y$ coordinate has descending rank at least
  $\r_{D, G}(1) + 1$ and at most $\r_{D, G}(1) + 1 + \r_{D, G}(2)$,
  and whose $x$ coordinate has ascending rank at most $\c_{D, G}(1)$, etc.

  Thus for a fixed $D$, the entries of $\A_{D, G}$ can be computed
  as a function of $D$ and the vectors $\r_{D, G}$ and $\c_{D, G}$,
  but without a direct dependence on the positions of the column and row
  dividers of $G$.
  It then follows that for a fixed $D$ and any two grids $G$ and $G'$
  of the same size, if $\r_{D, G} = \r_{D, G'}$ and
  $\c_{D, G} = \c_{D, G'}$, then $\A_{D, G} = \A_{D, G'}$, which
  completes the proof. 
\end{proof}

Observe that for a fixed $n$ and $k$ and $\ell$, for every
dataset $D$ of size $n$, and for every $G \in \mathcal{E}(D, c, k, [\ell])$,
the vectors $\c_{D, G} \in \mathbb{Z}^\ell$ are all identical. 
This follows from the fact that for any $D$ of size $n$, the count
matrices $\A_{D, G}$ for any $G \in \mathcal{E}(D, c, k, [\ell])$ have row sums
that depend only on $n$ and $\ell$, and not on $D$. 
Let $\c^{n, [\ell]} \in \mathbb{Z}^\ell$ denote this unique vector.

Additionally, observe that there are only finitely many vectors
$\r \in \mathbb{Z}^k$ satisfying $\| \r \|_{1} = n$.
Let $\mathcal{R}(n, k)$ denote the set of all such vectors
where $T(n, k) = |\mathcal{R}(n, k)|$, and fix some enumeration
$\mathcal{R}(n, k) = \{\r^1, \r^2, \dots, \r^{T(n, k)}\}$.

So for a fixed dataset $D$ of size $n$,  
\begin{equation}
  \forall \; G \in \mathcal{E}(D, c, k, [\ell]) \;\;
  \left(\;
    \exists \; \text{unique }
    i \in [T(n, k)]
    \;\text{such that}\;
    \r_{D, G} = \r^i
    \;\right)
  \;,
  \label{eq:mice-sens:fact1}
\end{equation}
which follows directly from the definition of $\r_{D, G}$.
So for any $i \in [T(n,k)]$, for any two $k \times \ell$
grids $G$ and $G'$ such that $\r_{D, G} = \r_{D, G'} = \r^i$,
it follows by Lemma~\ref{lem:mice-sens:framework} that
$D|_G = D|_{G'}$
(since we also have $\c_{D, G} = \c_{D, G'} = \c^{n, [\ell]}$),
and thus $I(D|_G) = I(D|_{G'})$.

In the other direction, it is easy to see that
\begin{equation}
  \forall \; i \in [T(n, k)]\;\;
  \left(\;
    \exists \;
    G \in \mathcal{E}(D,c, k, [\ell])
    \;\text{such that}\;
    \r_{D, G} = \r^i
    \;\right)
  \;,
  \label{eq:mice-sens:fact2}
\end{equation}
since for any $i$, such a grid $G$ can be constructed
by sorting the points of $D$ by their $y$ coordinate
and drawing a row divider between certain numbers of
consecutive points as specified by $\r^i$.
  
For a fixed dataset $D$ of size $n$, and for any $\r^i \in \mathcal{R}(n, k)$,
let $\P\left(D,\r^i, \c^{n, [\ell]}\right) \in \mathbb{R}^{k \times \ell}$
be the probability matrix derived from the vectors
$\r^i$ and $\c^{n, [\ell]}$ (i.e., as described in the
proof of Lemma~\ref{lem:mice-sens:framework}).
Then by \eqref{eq:mice-sens:fact2}, for every $i \in [T(n, k)]$,
there exists some $G \in \mathcal{E}(D, k, [\ell])$ such that
$\P(D, \r^i, \c^{n, [\ell]}) = \P_{D, G}$ and thus
$I(\P(D, \r^i, \c^{n, [\ell]})) = I(D|_G)$. 

Finally, by combining the preceding arguments, we state
the following reformulation of entries $(\M^{\calE(D, c)}_D)_{k, \ell}$.

\begin{restatable}{lem}{micesensreform}
  \label{lem:mice-sens:reform}
  For any fixed dataset $D$ of size $n$,
  and for any $k \le \ell$:
  \begin{equation}
    \left(\M^{\calE(D, c)}_D\right)_{k, \ell}
    \;\;=\;
    \max_{G \in \mathcal{E}(D,c,k, [\ell])}
    \;\frac{I(D|_G)}{\log_2 k}
    \;\;=\;\;
    \max_{i \in T(n, k)}
    \;\frac{I\left(
        \P\left(
          D, \r^i, \c^{n, [\ell]}
        \right)
      \right)
    }{\log_2 k} \;.
    \label{eq:mice-sens:new-form}
  \end{equation}
\end{restatable}

Crucially, for any dataset $D$ of size $n$ and a fixed $k\le\ell$,
the constraint set in the new formulation of
Lemma~\ref{lem:mice-sens:reform} depends only on $n$ and $k$,
and \textit{not} on $D$.
This yields the following useful observation, which we state as a Corollary:

\begin{cor}
  \label{cor:mice-sens:M-ineq}
  For any $n$, for any $2 \le k \le \ell$, and for any two datasets $D, D'$,
  both of size $n$:
  \begin{equation}
    \left|\;
      \left(\M^{\calE(D, c)}_D\right)_{k, \ell}
      - \left(\M^{\calE(D', c)}_{D'}\right)_{k, \ell}
    \;\right|
    \;\;\le\;\;
    \max_{i \in T(n, k)}\;
    \left|\;
    I\left(
      \P(D, \r^i, \c^{n, [\ell]})
    \right)
    \;-\;
    I\left(
      \P(D', \r^i, \c^{n, [\ell]})
    \right)
    \;\right| \;\;.
  \end{equation}
\end{cor}

\begin{proof}
  By Lemma~\ref{lem:mice-sens:reform}, we have
  \begin{align*}
    \left|\;
    \left(\M^{\calE(D, c)}_D\right)_{k, \ell}
      - \left(\M^{\calE(D', c)}_{D'}\right)_{k, \ell}
    \;\right|
    &\;=\;
    \left|\;
    \max_{i \in T(n, k)}\;
    \frac{I\left(
      \P(D, \r^i, \c^{n, [\ell]})
    \right)}{\log_2 k}
    \;\;-\;\;
    \max_{i \in T(n, k)}\;    
    \frac{I\left(
      \P(D', \r^i, \c^{n, [\ell]})
    \right)}{\log_2 k}
            \;\right| \\
    &\;\le\;
    \left|\;
    \max_{i \in T(n, k)}\;
    I\left(
      \P(D, \r^i, \c^{n, [\ell]})
    \right)
    \;\;-\;\;
    \max_{i \in T(n, k)}\;    
    I\left(
      \P(D', \r^i, \c^{n, [\ell]})
    \right)
    \;\right| \;,
  \end{align*}
  where the inequality follows from factoring out the common $\log_2 k$ term
  and noting that $\log_2 k \ge \log_2 2 = 1$ by assumption.

  Given that the constraint sets in each maximization term are equal,
  we then have
  \begin{align*}
    \left|\;
      \max_{i \in T(n, k)}\;
      I\left(
        \P(D, \r^i, \c^{n, [\ell]})
      \right)
      \;-
      \max_{i \in T(n, k)}\;    
      I\left(
        \P(D', \r^i, \c^{n, [\ell]})
      \right)
      \;\right| \\
    \;\le\;
    \max_{i \in T(n, k)}\;
    \left|\;
    I\left(
      \P(D, \r^i, \c^{n, [\ell]})
    \right)
    \;-\;
    I\left(
      \P(D', \r^i, \c^{n, [\ell]})
    \right)
    \;\right|
  \end{align*}
  which completes the proof.   
\end{proof}

\subsubsection{Sensitivity of $\mice$ via Sensitivity of Mutual Information}
\label{sec:mice-sens:via-mi}

Using the new formulation from Lemma~\ref{lem:mice-sens:reform}
we will derive an upper bound on the sensitivity of $\mice$ by
deriving an upper bound on the sensitivity of mutual information
with respect to a fixed $i \in T(n, k)$.

For this, note that for any pair of neighboring datasets $D \sim D'$
of size $n$, and for any $B := B(n)$ and $c > 0$:
\begin{align}
  |\; \mice(D, B, c) - \mice(D', B, c) \;|
  &\;=\;
    \left|\;
    \max_{k,\ell \;:\;  k\cdot \ell \le B}
    \Big(\M^{\calE(D, c)}_D\Big)_{k, \ell}
    - \max_{k,\ell \;:\;  k\cdot \ell \le B}
    \left(\M^{\calE(D', c)}_{D'}\right)_{k, \ell}
    \;\right| \nonumber \\
  &\;\le\;
    \max_{k,\ell \;:\;  k\cdot \ell \le B}
    \; \left|\; 
    \left(\M^{\calE(D, c)}_D\right)_{k, \ell}
    - 
    \left(\M^{\calE(D', c)}_{D'}\right)_{k, \ell}
    \;\right| .
    \label{eq:mice-sens:ub-main:1}
\end{align}

Now by Corollary~\ref{cor:mice-sens:M-ineq}, we know (assuming wlog that
$k \le \ell$) that
\begin{equation}
  \left|\; 
    \left(\M^{\calE(D, c)}_D\right)_{k, \ell}
    - 
    \left(\M^{\calE(D', c)}_{D'}\right)_{k, \ell}
    \;\right|
  \;\le\;
    \max_{i \in T(n, k)}\;
    \left|\;
      I\left(
        \P(D, \r^i, \c^{n, [\ell]})
      \right)
      \;-\;
    I\left(
      \P(D', \r^i, \c^{n, [\ell]})
    \right)
    \;\right| .
  \label{eq:mice-sens:ub-main:2}
\end{equation}
We further claim the following bound:
\begin{restatable}{lem}{micesensmibound}
  \label{lem:mice-sens:mi-bound}
  For any $n \ge 6$ and any $2 \le k \le \ell$,
  and any neighboring datasets $D \sim D'$ both of size
  $n$:
  \begin{equation*}
    \max_{i \in T(n, k)}\;
    \left|\;
      I\left(
        \P(D, \r^i, \c^{n, [\ell]})
      \right)
      \;-\;
      I\left(
        \P(D', \r^i, \c^{n, [\ell]})
      \right)
      \;\right|
    \;\le\;
    k\ell \cdot
    \left(
      \frac{2 \log_2 n}{n}
      + \frac{4.8}{n}
    \right) \;.
  \end{equation*}
\end{restatable}

Let us grant Lemma~\ref{lem:mice-sens:mi-bound} as true
for now. The proof of Theorem~\ref{thm:mice-sens} then follows
easily:
\begin{proof}[Proof (of Theorem~\ref{thm:mice-sens})]
  Applying Lemma~\ref{lem:mice-sens:mi-bound}
  to \eqref{eq:mice-sens:ub-main:2} and
  substituting into \eqref{eq:mice-sens:ub-main:1} gives us
  \begin{equation}
    |\; \mice(D, B, c) - \mice(D', B, c) \;|
    \;\le\;
    \max_{ k, \ell \;:\;k\cdot\ell \le B}\;\;
    k\ell \cdot
    \left(
      \frac{2 \log_2 n}{n}
      + \frac{4.8}{n}
    \right) \;,
    \label{eq:mice-sens:ub-main:3}
  \end{equation}
  which holds for any $D \sim D'$ of size $n$. 
  Observing that $k\cdot\ell \le B$ completes the proof. 
\end{proof}

It remains to prove Lemma~\ref{lem:mice-sens:mi-bound}, which
we proceed to do in the following subsection.

\subsubsection{Proof of Lemma~\ref{lem:mice-sens:mi-bound}}

Fix any $n \ge 6$ and any $2 \le k \le \ell$, and
consider any fixed pair of neighboring datasets $D \sim D'$, both
of size $n$.

Fix any $\r^i \in \mathcal{R}(n, k)$, and let
\begin{align*}
  \P &\;:=\; \P(D, \r^i, \c^{n, [\ell]})
       \;\;\;\text{with entries $p(i, j)$ and column sums $p(*, j)$}\\
  \P' &\;:=\; \P(D', \r^i, \c^{n, [\ell]})
        \;\;\text{with entries $p'(i, j)$ and column sums $p'(*, j)$}
        \;.
\end{align*}
Let $\A$ and $\A'$ denote the corresponding count matrices for
$\P$ and $\P'$, respectively. 
Because $\P$ and $\P'$ are non-negative matrices representing
two-dimensional, discrete joint distributions, we define
information-theoretic properties of $\P$ and $\P'$ in the natural way.
Specifically, let
\begin{align*}
  I(\P)
  &\;=\;
    H_x(\P) - H_{x|y}(\P) \\
  H_x(\P)
  &\;=\;
    \sum_{j \in [\ell]}\;
    p(*, j) \log_2 \left(\frac{1}{p(*, j)}\right) \\
  H_{x|y}(\P)
  &\;=\;
    \sum_{i \in [k]} \sum_{j \in [\ell]}\;
    p(i, j) \log_2 \left(\frac{p(*, j)}{p(i, j)}\right) \;.
\end{align*}
The analogous definitions for $\P'$ are defined in terms of
$p'(i, j)$ and $p'(*, j)$.

Our goal is to give an upper bound on
$|I(\P) - I(\P')|$, which we can write as
\begin{align*}
  |I(\P) - I(\P')|
  &\;=\;
    \left|\;
    H_x(\P) - H_{x|y}(\P)
    \;-\;
    \left(
    H_x(\P') - H_{x|y}(\P')
    \right)
    \;\right| \\
  &\;=\;
    \left|\;
    H_x(\P) - H_x(\P')
    \;+\;
    H_{x|y}(\P') - H_{x|y}(\P)
    \;\right| \;.
\end{align*}
Because $\P$ and $\P'$ are both generated using
$\c^{n, [\ell]}$, it follows by definition that
$p(*, j) = p'(*, j)$ for all $j \in [\ell]$,
and thus $H_x(\P) = H_x(\P')$.
Then
\begin{align}
  |I(\P) - I(\P')|
  &\;=\;
    \left|\;
    H_{x|y}(\P') - H_{x|y}(\P)
    \;\right| \nonumber\\
  &\;=\;
    \left|\;
    \sum_{i \in [k]} \sum_{j \in [\ell]}\;
    p'(i, j) \log_2 \left(\frac{p'(*, j)}{p'(i, j)}\right)
    \;-\;
    \sum_{i \in [k]} \sum_{j \in [\ell]}\;
    p(i, j) \log_2 \left(\frac{p(*, j)}{p(i, j)}\right)
    \;\right| \nonumber\\
  &\;\le\;
    \sum_{i \in [k]} \sum_{j \in [\ell]}\;
    \left|\;
    p'(i, j) \log_2 \left(\frac{p'(*, j)}{p'(i, j)}\right)
    \;-\;
    p(i, j) \log_2 \left(\frac{p(*, j)}{p(i, j)}\right)
    \;\right| \nonumber\\
  &\;=\;
    \sum_{i \in [k]} \sum_{j \in [\ell]}\;
    \left|\;
    p'(i, j) \log_2 p'(*, j)
    - p'(i, j) \log_2 p'(i, j)
    - p(i, j) \log_2 p(*, j)
    + p(i,j) \log_2 p(i, j)
    \;\right| \nonumber\\
  &\;=\;
    \sum_{i \in [k]} \sum_{j \in [\ell]}\;
    \left|\;
    (p'(i, j) - p(i,  j)) \log_2 p(*, j)
    \;+\;
    p(i,j) \log_2 p(i, j)
    - p'(i, j) \log_2 p'(i, j)
    \;\right| \;.
    \label{eq:mice-sens-ub:alpha-sum}
\end{align}
Here, the third line is due to the triangle inequality, and
the final line is due to $p(*, j) = p'(*, j)$ for all $j$. 

For any $(i, j) \in [k] \times [\ell]$, let
\begin{equation*}
  \alpha(i, j)
  \;=\;
  \left|\;
    (p'(i, j) - p(i,  j)) \log_2 p(*, j)
    \;+\;
    p(i,j) \log_2 p(i, j)
    - p'(i, j) \log_2 p'(i, j)
    \;\right| \;.
\end{equation*}

The following lemma gives a uniform upper bound on $\alpha(i, j)$
for any $(i, j)$:

\begin{lem}
  \label{lem:mice-sens:alpha-bound}
  For any $n \ge 6$, any $2 \le k \le \ell$, and any $\P$ and $\P'$
  as described above,
  \begin{equation*}
    \alpha(i, j)
    \;\le\;
    \frac{2 \log_2 n}{n} + \frac{4.8}{n}
  \end{equation*}
  for any $(i, j) \in [k] \times [\ell]$. 
\end{lem}

Granting this lemma true for now and applying it to expression
\eqref{eq:mice-sens-ub:alpha-sum} then implies
\begin{equation*}
  |\; I(\P) - I(\P') \;|
  \;\le\;
  \sum_{i \in [k]}
  \sum_{j \in [\ell]}
  \; \frac{2 \log_2 n}{n} + \frac{4.8}{n}
  \;\le\;
  k\ell \cdot
  \left(
    \frac{2 \log_2 n}{n} + \frac{4.8}{n}
  \right) \;. 
\end{equation*}
Because we have considered an arbitrary
vector $\r^i \in \mathcal{R}(n, k)$, the
statement of Lemma~\ref{lem:mice-sens:mi-bound} follows.

Thus it only remains to prove Lemma~\ref{lem:mice-sens:alpha-bound}. 

\begin{proof}[Proof (of Lemma~\ref{lem:mice-sens:alpha-bound})]
  For a fixed grid $G \in \mathcal{E}(D, c, k, [\ell])$,
  let $\phi_x$ and $\phi'_x$ denote the
  $x$-coordinate point mapping functions wrt $G$
  for $D$ and $D'$ respectively
  (i.e., for $d \in D$, the function $\phi_x(d)$ returns
  the column partition index of $d$ wrt $D|_G$).
  Then observe that for any $(i, j) \in [k] \times [\ell]$,
  we have
\begin{equation*}
  |\; p(i, j) - p'(i, j) \;| \le 2/n \;.
\end{equation*}
This is because for every column $j \in [\ell]$, there is at most one point
$d \in D$  where $\phi_x(d) = j$ but $\phi'_x(d) \in \{j-1, j + 1\}$
(equivalently, at most one point $d' \in D'$ where $\phi'_x(d') \in \{j-1, j+1\}$
but $\phi_x(d) = j$),
The same argument holds for every row $i \in [k]$.
Thus $\A$ and $\A'$ differ by at most $2$ at any entry $(i, j)$,
and so $|p(i, j) - p'(i,j)| \le 2/n$ for all $(i, j)$. 

Using this fact, we will derive an upper bound on $\alpha(i, j)$ for
any fixed $(i, j)$ via three cases:
(1) when $p(i, j) = p'(i, j)$,
(2) when $p(i, j) > p'(i, j)$,
and (3) when $p(i, j) < p'(i, j)$.
For ease of readability, and when $(i, j)$ is fixed and clear from context,
we we will write $p := p(i, j)$ and $p' := p'(i, j)$.
Thus for fixed $(i,j)$ we seek to bound
\begin{equation*}
  \alpha
  \;=\;
  |\;
  (p' - p) \log_2 p(*, j)
  + p \log_2 p
  - p' \log_2 p'
  \;| \;.
\end{equation*}
We have the following cases:
\begin{enumerate}
\item
  \textit{Case when $p = p'$:}
  
  We trivially have that $\alpha = 0$.

\item
  \textit{Case when $p > p'$:}

  \begin{enumerate}
  \item
    \textit{Case when $p - p' = 1/n$}: \medskip

    Note that $p - p' = 1/n$ implies $p(*, j) \ge p \ge 1/n$.
    If $p = 1/n$ (and $p' = 0$), then
    \begin{align*}
      \alpha
      \;=\;
      \left|\;
      -\frac{1}{n} \log_2 p(*, j)
      + \frac{1}{n} \log_2 \frac{1}{n}
      \;\right|
      &\;=\;
        \left|\;
        \frac{1}{n}
        \log_2 \frac{1/n}{p(*, j)}
        \;\right| \\
      &\;=\;
        \left|\;
        \frac{1}{n}
        \log_2 \frac{p(*, j)}{1/n}
        \;\right| \;.
    \end{align*}
    Since $1/n \le p(*, j) \le 1$, it follows that
    $\alpha \le (\log_2 n) /n$. \medskip

    On the other hand, say $p \ge 2/n$. Then
    \begin{align}
      \alpha
      &\;=\;
        \left|\;
        -\frac{1}{n} \log_2 p(*, j)
        + p \log_2 p - (p - 1/n) \log_2 (p - 1/n)
        \;\right| \nonumber\\
      &\;=\;
        \left|\;
        p \log_2 \frac{p}{p - 1/n}
        + \frac{1}{n} \log_2 \frac{p - 1/n}{p(*, j)}
        \;\right| \nonumber\\
      &\;\le\;
        \left|\;
        p \log_2 \frac{p}{p - 1/n}
        \;\right|
        +
        \left|\;
        \frac{1}{n} \log_2 \frac{p - 1/n}{p(*, j)}
        \;\right| \;,
        \label{eq:case2a:main}
    \end{align}
    where the last line is due to the triangle inequality.
    Observe that the left hand term in \eqref{eq:case2a:main}
    is decreasing in $p$, and since $p \ge 2/n$ by assumption:
    \begin{align}
      \left|\;
        p \log_2 \frac{p}{p - 1/n}
        \;\right|
      \;\le\; 
      \left|\;
        \frac{2}{n} \log_2 \frac{2/n}{2/n - 1/n}
        \;\right|
      \;=\;
      \frac{2}{n}
      \label{eq:case2a:lhs}
    \end{align}
    For the right hand term in \eqref{eq:case2a:main}, note that
    $p - 1/n = p' \le p(*, j)$ (since the joint mass in any cell $(i, j)$
    is at most the marginal mass of that cell's column), so
    this term is also decreasing in $p$. Again since $p \ge 2/n$ by
    assumption, we have
    \begin{align}
      \left|\;
      \frac{1}{n} \log_2 \frac{p - 1/n}{p(*, j)}
      \;\right|
      \;\le\;
      \left|\;
      \frac{1}{n}
      \log_2 \frac{2/n - 1/n}{p(*, j)}
      \right|\;
      \;=\;
        \left|\;
        \frac{1}{n}
        \log_2 \left(n \cdot p(*, j)\right)
        \right|\;
      \;\le\;
        \frac{\log_2 n}{n} \;,
        \label{eq:case2a:rhs}
    \end{align}
    where the final inequality holds because $2/n \le p(*, j) \le 1$.\medskip
    
    Substituting \eqref{eq:case2a:lhs} and \eqref{eq:case2a:rhs}
    back into \eqref{eq:case2a:main} gives
    $\alpha \le (2/n) + (\log_2 n)/n$ when $p \ge 2/n$.
    Thus this bound holds for Case (2a) in general. \medskip

  \item
    \textit{Case when $p - p' = 2/n$:} \medskip

    The condition $p - p' = 2/n$ implies $p(*, j) \ge p \ge 2/n$.
    First consider when $p = 2/n$ and $p'=0$. Then
    \begin{align*}
      \alpha
      \;=\;
      \left|\;
      -\frac{2}{n}
      \log_2 p(*, j)
      + \frac{2}{n} \log_2 \frac{2}{n}
      \;\right|
      &\;=\;
        \left|\;
        \frac{2}{n}
        \log_2 \frac{p(*, j)}{2/n}
        \;\right| \\
      &\;=\;
        \left|\;
        \frac{2}{n}
        \log_2 \frac{n}{2}
        \;\right|
        \;\le\;
        \frac{\log_2 n}{n} \;,
    \end{align*} 
    where the penultimate inequality follows from
    the assumption that $2/n \le p(*, j) \le 1$. \medskip

    On the other hand, consider when $p \ge 3/n$. Then
    \begin{align}
      \alpha
      &\;=\;
        \left|\;
        -\frac{2}{n} \log_2 p(*, j)
        + p \log_2 p - (p - 2/n) \log_2 (p - 2/n)
        \;\right| \nonumber\\
      &\;=\;
        \left|\;
        p \log_2 \frac{p}{p - 2/n}
        + \frac{2}{n} \log_2 \frac{p - 2/n}{p(*, j)}
        \;\right| \nonumber\\
      &\;\le\;
        \left|\;
        p \log_2 \frac{p}{p - 2/n}
        \;\right|
        +
        \left|\;
        \frac{2}{n} \log_2 \frac{p - 2/n}{p(*, j)}
        \;\right| \;.
        \label{eq:case2b:main}
    \end{align}
    As in Case (2a), the left hand term in \eqref{eq:case2b:main}
    is decreasing in $p \ge 3/n$, meaning
    \begin{align}
      \left|\;
        p \log_2 \frac{p}{p - 2/n}
        \;\right|
      \;\le\; 
      \left|\;
        \frac{3}{n} \log_2 \frac{3/n}{3/n - 2/n}
        \;\right|
      \;=\;
      \frac{3}{n} \log_2 3
      \;\le\;
      \frac{4.8}{n} .
      \label{eq:case2b:lhs}
    \end{align}
    Since $p \le p(*, j)$, the right hand term in \eqref{eq:case2b:main}
    is also decreasing in $p \ge 3/n$. So
    \begin{align}
      \left|\;
      \frac{2}{n} \log_2 \frac{p - 2/n}{p(*, j)}
      \;\right|
      \;\le\;
      \left|\;
      \frac{2}{n} \log_2 \frac{3/n - 2/n}{p(*, j)}
      \;\right|
      \;\le\;
      \left|\;
      \frac{2}{n} \log_2 (n\cdot p(*, j))
      \;\right|
      \;\le\;
      \frac{2\log_2 n}{n},
      \label{eq:case2b:rhs}
    \end{align}
    where the finaly inequality holds since $p(*, j) \le 1$. \medskip

    Substituting \eqref{eq:case2b:lhs} and \eqref{eq:case2b:rhs}
    back into \eqref{eq:case2b:main} gives
    $\alpha \le (4.8/n) + (2\log_2 n) /n$, which holds in general
    for Case (2b). 
  \end{enumerate}

\item
  \textit{Case when $p' >  p$}:
  
  \begin{enumerate}
  \item
    \textit{Case when $p'-p = 1/n$}:\medskip

    This condition implies $0 \le p \le 1 - 1/n$.
    First consider when $p = 0$ and $p' = 1/n$.
    Then
    \begin{align*}
      \alpha \;=\;
      \left|\;
      \frac{1}{n} \log_2 p(*, j)
      - \frac{1}{n} \log_2 \frac{1}{n}
      \;\right|
      \;=\;
      \left|\;
      \frac{1}{n} \log_2 \frac{p(*, j)}{1/n}
      \;\right|
      \;\le\;
      \frac{\log_2 n}{n} \;,
    \end{align*}
    where the final inequality is due to
    $ 1/n = p' \le p(*, j) \le 1$. \medskip

    Now consider when $p \ge 1/n$. Then
    \begin{align}
      \alpha
      &\;=\;
        \left|\;
        \frac{1}{n} \log_2 p(*, j)
        + p \log_2 p
        - (p + 1/n) \log_2 (p + 1/n)
        \;\right| \nonumber \\
      &\;=\;
        \left|\;
        p \log_2 \frac{p}{p + 1/n}
        + \frac{1}{n} \log_2 \frac{p(*, j)}{p + 1/n}
        \;\right| \nonumber \\
      &\;\le\;
        \left|\;
        p \log_2 \frac{p}{p + 1/n}
        \;\right|
        +
        \left|\;
        \frac{1}{n} \log_2 \frac{p(*, j)}{p + 1/n}
        \;\right| \;.
        \label{eq:case3a:main} 
    \end{align}
    The left hand term of \eqref{eq:case3a:main} is increasing with
    $p$, and by the assumption that $p \le 1-1/n$,
    \begin{equation*}
      \left|\;
        p \log_2 \frac{p}{p + 1/n}
        \;\right|
      \;\le\;
      \left|\;
        \left(\frac{n-1}{n}\right) \log_2 \left(\frac{n-1}{n}\right)
        \;\right|
      \;\le\;
      \left|\;
        \log_2 \frac{n-1}{n}
        \;\right|
      \;=\;
      \left|\;
        \log_2 \frac{n}{n-1}
        \;\right|\;.
    \end{equation*}
    Because $(n+2)/n \ge n/(n-1)$ for all $n \ge 2$, we have
    \begin{equation*}
      \left|\;
      p \log_2 \frac{p}{p + 1/n}
      \;\right|
      \;\le\;
      \left|\;
      \log_2 \frac{n+2}{n}
      \;\right|
      \;=\;
      \left|\;
      \log_2 (1 + 2/n)
      \;\right| \;.
    \end{equation*}

    Note that because $e \le 2^{(3/2)}$, then
    $1 + x \le e^x \le 2^{(3/2)x}$ for all $x$.
    This implies $\log_2 (1+x) \le (3/2) x$ for all $x$, and so
    \begin{equation}
      \left|\;
        p \log_2 \frac{p}{p + 1/n}
        \;\right|
      \;\le\;
      \left|\;
        \log_2 (1 + 2/n)
        \;\right|
      \;\le\;
      \frac{3}{2} \cdot \frac{2}{n}
      \;=\;
      \frac{3}{n} \;.
      \label{eq:case3a:lhs}
    \end{equation}
    The right hand term of \eqref{eq:case3a:main}
    is decreasing in $p$ since $p + 1/n = p' \le p(*, j)$.
    Since $p \ge 1/n$ by assumption,
    \begin{align}
      \left|\;
        \frac{1}{n} \log_2 \frac{p(*, j)}{p + 1/n}
        \;\right|
      &\;\le\;
      \left|\;
        \frac{1}{n} \log_2 \frac{p(*, j)}{1/n + 1/n}
        \;\right| \nonumber \\
      &\;=\;
      \left|\;
        \frac{1}{n} \log_2 \left((n/2) \cdot p(*, j)\right)
        \;\right| 
      \;\le\;
        \frac{\log_2 (n/2)}{n}
      \;\le\;
        \frac{\log_2 n}{n} \;,
        \label{eq:case3a:rhs}
    \end{align}
    where the penultimate inequality holds because $p(*, j) \le 1$. \medskip
    
    Substituting \eqref{eq:case3a:lhs} and \eqref{eq:case3a:rhs}
    into \eqref{eq:case3a:main} gives
    $\alpha \le (3/n) + (\log_2 n)/n$, which holds in general
    for Case (3a). \medskip

  \item
    \textit{Case when $p-p' = 2/n$}: \medskip

    This condition implies $0 \le p \le 1 - 2/n$.
    When $p = 0$ and $p' = 2/n$, we have
    \begin{equation*}
      \alpha \;=\;
      \left|\;
        \frac{2}{n} \log_2 p(*, j)
        - \frac{2}{n} \log_2 \frac{2}{n}
        \;\right|
      \;=\;
      \left|\;
        \frac{2}{n} \log_2 \frac{p(*, j)}{2/n}
        \;\right|
      \;\le\;
      \left|\;
        \frac{2}{n} \log_2 (n/2)
        \;\right|
      \;\le\;
      \frac{2\log_2 n}{n} \;,
    \end{equation*}
    where the penultimate inequality is due to
    $2/n = p' \le p(*, j) \le 1$. \medskip

    Now consider when $p \ge 1/n$. We have
    \begin{align}
      \alpha
      &\;=\;
        \left|\;
        \frac{2}{n} \log_2 p(*, j)
        + p \log_2 p
        - (p + 2/n) \log_2 (p + 2/n)
        \;\right| \nonumber \\
      &\;=\;
        \left|\;
        p \log_2 \frac{p}{p + 2/n}
        + \frac{2}{n} \log_2 \frac{p(*, j)}{p + 2/n}
        \;\right| \nonumber \\
      &\;\le\;
        \left|\;
        p \log_2 \frac{p}{p + 2/n}
        \;\right|
        +
        \left|\;
        \frac{2}{n} \log_2 \frac{p(*, j)}{p + 2/n}
        \;\right| \;.
        \label{eq:case3b:main} 
    \end{align}

    Similar to Case (3a), the left hand term of \eqref{eq:case3b:main}
    is increasing with $p$, and by the assumption that $p \le 1-2/n$,
    \begin{equation*}
      \left|\;
        p \log_2 \frac{p}{p + 2/n}
        \;\right|
      \;\le\;
      \left|\;
        \left(\frac{n-2}{n}\right) \log_2 \left(\frac{n-2}{n}\right)
        \;\right|
      \;\le\;
      \left|\;
        \log_2 \frac{n-2}{n}
        \;\right|
      \;=\;
      \left|\;
        \log_2 \frac{n}{n-2}
        \;\right|\;.
    \end{equation*}
    Because $(n+3)/n \ge n/(n-2)$ for all $n \ge 6$, we have
    \begin{equation*}
      \left|\;
        p \log_2 \frac{p}{p + 2/n}
        \;\right|
      \;\le\;
      \left|\;
        \log_2 \frac{n+3}{n}
        \;\right|
      \;=\;
      \left|\;
        \log_2 (1 + 3/n)
        \;\right| \;.
    \end{equation*}
    Then applying the identity $\log_2 (1+x) \le (3/2)x$ yields
    \begin{equation}
      \left|\;
        p \log_2 \frac{p}{p + 2/n}
        \;\right|
      \;\le\;
      \left|\;
        \log_2 (1 + 3/n)
        \;\right|
      \;\le\;
      \frac{3}{2} \cdot \frac{3}{n}
      \;=\;
      \frac{4.5}{n} \;.
      \label{eq:case3b:lhs}
    \end{equation}

    Now the right hand term of \eqref{eq:case3b:main} is decreasing
    in $p$ since $p + 2/n = p' \le p(*, j)$. So by the assumption
    $p \ge 1/n$, 
    \begin{align}
      \left|\;
      \frac{2}{n} \log_2 \frac{p(*, j)}{p + 2/n}
      \;\right|
      &\;\le\;
        \left|\;
        \frac{2}{n} \log_2 \frac{p(*, j)}{1/n + 2/n}
        \;\right| \nonumber \\
      &\;\le\;
        \left|\;
        \frac{2}{n} \log_2 ((n/3) \cdot p(*, j))
        \;\right|
        \;\le\;
        \frac{2 \log_2 (n/3)}{n}
        \;\le\;
        \frac{2 \log_2 (n)}{n} \;,
        \label{eq:case3b:rhs}
    \end{align}
    where the penultimate inequality is due to $p(*, j) \le 1$.\medskip

    Now substituting expressions \eqref{eq:case3b:lhs} and
    \eqref{eq:case3b:rhs} into \eqref{eq:case3b:main} gives the
    bound $\alpha \le (4.5/n) + (2 \log_n)/n$, which holds
    in general for Case (3b). 
  \end{enumerate}
\end{enumerate}

To summarize the bounds on $\alpha$ for each of the cases:
\begin{itemize}
\item
  \textit{Case when $p = p'$}: $\alpha = 0$.
\item
  \textit{Cases when $p \neq p'$}:
  \begin{center}
    \def\arraystretch{2}%
    \begin{tabular}{ c | c | c }

      \;
      & $|p - p'| = 1/n$
      & $|p-p'| = 2/n$   \\
      \hline
      $p > p'$
      & Case (2a): $\alpha \le (\log_2 n)/n + 2/n$
      & Case (2b): $\alpha \le (2 \log_ 2n)/n + 4.8/n$ \\
      \hline
      $p' > p$
      & Case (3a): $\alpha \le (\log_2 n)/n + 3/n$
      & Case (3b): $\alpha \le (2 \log_ 2n)/n + 4.5/n$ \\
    \end{tabular}
  \end{center}

\end{itemize}

It follows that for any $(i, j)$,
\begin{equation*}
  \alpha  \;\le\; \frac{2 \log_2 n}{n} + \frac{4.8}{n} \;,
\end{equation*}
which completes the proof.

\end{proof}

\subsection{MICe-Lap Privacy}

Here we prove the privacy guarantee of the $\micelap$ mechanism
from Theorem~\ref{thm:micelap-dp}.
This requires first stating the post-processing principle,
which says that the result of performaing  any additional computation
on the output of an $\eps$-DP mechanism
is still private.


\begin{thm}[DP post-processing, \citet{dwork-roth}]
  \label{thm:dp-post}
  Let $A: \mathcal{X} \to \mathcal{Y}$
  be an $\eps$-DP mechanism.
  Then for any function $f: \mathcal{Y} \to \mathcal{Z}$,
  the composition $(f \circ A): \mathcal{X} \to \mathcal{Z}$
  is also $\eps$-DP.
\end{thm}

The privacy of the MICe-Lap mechanism then follows directly:

\micelapdp*
\begin{proof}\vspace*{-1.5em}
  This follows by the $\eps$-DP privacy of the Laplace mechanism,
  and by applying the post-processing property from
  Theorem~\ref{thm:dp-post} with truncation to the $[0, 1]$ interval.
\end{proof}


%
%

\section{MICr Details}
\label{sec:appendix:micr}

In this section, we give more details
on the $\micr$ statistic and develop the proofs of
Theorem~\ref{thm:micrcons:informal} (MICr consistency)
and Theorem~\ref{thm:micr-sens} (MICr sensitivity).
Some of the notation and definitions were introduced previously
in Sections~\ref{sec:prelims} and \ref{sec:micr},
but we will restate them here for completeness
and readability. 

\subsection{Computing MICr}
\label{sec:appendix:micr-overview}

\subsubsection{The set of grids $\;\calR(L, c, k, \ell)$}

Recall for an interval $I$ that $R_{I, \ell}$ is a size-$\ell$
range-equipartition of $I$, and $\calP(I, k, [c\ell])$ is the
set of all size-$k$ subpartitions of a size-$(c\ell)$
range-equipartition of $I$.
Then for fixed range bounds $L = L_x \times L_y$ and any finite $c > 0$:
\begin{itemize}
\item
  for any $2 \le k \le \ell$,
  the set $\calR(L, c, k, [\ell])$ contains all grids
  $G = (P, Q)$ where $Q = R_{L_x, \ell}$ and
  $P \in \calP(L_y, k, [c\ell])$.
\item
  for any $2 \le \ell < k$,
  the set $\calR(L, c, [k], \ell)$ contains all grids
  $G = (P, Q)$ where $Q \in \calP(L_x, \ell, [ck])$ and
  $P = R_{L_y, k}$. 
\end{itemize}

For any $L$ and $c$, and any $k, \ell \ge 2$,
we overload the defintion of $\calR$ and write
\begin{align}
  \calR(L, c, k, \ell)
  &\;=\;       
  \begin{cases}
    \begin{split}
      &\calR(L, c, k, [\ell])
      \;\;\;\; \text{if $k \le \ell$}  \\
      &\calR(L, c, [k], \ell)
      \;\;\;\; \text{if $k > \ell$}
      \;\;.
    \end{split}
    \label{eq:micr-cons:calR-def}
  \end{cases}
\end{align}

\subsubsection{Computing $\micr$ on finite datasets}

Recall now that for a fixed $L$ and $c$, and
for any dataset $D$ of size $n$ with range restricted to $L$, 
let $\M^{\calR(L, c)}_D$ denote the
sample range-equicharacteristic matrix for that dataset.
For any $k, \ell \ge 2$, its entries are given by
\begin{equation}
  \left(
    \M^{\calR(L, c)}_D
  \right)_{k, \ell}
  \;\;=\;
  \max_{G \in \calR(L, c, k, \ell)}\;
  \frac{I(D|_G)}{\log_2 \min \{k, \ell\}} \;.
  \label{eq:micr:m-D-entries}
\end{equation}
Then for a maximum grid parameter $B := B(n)$,
we define
\begin{equation}
  \micr(D, B, L, c)
  \;=\;
  \max_{k, \ell \;:\; k\cdot \ell \le B}\;
  \left(
    \M^{\calR(L, c)}_D
  \right)_{k, \ell} \;.
  \label{eq:micr:def}
\end{equation}

Note that because $\micr$ and $\mice$ differ only
in the \textit{definition} of the respective master grids,
but not in their size, $\micr$ can be computed in time
asympotically equivalent to $\mice$ again using the
$\optimizeaxis$ routine described in Appendix~\ref{sec:app:mice-alg}.
Stated formally (and analogous to Theorem~\ref{thm:mice-runtime}):

\vspace*{0.3em}
\begin{thm}
  \label{thm:micr-runtime}
  For any dataset $D$ of size $n$ restricted to range $L$, $B := B(n)$, and $c > 0$,
  $\micr(D, L, B, c)$ can be computed using $\optimizeaxis$
  in $O(c^2 B^4)$ time. 
\end{thm}

\subsubsection{Computing $\micr$ on joint distributions}

Because the set $\calR(L, c, k, \ell)$
depends only on $L$ and $c$, and not on the dataset $D$,
we can define a range-equicharacteristic matrix
for joint distributions analogously to
\eqref{eq:micr:m-D-entries}. 

Again for a fixed $L$ and $c$, and for any
joint distribution $(X, Y)$ with range restricted to $L$,
let  $\M^{\calR(L, c)}_{(X, Y)}$
denote the range-equicharacteristic matrix
for that distribution.
For any $k, \ell \ge 2$, its entries are given by
\begin{equation}
  \left(
    \M^{\calR(L, c)}_{(X, Y)}
  \right)_{k, \ell}
  \;\;=\;
  \max_{G \in \calR(L, c, k, \ell)}\;
  \frac{I((X, Y)|_G)}{\log_2 \min \{k, \ell\}} \;.
  \label{eq:micr:m-pop-entries}
\end{equation}

\subsubsection{The unconstrained MICr statistic}

For theoretical purposes, we also define the
following \textit{unconstrained} variant of the $\micr$ statistic
with equicharacteristic matrix $\M^{\calR(L, \infty)}_D$.
Compared to $\calR(L, c, k, \ell)$, the set $\calR(L, \infty, k, \ell)$
(wlog when $k \le \ell$) contains grids $G$ with \textit{any}
row partition of $[y_0, y_1]$ of size $k$ (i.e., not just those
restricted to subpartitions of some finite-size master
range-equipartition).
Thus $\calR(L, \infty, k, \ell)$ is exactly the set of grids
$G$ obtained from $\calR(L, c, k, \ell)$ when taking $c \to \infty$.

More formally, for an interval $I$, let $\calP(I, k)$ denote
the set of \textit{all} partitions of $I$ of size $k$.
And recall that the grid $R_{I, \ell}$ is a range equipartion
of $I$ of size $\ell$. 
Then for a fixed $L = L_x \times L_y$:
\begin{itemize}
\item
  for $2 \le k \le \ell$, the set $\calR(L, \infty, k, [\ell])$
  contains all grids $G = (P, Q)$ where
  $Q = R_{L_x, \ell}$ and $P \in \mathcal{P}(L_y, k)$.
\item
  for any $2 \le \ell < k$, the set $\calR(L, \infty, [k], \ell)$
  contains all grids $G = (P, Q)$ where
  $Q \in \mathcal{P}(L_x, \ell)$ and $P = R_{L_y, k}$. 
\end{itemize}
The set $\calR(L, \infty, k, \ell)$ is then defined analogously
to $\calR(L, c, k, \ell)$ from \eqref{eq:micr-cons:calR-def}. 

For any dataset $D$ of size $n$ restricted to the range $L = L_x \times L_y$,
let $\M^{\calR(L, \infty)}_D$ denote its \textit{unconstrained}
range-equicharacteristic matrix. For $k, \ell \ge 2$,
it entries are given by
\begin{equation}
  \left(
    \M^{\calR(L, \infty)}_D
  \right)_{k, \ell}
  \;=\;
  \max_{G \in \calR(L, \infty, k, \ell)}\;
  \frac{I(D|_G)}{\log_2 \min\{k, \ell\}} \;.
  \label{eq:micru:M-entries-D}
\end{equation}
Then for a maximum grid size parameter $B = B(n)$, define
\begin{equation*}
  \micr(D, L, B, c)
  \;=\;
  \max_{k\cdot \ell \le B}\;
  \left(
    \M^{\calR(L, \infty)}_D
  \right)_{k, \ell} \;.
\end{equation*}

For a jointly-distributed pair of random variables
$\Pi = (X, Y)$ with range restricted to $L$,
we similarly define the matrix $\M^{\calR(L, \infty)}_\Pi$
with entries
\begin{equation}
  \left(
    \M^{\calR(L, \infty)}_\Pi
  \right)_{k, \ell}
  \;=\;
  \max_{G \in \calR(L, \infty, k, \ell)}\;
  \frac{I(\Pi|_G)}{\log_2 \min\{k, \ell\}} \;.
  \label{eq:micru:M-entries-Pi}
\end{equation}

\subsection{MICr Consistency}
\label{sec:appendix:micr-cons}

Using the notation introduced in the previous section,
we now develop the proof of Theorem~\ref{thm:micrcons:informal},
which shows that the $\micr(\cdot, L, B, c)$ statistic is a
consistent estimator of $\micstar$.
However, our first step is to show that
the \textit{unconstrained} $\micr(\cdot, L, B, \infty)$ statistic
is a consistent estimator of $\micstar$.
Once this step is established,
we leverage the relationship between
entries of $\M^{\calR(L, c)}$ and $\M^{\calR(L, \infty)}$
to prove the consistency of $\micr(\cdot, L, B, c)$. 

\subsubsection{Consistency of the $\micr(\cdot, L, B, \infty)$ estimator}

We will prove the following Theorem:

\begin{restatable}{thm}{micrucons}
  \label{thm:micru-cons}
  Let $\Pi = (X, Y)$ be a jointly-distributed pair of random
  variables with range bounded by $L$,
  and let $D_n$ be a dataset of $n$ points sampled i.i.d. from $\Pi$. 
  For every $0 < \alpha < 0.5$, and for every
  $\omega(1) \le B(n) = O(n^\alpha)$,
  \begin{equation*}
    \micr(D_n, L, B(n), \infty) \longrightarrow \micstar(\Pi)
  \end{equation*}
  in probability as $n \to \infty$.   
\end{restatable}

The proof of this theorem uses the following two lemmas,
which use arguments adapted from \citet{reshef16-mice} and
\citet{arxivnote}, respectively.

The first lemma shows that, in the distributional setting,
the supremum of the range-equicharacteristic matrix
$\M^{\calR(L, \infty)}_\Pi$ and the standard
characteristic matrix $\M^{\G}_\Pi$ are equal,
the latter of which is equal to $\micstar(\Pi)$ by definition.

\begin{lem}
  \label{lem:micru-cons:boundary}
  Let $\Pi = (X, Y)$ be a pair of jointly-distributed
  random variables with range bounded by $L = L_x \times L_y$. Then
  \begin{equation*}
    \sup\; \M^{\calR(L, \infty)}_\Pi
    \;=\;  \sup \M^{\G}_\Pi
    \;=\; \micstar(\Pi) \;.
  \end{equation*}
\end{lem}

\begin{proof}
  The statement follows by an argument identical to
  the proof of Theorem 21 from \cite{reshef16-mice}.
  In particular, the mutual information of $I(X, Y)$
  is the supremum of $I(X|_Q, Y)$ over all finite partitions $Q$.
  By definition, the $(k, \ell)$'th entry of $\M^{\calR(L, \infty)}_\pi$
  uses the column partition $R_{L_x, \ell}$, which is
  a range-equipartition of $L_x$ of size $\ell$,
  to partition $X$. Thus taking $\ell$ in the limit,
  $I(X|_{R_{L_x, \ell}}, Y) = I(X, Y)$.
  The remainder of the argument in the original proof carries
  over unchanged. 
\end{proof}

The next lemma shows that, when $n$ is sufficiently large,
for a dataset $D_n$ of $n$ points sampled i.i.d. from $\Pi$,
a subset of corresponding entries $(k, \ell)$ in
$\M^{\calR(L, \infty)}_{D_n}$ and $\M^{\calR(L, \infty)}_\Pi$
have similar values with high probability. 

\begin{lem}
  \label{lem:micru-cons:points}
  Let $\Pi = (X, Y)$ be a pair of jointly-distributed
  random variables with range bounded by $L$,
  and let $D_n$ be a dataset of $n$ points sampled i.i.d. from $\Pi$.
  For all $n$, and for every $0 < \alpha < 0.5$, there
  exists a constant $u > 0$ such that
  \begin{equation*}
    \left|\;
      \left(
        \M^{\calR(L, \infty)}_{D_n}
      \right)_{k, \ell}
      \;-\;
      \left(
        \M^{\calR(L, \infty)}_{\Pi}
      \right)_{k, \ell}
      \;\right|
    \;=\;
    O\left(
      \frac{1}{n^u}
    \right)
  \end{equation*}
  simultaneously for every $k\ell \le B(n) = O(n^\alpha)$
  with probability at least $1 - O(n^{-1.5})$. 
\end{lem}

\begin{proof}
  Because the the $(k, \ell)$ entries of
  $\M^{\calR(L, \infty)}_\Pi$ and $\M^{\calR(L, \infty)}_{D_n}$
  are both maximizations over the same set of grids
  $\calR(L, \infty, k, \ell)$, the proof follows identically
  to that of Theorem 5 of \citet{arxivnote}. 
\end{proof}

The proof of Theorem~\ref{thm:micru-cons} then follows.

\begin{proof}[Proof (of Theorem~\ref{thm:micru-cons})]
  The proof is identical to that of Theorem 6 from \citet{reshef16-mice}.
  The exact argument goes through so long as
  (1),
  $\sup \M^{\calR(L, \infty)}_\Pi = \sup \M^{\G}_\Pi$,
  and (2), the $(k, \ell)$'th entry of 
  $\M^{\calR(L, \infty)}_{D_n}$ converges in probability to
  the $(k, \ell)$'th entry of $\M^{\calR(L, \infty)}_{D_n}$
  as $n \to \infty$ 
  for all $k\ell \le B(n) = O(n^\alpha)$.
  These arguments follow from
  Lemmas~\ref{lem:micru-cons:boundary}
  and \ref{lem:micru-cons:points}, respectively. 
\end{proof}

\subsubsection{Consistency of the $\micr(\cdot, L, B, c)$ estimator}

We now use the consistency of $\micr(\cdot, L, B, \infty)$
to prove the consistency of $\micr(\cdot, L, B, c)$ for finite $c$.
This was stated informally in Theorem~\ref{thm:micrcons:informal}
and is presented in more detail here. 

\begin{restatable}{thm}{micrcons}
  \label{thm:micr-cons}
  Let $\Pi = (X, Y)$ be a joint distribution
  with range bounded by $L$, and let $D_n$ be a dataset
  of $n$ points sampled i.i.d. from $\Pi$.
  For every finite $c > 0$, $\alpha \in (0, 0.5)$,
  and $\omega(1) \le B(n) = O(n^\alpha)$,
  the statistic $\micr(D_n, L, B, c) \to \micstar(\Pi)$
  in probability as $n \to \infty$.
\end{restatable}

To prove this theorem, we use (in addition to Theorem~\ref{thm:micru-cons})
the following two analogs to
Lemmas~\ref{lem:micru-cons:boundary} and \ref{lem:micru-cons:points}. 

\begin{lem}
  \label{lem:micr-cons:boundary}
  Let $\Pi = (X, Y)$ be a pair of jointly-distributed
  random variables with range bounded by $L = L_x \times L_y$. Then
  for finite $c > 0$:
  \begin{equation*}
    \sup\; \M^{\calR(L, c)}_\Pi
    \;=\;  \sup \M^{\G}_\Pi
    \;=\; \micstar(\Pi) \;.
  \end{equation*}
\end{lem}

\begin{proof}
  The proof follows identically to that of Proposition 8 from \citet{reshef16-mice},
  which gives an analogous argument for the
  $\mice(\Pi, c)$ case. The key observation is that, wlog for $k \le \ell$,
  the $(k, \ell)$'th entry of $\M^{\calR(L, c)}_\Pi$ is a maximization over
  row partitions $P \in \calP(L_y, k, [c\ell])$, which recall
  is the set of all size-$k$ sub-partitions of a size-($c\ell$)
  master range-equipartition of $L_y$. So as $\ell \to \infty$, the set
  $\calP(L_y, k, [c\ell])$ approaches $\calP(L_y, k)$. 
  The remainder of the original argument then follows identically.
\end{proof}

\begin{lem}
  \label{lem:micr-cons:points}
  Let $\Pi = (X, Y)$ be a pair of jointly-distributed
  random variables with range bounded by $L$, fix a finite $c > 0$, 
  and let $D_n$ be a dataset of $n$ points sampled i.i.d. from $\Pi$.
  For all $n$, and for every $0 < \alpha < 0.5$, there
  exists a constant $u > 0$ such that
  \begin{equation*}
    \left|\;
      \left(
        \M^{\calR(L, c)}_{D_n}
      \right)_{k, \ell}
      \;-\;
      \left(
        \M^{\calR(L, c)}_{\Pi}
      \right)_{k, \ell}
      \;\right|
    \;=\;
    O\left(
      \frac{1}{n^u}
    \right)
  \end{equation*}
  simultaneously for every $k\ell \le B(n) = O(n^\alpha)$
  with probability at least $1 - O(n^{-1.5})$. 
\end{lem}

\begin{proof}
  Similar to Lemma~\ref{lem:micru-cons:points}, because
  the $(k, \ell)$ entries of $\M^{\calR(L, c)}_\Pi$ and
  $\M^{\calR(L, c)}_{D_n}$
  are both maximizations over the same set of grids
  $\calR(L, c, k, \ell)$, the proof follows identically
  to that of Theorem 5 of \citet{arxivnote}. 
\end{proof}

In addition to these two lemmas, we state the
following additional inequality that relates
corresponding entries of $\M^{\calR(L, c)}_D$
and $\M^{\calR(L, \infty)}_D$ for some dataset $D$
and finite $c$.

\begin{lem}
  \label{lem:micr-cons:sample-entry-bound}
  Fix any dataset $D$ with range restricted to $L$ and 
  any finite $c > 0$. Then for all $k, \ell \ge 2$
  \begin{equation}
    \left(
      \M^{\calR(L, c)}_D
    \right)_{k,\ell}
    \;\le\;
    \left(
      \M^{\calR(L, \infty)}_D
    \right)_{k, \ell} \;.
  \end{equation}
\end{lem}

\begin{proof}
  By definition, for any finite $c > 0$ and any $k, \ell \ge 2$
  the set $\calR(L, c, k, \ell)$ is a subset of
  $\calR(L, \infty, k, \ell)$. For a
  fixed dataset $D$, the $(k, \ell)$ entries of
  $\M^{\calR(L, c)}_D$ and $\M^{\calR(L, \infty)}_D$
  are maximizations of a common function over the sets
  $\calR(L, c, k, \ell)$ and $\calR(L, \infty, k, \ell)$,
  respectively. Increasing the size of the constraint set
  can never decrease the maximum function value subject to
  the constraints, so the statement of the lemma follows. 
\end{proof}

Using these lemmas, we are now ready to prove
Theorem~\ref{thm:micr-cons}, which we do via a more
direct adaptation of the proof of Lemma H.4 from
\citet{reshef16-mice}.

\begin{proof}[Proof (of Theorem~\ref{thm:micr-cons})]
  For any $0 < \alpha < 0.5$, fix $B(n) = O(n^\alpha)$.
  Our goal is to show that for a dataset $D_n$ of
  $n$ points sampled i.i.d. from $\Pi$, that
  \begin{equation*}
    \micr(D_n, L, B(n), c)
    \longrightarrow
    \micstar(\Pi)
  \end{equation*}
  in probability as $n \to \infty$.

  Recall that by definition, $\micstar(\Pi) = \sup \M^\G_\Pi$,
  so this means that for every $\epsilon > 0$ and every
  $0 < p \le 1$, we must show that there exists some
  $N$ such that
  \begin{equation}
    \Pr
    \left(\;
      \left|\;
        \max_{k, \ell \;:\; k\ell \le B(n)}\;
        \left(\M^{\calR(L, c)}_{D_n}\right)_{k, \ell}
        -
        \sup \M^\G_\Pi
        \;\right|
      \;<\;
      \epsilon
      \;\right)
    \;>\;
    1 - p 
    \label{eq:micr-cons:goal}
  \end{equation}
  for all $n \ge N$.

  Fix $\epsilon$ and $p$. By Lemma~\ref{lem:micr-cons:boundary},
  we know $\sup M^{\calR(L, c)}_\Pi = \sup\M^\G_\Pi$.
  So by definition, there must exist an entry $(k, \ell)$ such that
  \begin{equation}
    \left|\;
      \left(\M^{\calR(L, c)}_\Pi\right)_{k, \ell}
      \;-\;
      \sup\; \M^{\G}_\Pi
      \;\right|
    \;<\;
    (\epsilon / 2) \;.
    \label{eq:micr-cons:1}
  \end{equation}
  Denote such an entry by $(k_\epsilon, \ell_\epsilon)$.

  Now by Lemma~\ref{lem:micr-cons:points}, there exists some
  $N_\epsilon$ such that, for all $n \ge N_\epsilon$:
  \begin{equation}
    \Pr
    \left(\;
      \left|\;
        \left(\M^{\calR(L, c)}_{D_n}\right)_{k_\epsilon, \ell_\epsilon}
        -
        \left(\M^{\calR(L, c)}_{\Pi}\right)_{k_\epsilon, \ell_\epsilon}
        \;\right|
      \;<\;
      (\epsilon/2)
      \;\right)
    \;>\;
    1 - p \;.
    \label{eq:micr-cons:2}
  \end{equation}
  So with probability greater than $1-p$, for all $n \ge N_\epsilon$,
  \begin{align}
    \left|\;
    \left(\M^{\calR(L, c)}_{D_n}\right)_{k_\epsilon, \ell_\epsilon}
    \;-\;
    \sup\; \M^\G_\Pi
    \;\right|
    &\;\le\;
      \left|\;
      \left(\M^{\calR(L, c)}_{D_n}\right)_{k_\epsilon, \ell_\epsilon}
      -
      \left(\M^{\calR(L, c)}_{\Pi}\right)_{k_\epsilon, \ell_\epsilon}
      \;\right|
      \;+\;
      \left|\;
      \left(\M^{\calR(L, c)}_{\Pi}\right)_{k_\epsilon, \ell_\epsilon}
      \;-\;
      \sup\; \M^\G_\Pi
      \;\right| \nonumber \\
    &\;<\;
      (\epsilon/2) + (\epsilon/2) \nonumber \\
    &\;=\;
      \epsilon \;.
      \label{eq:micr-cons:3}
  \end{align}
  Here, the first line is due to the triangle inequality, and
  the second line is due to the bounds from \eqref{eq:micr-cons:1}
  and \eqref{eq:micr-cons:2}.

  Now let $N_\alpha$ be some integer such that,
  $B(N_\alpha) \ge k_\epsilon \cdot \ell_\epsilon$.
  Since $B(n) = O(n^\alpha)$ is increasing in $n$,
  this implies that
  $k_\epsilon \cdot \ell_\epsilon \le B(N_\alpha) \le B(n)$
  for all $n \ge N_\alpha$, which means
  \begin{equation}
    \left(\M^{\calR(L, c)}_{D_n}\right)_{k_\epsilon, \ell_\epsilon}
    \;\le\;
    \max_{k, \ell:\; k\ell \le B(n)}\;
    \left(\M^{\calR(L, c)}_{D_n}\right)_{k, \ell} \;.
    \label{eq:micr-cons:4}
  \end{equation}
  Then for all $n \ge \max\{N_\epsilon, N_\alpha\}$, it follows that
  \begin{equation}
    \max_{k, \ell:\; k\ell \le B(n)}\;
    \left(\M^{\calR(L, c)}_{D_n}\right)_{k, \ell}
    \;\ge\;
    \left(\M^{\calR(L, c)}_{D_n}\right)_{k_\epsilon, \ell_\epsilon}
    \;>\;
    \sup\; \M^\G_\Pi - \epsilon \;
    \label{eq:micr-cons:5}
  \end{equation}
  with probability greater than $1 - p$,
  where the first inequality holds for $n \ge N_\alpha$
  by \eqref{eq:micr-cons:4},
  and the second inequality holds for $n \ge N_\epsilon$
  by \eqref{eq:micr-cons:3}.
  This implies one direction of our goal from \eqref{eq:micr-cons:goal}.

  For the other direction, consider any $k, \ell \ge 2$.
  By Lemma~\ref{lem:micr-cons:sample-entry-bound}, we have
  for any $n$ and fixed $D_n$ that 
  \begin{equation*}
    \left(\M^{\calR(L, c)}_{D_n}\right)_{k, \ell}
    \;\le\;
    \left(\M^{\calR(L, \infty)}_{D_n}\right)_{k, \ell} \;.
  \end{equation*}
  This implies that
  \begin{equation}
    \max_{k, \ell:\; k\cdot\ell \le B(n)}\;
    \left(\M^{\calR(L, c)}_{D_n}\right)_{k, \ell}
    \;\le\;
    \max_{k, \ell:\; k\cdot\ell \le B(n)}\;
    \left(\M^{\calR(L, \infty)}_{D_n}\right)_{k, \ell} \;.
    \label{eq:micr-cons:6}
  \end{equation}
  
  Now by Theorem~\ref{thm:micru-cons}, we know that
  \begin{equation*}
     \max_{k, \ell:\; k\cdot\ell \le B(n)}\;
     \left(\M^{\calR(L, \infty)}_{D_n}\right)_{k, \ell}
     \;\longrightarrow\;
     \sup\; \M^\G_\Pi
  \end{equation*}
  in probability as $n \to \infty$. 
  So there must exist some $N_\infty$ such that
  for all $n \ge N_\infty$
  \begin{align}
    \left|\;
      \max_{k, \ell:\; k\cdot\ell \le B(n)}\;
      \left(\M^{\calR(L, \infty)}_{D_n}\right)_{k, \ell}
      \;-\;
      \sup\; \M^\G_\Pi
      \;\right|
    &\;<\; \epsilon \nonumber \\
    &\;\;\implies\;\;
      \max_{k, \ell:\; k\cdot\ell \le B(n)}\;
      \left(\M^{\calR(L, \infty)}_{D_n}\right)_{k, \ell}
      \;<\;
      \sup\; \M^\G_\Pi + \epsilon 
      \label{eq:micr-cons:7}
  \end{align}
  with probability greater than $1-p$.

  Then combining \eqref{eq:micr-cons:6} and \eqref{eq:micr-cons:7}
  says that for all $n \ge N_\infty$, with probability greater
  than $1-p$:
  \begin{equation}
    \max_{k, \ell:\; k\cdot\ell \le B(n)}\;
    \left(\M^{\calR(L, c)}_{D_n}\right)_{k, \ell}
    \;\;\le\;\;
    \max_{k, \ell:\; k\cdot\ell \le B(n)}\;
    \left(\M^{\calR(L, \infty)}_{D_n}\right)_{k, \ell}
    \;\;<\;\;
    \sup\; \M^\G_\Pi + \epsilon  \;.
  \end{equation}
  This implies the second direction of our goal from
  \eqref{eq:micr-cons:goal}.

  Setting $N \ge \max\{N_\epsilon, N_\alpha, N_\infty\}$ ensures
  that for all $n \ge N$, expression \eqref{eq:micr-cons:goal}
  always holds, which completes the proof. 
\end{proof}


%
%
\subsection{MICr Sensitivity}
\label{sec:appendix:micr-sens}

In this section, we develop the proof of
Theorem~\ref{thm:micr-sens}, which gives
an upper bound on the sensitivity of the
$\micr$ statistic.
The high-level strategy is to
reduce the sensitivity of the statistic to the
sensitivity of mutual information with respect to
a fixed grid, similar to the arguments developed in
Section~\ref{sec:mice-sens:via-mi} for the bound on
the $\mice$ sensitivity. 

For a fixed dataset $D$ of $n$ points, a set of range limits $L$,
a constant $c > 0$, and a maximum grid parameter $B := B(n)$,
recall that
\begin{equation}
  \micr(D, B, L, c)
  \;=\;
  \left(\M^{\calR(L,c)}_D\right)_{k, \ell} \;,
  \label{eq:micr-sens:intro:stat-def}
\end{equation}
where $\M^{\calR(L, c)}_D$ is the range-equicharacteristic
matrix for $D$.

Now for any pair of neighboring datasets $D \sim D'$ of size $n$
and fixed $B := B(n)$, $L$, and $c$:
\begin{align}
  |\; \micr(D, B, L, c) - \micr(D', B, L, c) \;|
  &\;=\;    
    \left|\;
    \max_{k, \ell \;:\; k\cdot\ell \le B}\;
    \left(\M^{\calR(L,c)}_D\right)_{k, \ell}
    \;-\;
    \max_{k, \ell \;:\; k\cdot\ell \le B}\;
    \left(\M^{\calR(L,c)}_{D'}\right)_{k, \ell}
    \;\right| \nonumber \\
  &\;\le\;
    \max_{k, \ell \;:\; k\cdot\ell \le B}\;
    \left|\;
    \left(\M^{\calR(L,c)}_D\right)_{k, \ell}
    \;-\;
    \left(\M^{\calR(L,c)}_{D'}\right)_{k, \ell}
    \;\right| \;.
    \label{eq:micr-sens:ub-main:1}
\end{align}

As before, we will assume wlog that
$k \le \ell$, as the corresponding arguments for the $k > \ell$
case follow by symmetry.
So by the definition from \eqref{eq:micr-cons:calR-def},
for any fixed $2 \le k \le \ell$, observe that
\begin{align}
  \left|\;
  \left(\M^{\calR(L,c)}_D\right)_{k, \ell}
  \;-\;
  \left(\M^{\calR(L,c)}_{D'}\right)_{k, \ell}
  \;\right|
  &\;=\;
  \left|\;
  \max_{G \in \calR(L, c, k, [\ell])}
  \frac{I\left(D|_G\right)}{\log_2 k}
  \;-\;
  \max_{G \in \calR(L, c, k, [\ell])}
  \frac{I\left(D'|_G\right)}{\log_2 k}
    \;\right| \nonumber \\
  &\;\le\;
    \left|\;
    \max_{G \in \calR(L, c, k, [\ell])}
    I\left(D|_G\right)
    \;-\;
    \max_{G \in \calR(L, c, k, [\ell])}
    I\left(D'|_G\right)
    \;\right| \nonumber \\
  &\;\le\;
    \max_{G \in \calR(L, c, k, [\ell])}
    \left|\;
    I\left(D|_G\right)
    \;-\;
    I\left(D'|_G\right)
    \;\right| \;. 
    \label{eq:micr-sens:ub-main:2}
\end{align}
Here, the first inequality is due to factoring out the $\log_2 k$ term
and noting that $\log_2 k \ge 1$ when $k \ge 2$, and the final
inequality follows from the fact that the
set $\calR(L, c, k, [\ell])$ is the same in both maximization terms.

We then claim the following upper bound on \eqref{eq:micr-sens:ub-main:2}
that holds for any $D \sim D'$ of size $n$:

\begin{lem}
  \label{lem:micr-sens:mi-sens}
  For any $n \ge 4$ and any set of range bounds $L$, consider
  any pair of neighboring datasets $D \sim D'$, both of size $n$,
  and fix a pair $2 \le k \le \ell$. Then for any
  grid $G \in \calR(L, c, k, [\ell])$:
  \begin{equation*}
    |\; I(D|_G) - I(D'|_G) \;|
    \;\le\;
    \frac{4 \log_2 n}{n} + \frac{6}{n} \;.
  \end{equation*}
\end{lem}

Observe that this lemma holds for any $G \in \calR(L, c, k, [\ell])$,
and thus it holds for the maximizing grid in \eqref{eq:micr-sens:ub-main:2}.
Granting the lemma true for now then implies the upper bound on the
$\micr$ statistic (i.e., the proof of Theorem~\ref{thm:micr-sens}):

\begin{proof}[Proof (of Theorem~\ref{thm:micr-sens})]
  Apply Lemma~\ref{lem:micr-sens:mi-sens} to \eqref{eq:micr-sens:ub-main:2}.
  Subtituting back into \eqref{eq:micr-sens:ub-main:1} implies that
  for any $D \sim D'$ of size $n \ge 4$ and any $B$:
  \begin{equation*}
    |\; \micr(D, B, L, c) - \micr(D', B, L, c) \;|
    \;\le\;
    \max_{k, \ell \;:\; k\cdot\ell \le B}\;\;
    \frac{4 \log_2 n}{n} + \frac{6}{n}
    \;=\;\;
    \frac{4 \log_2 n}{n} + \frac{6}{n} \;.
  \end{equation*}
  The final equality follows by observing that the term from
  Lemma~\ref{lem:micr-sens:mi-sens} does not depend on $k$ or $\ell$. 
\end{proof}

It now remains to prove Lemma \ref{lem:micr-sens:mi-sens}.

\subsubsection{Proof of Lemma~\ref{lem:micr-sens:mi-sens}}

The proof follows similarly to that of Lemma~\ref{lem:mice-sens:mi-bound}.
Fix any $n \ge 4$, any $2 \le k \le \ell$, and any set of range boundaries $L$,
and consider any pair of neighboring datasets $D \sim D'$,
both of size $n$.
Now consider any grid $G \in \calR(L, c, k, [\ell])$. 
Let
\begin{align*}
  \P &\;:=\; \P_{D, G}
       \;\;\;\text{with entries $p(i, j)$, row sums $p(i, *)$,
       and column sums $p(*, j)$}\\
  \P' &\;:=\; \P_{D',G}
        \;\;\text{with entries $p'(i, j)$, row sums $p'(i, *)$,
        and column sums $p'(*, j)$}
        \;,
\end{align*}
and let $\A$ and $\A'$ denote the corresponding
count matrices for $\P$ and $\P'$ respectively.
Because $\P$ and $\P'$ are non-negative matrices representing
two-dimensional, discrete joint distributions, we define
the mutual information of $\P$ and $\P'$ in the natural way.
Specifically, let
\begin{align*}
  I(\P)
  \;=\;
  \sum_{i \in [k]}
  \sum_{j \in [\ell]}\;
  p(i, j) \log_2 \left(  \frac{p(i, j)}{p(i, *) p(*, j)} \right) \;.
\end{align*}  

Then to upper bound $|I(\P) - I(\P')|$, we have
\begin{align}
  |\; I(\P) - I(\P') \;|
  \;\le\;
  \sum_{i \in [k]}
  \sum_{j \in [\ell]}\;
  \left|\;
  p(i, j) \log_2 \left(  \frac{p(i, j)}{p(i, *) p(*, j)} \right)
  -   p'(i, j) \log_2 \left(  \frac{p'(i, j)}{p'(i, *) p'(*, j)} \right)
  \;\right| \;,
  \label{eq:micr-sens:gamma-sum}
\end{align}
which follows from the triangle inequality. 

For any $(i, j) \in [k] \times [\ell]$, let
\begin{equation*}
  \gamma(i, j) \;=\;
  \left|\;
  p(i, j) \log_2 \left(  \frac{p(i, j)}{p(i, *) p(*, j)} \right)
  -   p'(i, j) \log_2 \left(  \frac{p'(i, j)}{p'(i, *) p'(*, j)} \right)
  \;\right| \;.
\end{equation*}

Now because $G$ is fixed with respect to both $D$ and $D'$, observe
that there at most two entries $(i, j)$ such that $p(i, j) \neq p'(i, j)$,
and thus at most two pairs $(i, j)$ such that $\gamma(i, j) > 0$.
Specifically for any $D \sim D'$, there are only two scenarios
regarding how the entries of $\P$ and $\P'$ differ:
\begin{enumerate}
\item
  \textit{Scenario A}:
  there is at most 1 pair of points $d \in D$ and $d' \in D'$ with
  nonequal coordinates, but $\phi(d, G)$ = $\phi(d', G)$.  \medskip

  Then for all $(i, j) \in [k] \times [\ell]$, $p(i, j) = p'(i, j)$
  and thus $\gamma(i, j) = 0$. \medskip
  
  By \eqref{eq:micr-sens:gamma-sum}, this means
  \begin{equation*}
    |I(\P) - I(\P')| = 0 \;.
  \end{equation*}

\item
  \textit{Scenario B}:
  there is at most 1 pair of points $d \in D$ and $d' \in D'$ with
  nonequal coordinates, and $\phi(d, G) \neq \phi(d', G)$.  \medskip

  Then there is exactly one $(i, j) \in [k] \times [\ell]$
  such that $p'(i, j) = p(i, j) - 1/n$ and thus $\gamma(i, j) \ge 0$.
  For this unique $(i, j)$, we denote $\gamma(i, j)$ by
  $\gamma^-$. \medskip

  There is also exactly one $(i, j) \in [k] \times [\ell]$
  such that $p'(i, j) = p(i, j) + 1/n$ and thus $\gamma(i, j) \ge 0$.
  For this unique $(i, j)$, we similarly denote $\gamma(i, j)$
  by $\gamma^+$. \medskip

  By \eqref{eq:micr-sens:gamma-sum}, this means that
  \begin{equation}
    |I(\P) - I(\P')| \;\le\; \gamma^+ + \gamma^- \;.
    \label{eq:mcir-sens:scenario2:bound}
  \end{equation}

  Intuitively, $\gamma^-$ corresponds to the entry $(i, j)$
  that loses a point in $\A'$ compared to the corresponding
  entry in $\A$. Similarly, $\gamma^+$ corresponds to the
  entry $(i, j)$ that gains a point in $\A'$ compared
  to the correpsonding entry in $\A$. \medskip

  Because $|D| = |D'| = n$, both such entries must exist in
  this scenario. 
\end{enumerate}

Let us assume a $D \sim D'$ and $G$ from Scenario $B$
(otherwise the lemma follows trivially).

We claim the following bound on $\gamma^-$:

\begin{claim}
  \label{claim:micr-sens:alpha-minus}
  For any $n \ge 4$ and for $\P$ and $\P'$ as described above,
  let $(i, j) \in [k] \times [\ell]$ be the cell such that
  \begin{equation*}
    p'(i,j) = p(i, j) - \frac{1}{n} \;.
  \end{equation*}
  Then 
  \begin{equation*}
    \gamma^-
    \;:=\;
    \gamma(i, j)
    \;\le\;
    \frac{2 \log_2 n}{n} + \frac{4}{n} \;.
  \end{equation*}
\end{claim}

For $\gamma^+$, we have a similar claim:
\begin{claim}
  \label{claim:micr-sens:alpha-plus}
  For any $n \ge 4$ and for $\P$ and $\P'$ as described above,
  let $(i, j) \in [k] \times [\ell]$ be the cell such that
  \begin{equation*}
    p'(i,j) = p(i, j) + \frac{1}{n} \;.
  \end{equation*}
  Then 
  \begin{equation*}
    \gamma^+
    \;:=\;
    \gamma(i, j)
    \;\le\;
    \frac{2 \log_2 n}{n} + \frac{2}{n} \;.
  \end{equation*}
\end{claim}

If we grant these claims true for now, then by
expression \eqref{eq:mcir-sens:scenario2:bound} it follows
that
\begin{equation*}
  |\; I(\P) - I(\P') \;|
  \;\le\;
  \frac{2 \log_2 n}{n} + \frac{4}{n}
  + \frac{2 \log_2 n}{n} + \frac{2}{n}
  \;=\;
  \frac{4 \log_2 n}{n} + \frac{6}{n} \;,
\end{equation*}
which is the statement of the lemma. So it only
remains to prove the two claims.

\begin{proof}[Proof (of Claim~\ref{claim:micr-sens:alpha-minus})]
  For readability, let
  $p := p(i, j)$ and $p' := p'(i,j)$. Since we assume
  $p' = p - 1/n$, we have
  that
  \begin{align*}
    p(i, *) - 1/n &\le p'(i, *) \le p(i, *)  \\
    p(*, j) - 1/n &\le p'(*, j) \le p(*, j) \;\;.
  \end{align*}
  Moreover, we have
  \begin{align*}
    \gamma^-
    &\;=\;
      \left|\;
      p \log_2 \frac{p}{p(i,*) p(*, j)}
      \;-\;
      (p-1/n) \log_2 \frac{p-1/n}{p'(i,*) p'(*, j)}
      \;\right| \;.
  \end{align*}

  Observe that when $p = 1/n$ and $p' = 0$, then
  \begin{align*}
    \gamma^-
    &\;=\;
      \left|\;
      \frac{1}{n} \log_2 \frac{1/n}{p(i, *) p(*, j)}
      \;\right| \;.
  \end{align*}
  When $p = 1/n$, then $1/n \le p(i, *), p(*, j) \le 1$
  (and thus $1/n^2 \le p(i, *) \cdot p(*, j) \le 1$), so
  \begin{align*}
    \left|\;
    \frac{1}{n} \log_2 \frac{1/n}{p(i, *) p(*, j)}
    \;\right|
    \;\le\;
    \max\left\{\;
    \frac{1}{n} \log_2 \frac{1/n}{1/n^2},
    \;
    \frac{1}{n} \log_2 \frac{1}{1/n}
    \;\right\}
    \;=\; \frac{\log_2 n}{n}
  \end{align*}

  On the other hand, consider when $p \ge 2/n$
  and $1/n \le p' \le 1 - 1/n$. 
  We then have
  \begin{align}
    \gamma^-
    &\;=\;
      \left|\;
      p \log_2 \left(
      \frac{p}{p-1/n}
      \cdot
      \frac{p'(i,*) p'(*, j)}
      {p(i,*) p(*, j)}
      \right)
      \;+\;
      \frac{1}{n}\log_2 \frac{p-1/n}{p'(i,*) p'(*, j)}
      \;\right| \nonumber \\
    &\;\le\;
      \left|\;
      p \log_2 \left(
      \frac{p}{p-1/n}
      \cdot
      \frac{p'(i,*) p'(*, j)}
      {p(i,*) p(*, j)}
      \right)
      \;\right|
      \;+\;
      \left|\;
      \frac{1}{n}\log_2 \frac{p-1/n}{p'(i,*) p'(*, j)}
      \;\right| \;, \label{eq:gamma-minus:main}
  \end{align}
  where we use the triangle inequality. 

  For the left hand term in \eqref{eq:gamma-minus:main},
  say that 
  \begin{equation*}
    \frac{p}{p-1/n}
    \cdot
    \frac{p'(i,*) p'(*, j)}{p(i,*) p(*, j)}
    \;>\;
    1.
  \end{equation*}
  Then because $p'(i, *) \cdot p'(*, j) \le p(i, *) \cdot p(*, j)$,
  \begin{align*}
    \left|\;
      p \log_2 \left(
      \frac{p}{p-1/n}
      \cdot
      \frac{p'(i,*) p'(*, j)}
      {p(i,*) p(*, j)}
    \right)
    \;\right|
    \;\le\;
    \left|\;
      p \log_2 \left(
      \frac{p}{p-1/n}
    \right)
    \;\right|
    \;\le\;
    \left|\;
      \frac{2}{n} \log_2 \left(
      \frac{2/n}{2/n-1/n}
      \right)
    \;\right|
    \;=\;
      \frac{2}{n} \;.
  \end{align*}
  Here, the final inequality is because the function is decreasing
  in $p$, and $p\ge 2/n$ by assumption.
  
  On the other hand, consider when
  \begin{equation*}
    \frac{p}{p-1/n}
    \cdot
    \frac{p'(i,*) p'(*, j)}{p(i,*) p(*, j)}
    \;<\;
    1.
  \end{equation*}
  
  Then because $p \le p(i, *), p(*, j) \le 1$, and because
  $p'(i, *)\cdot p'(* ,j) \ge (p(i, *) - 1/n) \cdot (p(*,j) - 1/n)$:
  \begin{align*}
    \left|\;
      p \log_2 \left(
      \frac{p}{p-1/n}
      \cdot
      \frac{p'(i,*) p'(*, j)}
      {p(i,*) p(*, j)}
    \right)
    \;\right|
    &\;=\;
      \left|\;
      p \log_2 \left(
      \frac{p-1/n}{p}
      \cdot
      \frac{p(i,*) p(*, j)}
      {p'(i,*) p'(*, j)}
    \right)
    \;\right| \\
    &\;\le\;
      \left|\;
      p \log_2 \left(
      \frac{p(i,*) p(*, j)}
      {(p(i,*)-1/n)(p(*, j)-1/n)}
      \right)
      \;\right| \\
    &\;\le\;
      \left|\;
      p(i,*) \log_2 \left(
      \frac{p(i,*)}
      {p(i,*)-1/n}
      \right)
      \;+\;
      p(*, j) \log_2 \left(
      \frac{p(*, j)}
      {p(*, j)-1/n}
      \right)
      \;\right| \;.
  \end{align*}
  Here, the two terms are decreasing in $p(i, *)$ and $p(*, j)$, respectively.
  Since $p(i, *), p(*, j) \ge p \ge 2/n$ by assumption, we have
  \begin{align}
    \left|\;
    p(i,*) \log_2 \left(
    \frac{p(i,*)}
    {p(i,*)-1/n}
    \right)
    \;+\;
    p(*, j) \log_2 \left(
    \frac{p(*, j)}{p(*, j)-1/n}
    \right) 
    \;\right| \\
    \;\le\;
    \left|\;
    \frac{2}{n}\log_2 \left(
    \frac{2/n}{2/n-1/n}
    \right)
    \;+\;
    \frac{2}{n} \log_2 \left(
    \frac{2/n}{2/n-1/n}
    \right)
    \;\right| 
    &\;=\;
    \frac{4}{n} \;,
    \label{eq:gamma-minus:lhs}
  \end{align}
  which bounds the left hand term of \eqref{eq:gamma-minus:main}
  in general. 
  
  For the right hand term of \eqref{eq:gamma-minus:main}, if
  \begin{equation*}
    \frac{1}{n}
    \cdot
    \frac{p - 1/n}{p'(i,*) p'(*, j)}
    \;<\;
    1, 
  \end{equation*}
  then
  \begin{align*}
    \left|\;
      \frac{1}{n}\log_2 \frac{p-1/n}{p'(i,*) p'(*, j)}
      \;\right|    
    &\;=\;
    \left|\;
      \frac{1}{n}\log_2 \frac{p'(i,*) p'(*, j)}{p-1/n}
      \;\right| \\
    &\;\le\;
      \left|\;
      \frac{1}{n}\log_2 \frac{1}{2/n-1/n}
      \;\right| 
    \;=\;
      \frac{\log_2 n}{n} \;.
  \end{align*}
  Here, the penultimate inequality is due to
  $p'(i, *)\cdot p'(*, j) \le 1$ and $p \ge 2/n$ by assumption.
  
  On the other hand, consider when
  \begin{equation*}
    \frac{1}{n}
    \cdot
    \frac{p - 1/n}{p'(i,*) p'(*, j)}
    \;>\;
    1 \;.
  \end{equation*}
  Then for the right hand term of \eqref{eq:gamma-minus:main}, we have 
  \begin{equation*}
    \left|\;
      \frac{1}{n}\log_2 \frac{p-1/n}{p'(i,*) p'(*, j)}
      \;\right|
    \;\le\;
    \left|\;
      \frac{1}{n}\log_2 \frac{p-1/n}{(p(i,*)-1/n)(p(*, j)-1/n)}
      \;\right| \;.
  \end{equation*}
  Now because $p \le \min\{p(i, *), p(*, j)\}$, then
  \begin{align*}
    \left|\;
      \frac{1}{n}\log_2 \frac{p-1/n}{p'(i,*) p'(*, j)}
      \;\right|
    &\;\le\;
    \max\left\{
    \left|\;
      \frac{1}{n}\log_2 \frac{1}{p(i, *) - 1/n}
    \;\right| \;,
    \left|\;
    \frac{1}{n}\log_2 \frac{1}{p(*, j) - 1/n}
    \;\right|
    \right\} \\
    &\;\le\;
    \left|\;
      \frac{1}{n}\log_2 \frac{1}{p(i, *) - 1/n}
    \;\right| 
    \;+\;
    \left|\;
    \frac{1}{n}\log_2 \frac{1}{p(*, j) - 1/n}
    \;\right| \;,
  \end{align*}
  where the inequality is due to $\max\{x, y\} \le x + y$ for all $x, y \ge 0$. 
  Because the terms in the sum are decreasing in $p(i, *)$ and $p(*, j)$
  respectively, and because $p(i, *), p(*, j) \ge 2/n$ by assumption,
  we have
  \begin{align}
    \left|\;
      \frac{1}{n}\log_2 \frac{1}{p(i, *) - 1/n}
    \;\right| 
    \;+\;
    \left|\;
    \frac{1}{n}\log_2 \frac{1}{p(*, j) - 1/n}
    \;\right|
    &\;\le\;
      \frac{2\log_2 n}{n} \;.
      \label{eq:gamma-minus:rhs}
  \end{align}
  This bounds the right hand term of \eqref{eq:gamma-minus:main}
  in general. 

  Now substituting the bounds
  \eqref{eq:gamma-minus:lhs} and \eqref{eq:gamma-minus:rhs}
  back into \eqref{eq:gamma-minus:main}, we have
  \begin{equation*}
    \gamma^-
    \;\le\;
    \frac{2\log_2 n}{n} + \frac{4}{n} \;,
  \end{equation*}
  which completes the proof.  \qedhere
\end{proof}

The proof of Claim~\ref{claim:micr-sens:alpha-plus} follows similarly:
\begin{proof}[Proof (of Claim~\ref{claim:micr-sens:alpha-plus})]
  Again for readability let
  $p := p(i, j)$ and $p' := p'(i,j)$. Since we assume
  $p' = p + 1/n$, we have
  that
  \begin{align*}
    p(i, *) &\le p'(i, *) \le p(i, *) + 1/n  \le 1\\
    p(*, j) &\le p'(*, j) \le p(*, j)  + 1/n \le 1
  \end{align*}
  Then
  \begin{align*}
    \gamma^-
    &\;=\;
      \left|\;
      p \log_2 \frac{p}{p(i,*) p(*, j)}
      \;-\;
      (p+1/n) \log_2 \frac{p+1/n}{p'(i,*) p'(*, j)}
      \;\right| \;.
  \end{align*}
  Now observe that when $p = 0$ and $p' = 1/n$, then
  \begin{align*}
    \gamma^+
    \;=\;
      \left|\;
      \frac{1}{n} \log_2 \frac{1/n}{p'(i, *) p'(*, j)}
      \;\right|
    \;\le\;
      \frac{\log_2 n}{n}
  \end{align*}
  where the inequality is due to
  $1/n^2 \le p'(i, *) \cdot p'(*, j) \le 1$
  since $ p'(i, *), p'(*, j) \ge p' = 1/n$. 

  Consider now when $p \ge 1/n$. Then
    \begin{align}
    \gamma^+
    &\;=\;
      \left|\;
      p \log_2 \left(
      \frac{p}{p+1/n}
      \cdot
      \frac{p'(i,*) p'(*, j)}
      {p(i,*) p(*, j)}
      \right)
      \;+\;
      \frac{1}{n}\log_2 \frac{p+1/n}{p'(i,*) p'(*, j)}
      \;\right| \nonumber \\
    &\;\le\;
      \left|\;
      p \log_2 \left(
      \frac{p}{p+1/n}
      \cdot
      \frac{p'(i,*) p'(*, j)}
      {p(i,*) p(*, j)}
      \right)
      \;\right|
      \;+\;
      \left|\;
      \frac{1}{n}\log_2 \frac{p+1/n}{p'(i,*) p'(*, j)}
      \;\right| \;, \label{eq:gamma-plus:main}
  \end{align}
  which follows by the triangle inequality.

  For the left hand term in \eqref{eq:gamma-minus:main},
  say that 
  \begin{equation*}
    \frac{p}{p+1/n}
    \cdot
    \frac{p'(i,*) p'(*, j)}{p(i,*) p(*, j)}
    \;>\;
    1 \;.
  \end{equation*}
  Then 
  \begin{align*}
    \left|\;
      p \log_2 \left(
      \frac{p}{p+1/n}
      \cdot
      \frac{p'(i,*) p'(*, j)}
      {p(i,*) p(*, j)}
      \right)
    \;\right|
    &\;\le\;
      \left|\;
      p \log_2 \left(
      \frac{p}{p+1/n}
      \cdot
      \frac{(p(i,*)+1/n)(p(*, j)+1/n)}
      {p(i,*) p(*, j)}
      \right)
    \;\right| \\
    &\;\le\;
      \left|\;
      p \log_2 \left(
      \frac{p}{p+1/n}
      \cdot
      \frac{1}
      {((n-1)/n)^2}
      \right)
      \;\right|  \\
    &\;\le\;
      \left|\;
      \frac{n-1}{n} \log_2 \left(
      \frac{(n-1)/n}{1}
      \cdot
      \frac{n^2}
      {(n-1)^2}
      \right)
      \;\right| \\
    &\;\le\;
      \left|\;
      \log_2 \left(
      \frac{n}{n-1}
      \right)
      \;\right| \;.
  \end{align*}
  Here, the second inequality holds given that
  the expression is increasing in both
  $p(i, *)$ and $p(*, j)$, both of which
  are at most $(n-1)/n$. 
  The third inequality follows because the expression
  is increasing with $p \le (n-1)/n$.

  Now because $2^{(29/20)} \approx 2.73 \ge e$,
  it follows that for all numbers $x$,
  $x \le e^x \le 2^{(29/20)x}$, which
  implies $\log_2 x \le (29/20) x$.
  Because $n/(n-1) \le (n+40/29)/n$ for all $n \ge 4$,
  then applying this identity means
  \begin{equation}
    \log_2 \frac{n}{n-1}
    \;\le\;
    \log_2 \frac{n+40/29}{n}
    \;\le\;
    \frac{29}{20} \cdot \frac{40}{29n}
    \;=\;
    \frac{2}{n} 
    \label{eq:gamma-plus:lhs-0}
  \end{equation}
  for all $n \ge 4$, which holds by assumption of the claim.
  So in this case, the left hand term 
  \eqref{eq:gamma-plus:main} is at most $2/n$. 

  On the other hand, suppose
  \begin{equation*}
    \frac{p}{p+1/n}
    \cdot
    \frac{p'(i,*) p'(*, j)}{p(i,*) p(*, j)}
    \;<\;
    1 \;.
  \end{equation*}
  Then 
  \begin{align*}
    \left|\;
      p \log_2 \left(
      \frac{p}{p+1/n}
      \cdot
      \frac{p'(i,*) p'(*, j)}
      {p(i,*) p(*, j)}
      \right)
    \;\right|
    &\;=\;
      \left|\;
      p \log_2 \left(
      \frac{p+1/n}{p}
      \cdot
      \frac{p(i,*) p(*, j)}
      {p'(i,*) p'(*, j)}
      \right)
      \;\right| \\
    &\;\le\;
      \left|\;
      p \log_2 \left(
      \frac{p+1/n}{p}
      \right)
      \;\right| \\
    &\;\le\;
      \left|\;
      \frac{n-1}{n} \log_2 \left(
      \frac{1}{(n-1)/n}
      \right)
      \;\right|
      \;\le\;
      \left|\;
      \log_2
      \frac{n}{n-1}
      \;\right| \;.
  \end{align*}
  Here, the first inequality holds given
  that $p(i,*) \cdot p(*, j) \le p'(i, *) \cdot p'(j, *)$.
  The second inequality follows because the expression is
  increasing in $p$, which can be at most $(n-1)/n$ by assumption.

  Applying the bound from \eqref{eq:gamma-plus:lhs-0}
  then yields
  \begin{equation}
    \left|\;
      p \log_2 \left(
      \frac{p}{p+1/n}
      \cdot
      \frac{p'(i,*) p'(*, j)}
      {p(i,*) p(*, j)}
      \right)
      \;\right|
    \;\le\;
    \frac{2}{n} \;,
    \label{eq:gamma-plus:lhs}
  \end{equation}
  which bounds the left hand term of
  \eqref{eq:gamma-plus:main} in general.

  For the right hand term of \eqref{eq:gamma-plus:main},
  consider when
  \begin{equation*}
    \frac{p + 1/n}{p'(i, *) p'(*, j)} < 1 \;.
  \end{equation*}
  Then because $p'(i, *), p'(*, j) \le 1$
  and $p \ge 1/n$ by assumption, the right hand term
  of \eqref{eq:gamma-plus:main} becomes
  \begin{align*}
    \left|\;
    \frac{1}{n} \log_2 \left(
    \frac{p + 1/n}{p'(i, *) p'(*, j)}
    \right)
    \;\right|
    \;=\;
    \left|\;
    \frac{1}{n} \log_2 \left(
    \frac{p'(i, *) p'(*, j)}{p+1/n}
    \right)
    \;\right|
    \;\le\;
    \left|\;
    \frac{1}{n} \log_2 \left(
    \frac{1}{1/n+1/n}
    \right)
    \;\right|
    \;\le\;
    \frac{\log_2 n}{n} \;.
  \end{align*}

  On the other hand, consider when
  \begin{equation*}
    \frac{p + 1/n}{p'(i, *) p'(*, j)} > 1 \;.
  \end{equation*}
  Then because $p \le 1 - 1/n$ and because
  $p'(i, *) \cdot p'(*,j ) \ge 1/n^2$ by the
  assumption that $p'(i, *), p'(*, j) \ge p \ge 1/n$, 
  \begin{align}
    \left|\;
    \frac{1}{n} \log_2 \left(
    \frac{p + 1/n}{p'(i, *) p'(*, j)}
    \right)
    \;\right|
    \;\le\;
    \left|\;
    \frac{1}{n} \log_2 \left(
    \frac{1}{1/n^2}
    \right)
    \;\right|
    \;=\;
    \frac{2 \log_2 n }{n}
    \;,
    \label{eq:gamma-plus:rhs}
  \end{align}
  which bounds the right hand
  term of \eqref{eq:gamma-plus:main} in general. 

  Combining the bounds \eqref{eq:gamma-plus:lhs}
  and \eqref{eq:gamma-plus:rhs} and
  substituting back into \eqref{eq:gamma-plus:main}
  then shows
  \begin{align*}
    \alpha^+ \;\le\;
    \frac{2 \log_2 n}{n} + \frac{2}{n} \;,
  \end{align*}
  which completes the proof. 
\end{proof}


%
%
\section{MICr-Lap Details}
\label{sec:appendix:micr-lap}

In this section, we provide the proofs of
Theorem~\ref{thm:micrlap-dp} (MICr-Lap privacy) and
Theorem~\ref{thm:micrlap-cons} (MICr-Lap consistency).

\subsection{MICr-Lap Privacy}

We begin with the privacy guarantee of $\micrlap$ (restated for convenience),
whose proof follows identically to that of 
Theorem~\ref{thm:micelap-dp}, which gave the privacy guarantee for $\micelap$:

\vspace*{0.5em}
\micrlapdp*
\vspace*{-1em}
\begin{proof}
  This follows by the privacy of the Laplace mechanism,
  and because truncating to the $[0, 1]$ interval is a
  post-processing operation that, by
  Theorem~\ref{thm:dp-post}, preserves privacy.
\end{proof}

\subsection{MICr-Lap Consistency}

We now prove that $\micrlap$ is a consistent estimator
of $\micstar$.

\begin{restatable}{thm}{micrlapcons}
  \label{thm:micrlap-cons}
  For every finite $c > 0$, $\epsilon > 0$,
  $\alpha \in (0, 0.5)$, and $\omega(1) \le B(n) = O(n^\alpha)$,
  $\micrlap(\cdot, L, B, c, \epsilon)$
  is a consistent estimator of $\micstar(\cdot)$.
\end{restatable}

\begin{proof}
  Recall from Section~\ref{sec:micr} that
  the standard deviation of the $\micrlap$ mechanism
  is at most
  \begin{equation*}
    (\sqrt{2}/\eps)\cdot
    ((4 \log_2 n)/n + 6/n) \;,
  \end{equation*}
  which is decreasing with $n$ for a fixed $\eps$.
  Then coupled with the fact that (non-private) $\micr$
  is a consistent estimator of $\micstar$ when
  $\omega(1) \le B(n) = O(n^\alpha)$ for $\alpha \in (0, 0.5)$
  (Theorem~\ref{thm:micr-cons}),
  it follows that $\micrlap$ is \textit{also} a
  consistent estimator under the same parameter settings of $B$
  for any $c$ and $\eps$. 
\end{proof}


%
%

\section{MICr-Geom Details}
\label{sec:appendix:micr-geom}

In this section, we provide more details on the MICr-Geom mechanism
and give the proofs of its privacy (Theorem~\ref{thm:micrgeom-dp})
and consistency (Theorems~\ref{thm:micrgeom-err} and \ref{thm:micrgeom-cons}).

\subsection{MICr-Geom Privacy}
\label{sec:app:micrgeom-privacy}

We start with the privacy guarantee of $\micrgeom$
from Theorem~\ref{thm:micrgeom-dp}, which is restated as follows:

\micrgeomdp*
\vspace*{-0.5em}
The proof uses the privacy of the $\truncgeom$
mechanism from \cite{ghosh12}, as well as the
composition property of differentially private mechanisms.
We state these two tools here:

\begin{thm}[Privacy of $\truncgeom$, \cite{ghosh12}]
  \label{thm:truncgeom-dp}
  Assume that $f$ is the value of some function of
  a dataset $D \in \R^{2\times n}$
  with $\ell_1$ sensitivity 1, and range $\{0, \dots, n\}$.
  Then for any $\epsilon > 0$ and $n$,
  $\truncgeom(\epsilon, n, f)$ is $\epsilon$-DP.
\end{thm}

\begin{thm}[General Composition of DP Mechanisms, \cite{dwork-roth}]
  \label{thm:comp}
  For any mechanisms $\mathcal{A}_1: \mathcal{X}_1 \to \mathcal{Y}_1$
  and $\mathcal{A}_2: \mathcal{X}_2 \to \mathcal{Y}_2$,
  if $\mathcal{A}_1$ and $\mathcal{A}_2$ are $\eps_1$ and $\eps_2$-DP,
  respectively, then the mechanism
  $\mathcal{A}: (\mathcal{X}_1 \times \mathcal{X}_2) \to (\mathcal{Y}_1 \times \mathcal{Y}_2)$
  defined by $\mathcal{A}(x_1, x_2) = (\mathcal{A}_1(x_1), \mathcal{A}_2(x_2))$
  for $x_1 \in \mathcal{X}_1, x_2 \in \mathcal{X}_2$ 
  is $(\eps_1 + \eps_2)$-DP. 
\end{thm}

The differential privacy of $\micrgeom$ follows from
Theorems~\ref{thm:truncgeom-dp}, \ref{thm:comp}, and from the
post-processing principle from Theorem~\ref{thm:dp-post}:

\vspace*{-1em}
\begin{proof}[Proof (of Theorem~\ref{thm:micrgeom-dp})]
  For each $\ell \ge 2$, the counts of all entries of 
  the noisy count matrix $\widehat \A$ are all independently
  $(\epsilon/2)$-DP via the privacy of the $\truncgeom$
  mechanism from Theorem~\ref{thm:truncgeom-dp}.
  Since at most two corresponding entries between
  $\A_{D, G}$ and $\A_{D', G}$ can have non-zero difference
  for any neighboring $D \sim D'$ and any fixed $G$,
  it follows from Theorem~\ref{thm:comp} that
  the entire noisy count matrix $\widehat \A$
  is $\epsilon$-DP. The same argument holds
  independently for all $k \ge 2$. 
  Then by the post-processing principle, every
  $(k, \ell)$ entry of $\widehat \M^{\calR(L, c)}_{D, \epsilon}$
  is independently $\epsilon$-DP.
  Because the $\micrgeom(D, L, B, c, \epsilon)$ statistic
  only returns one such entry, the final output of
  the mechanism is $\epsilon$-DP.
\end{proof}

\subsection{Computing MICr-Geom}
\label{sec:app:micrgeom-comp}

Recall from the definition of $\micrgeom$ in Section~\ref{sec:micr}
that for a fixed $\ell$, we construct a $c\ell \times \ell$
matrix $\hatA_{D, \Gamma}$ where each entry is generated
using the $\truncgeom$ mechanism from Definition~\ref{def:truncgeom}.
Because the $\truncgeom$ outputs range from $0$ to $n$,
for any $n$ and $\epsilon$, an array representing the exact PMF function
of the $\truncgeom$ distribution can be constructed in linear time,
with subsequent samples done
in constant time. Because the subsequent steps of computing $\micrgeom$
do not differ from computing $\micr$, the run time of the former
is asymptotically equivalent to the latter (i.e., from Theorem~\ref{thm:micr-runtime}).
Stated formally:

\begin{thm}
  For any dataset $D$ of size $n$ restricted to $L$, $B := B(n)$, $c>0$, and $\eps >0$,
  $\micrgeom(D, L, B, c, \eps)$ can be computed using
  $\optimizeaxis$ and a pre-computed $\truncgeom$ distribution
  in $O(c^2 B^4)$ time.
\end{thm}

\subsection{MICr-Geom Consistency}
\label{sec:appendix:micr-geom-cons}

In this section, we develop the proof of
Theorem~\ref{thm:micrgeom-cons}, which shows that
the $\micrgeom$ statistic is a consistent estimator of $\micstar$.
Formally, we prove the following theorem:

\begin{restatable}{thm}{micrgeomcons}
  \label{thm:micrgeom-cons}
  For every finite $c > 0$, $\epsilon > 0$,
  $\alpha \in (0, 0.5)$, and $\omega(1) \le B(n) = O(n^\alpha)$,
  $\micrgeom(\cdot, L, B, c, \epsilon)$
  is a consistent estimator of $\micstar(\cdot)$.
\end{restatable}

The proof of the theorem leverages Theorem~\ref{thm:micrgeom-err},
which shows that the error introduced by the $\micrgeom$ mechanism
in the $(k, \ell)$'th entry of the range-equicharacteristic matrix
vanishes as $n$ grows. Again restated:

\micrgeomerr*

Granting Theorem~\ref{thm:micrgeom-err} true for now, the proof
of Theorem~\ref{thm:micrgeom-cons} is straightforward:

\begin{proof}[Proof (of Theorem~\ref{thm:micrgeom-cons})]
  Fix any jointly-distributed pair of random variables $\Pi = (X, Y)$
  with range bounded by $L$.
  For any $0 < \alpha < 0.5$, fix $B(n) = O(n^\alpha)$,
  and fix finite $c > 0$ and $\epsilon > 0$. 
  Our goal is to show that
  \begin{equation*}
    \micrgeom(D_n, L, B(n), c, \epsilon)
    \longrightarrow
    \micstar(\Pi) = \sup \M^\G_\Pi
  \end{equation*}
  in probability as $n \to \infty$.
  This means that for every $\tau > 0$ and every $0 < p \le 1$,
  we must show that there exists some $N$ such that
    \begin{equation}
    \Pr
    \left(\;
      \left|\;
        \max_{k, \ell:\; k\ell \le B(n)}\;
        \left(\widehat \M^{\calR(L, c)}_{D_n, \epsilon}\right)_{k, \ell}
        -
        \sup \M^\G_\Pi
        \;\right|
      \;<\;
      \tau
      \;\right)
    \;>\;
    1 - p 
    \label{eq:micrgeom-cons:goal}
  \end{equation}
  for all $n \ge N$.

  By the triangle inequality, observe that
  \begin{align}
    \left|\;
    \max_{k, \ell:\; k\ell \le B(n)}\;
    \left(\widehat \M^{\calR(L, c)}_{D_n, \epsilon}\right)_{k, \ell}
    -
    \sup \M^\G_\Pi
    \;\right|
    \;\le\;
      &\left|\;
      \max_{k, \ell:\; k\ell \le B(n)}\;
      \left(\widehat \M^{\calR(L, c)}_{D_n, \epsilon}\right)_{k, \ell}
      \;-\;
      \max_{k,\ell:\; k\ell \le B(n)}\;
      \left(\M^{\calR(L, c)}_{D_n}\right)_{k, \ell}
      \;\right|
      \;+\; \label{eq:micrgeomcons:1}\\
      &\left|\;
      \max_{k, \ell:\; k\ell \le B(n)}\;
      \left(\M^{\calR(L, c)}_{D_n}\right)_{k, \ell}
      \;-\;
      \sup \M^\G_\Pi
      \;\right| \;.      \label{eq:micrgeomcons:2}
  \end{align}

  By Theorem~\ref{thm:micr-cons} (consistency of the
  non-private $\micr$ statistic), 
  the term in \eqref{eq:micrgeomcons:2} converges in probability
  to 0 as $n \to \infty$.
  Similary, by Theorem~\ref{thm:micrgeom-err}, the
  term in \eqref{eq:micrgeomcons:1} converges in probability
  to $0$ as $n \to \infty$.
  Thus for any $\tau$ and $p$, there exist integers $N_1$ and $N_2$
  such that both terms are each bounded from above by $\tau/2$ with probability
  greater than $1 - p$ for all $n \ge \max\{N_1, N_2\}$,
  which completes the proof. 
\end{proof}

It reamins to prove Theorem~\ref{thm:micrgeom-err}, which we develop in the next
subsection.

\subsection{Proof of Theorem~\ref{thm:micrgeom-err}}

The proof of this theorem relies on deriving a bound on
the change in mutual information between the noisy distribution
$\widehat \P^{\epsilon}_{D, \Gamma}$ and the non-noisy
$\P_{D, \Gamma}$, where $\Gamma$ is the master range-equipartition
grid for a fixed (wlog) $k \le \ell$.

First, observe that because
\begin{align}
  \left|\;
      \max_{k, \ell:\; k\ell \le B(n)}\;
      \left(\widehat \M^{\calR(L, c)}_{D_n, \epsilon}\right)_{k, \ell}
      -
      \max_{k, \ell:\; k\ell \le B(n)}\;
      \left(\M^{\calR(L, c)}_{D_n}\right)_{k, \ell}
  \;\right|
  \;\le\;
  \max_{k, \ell:\; k\ell \le B(n)}\;
  \left|\;
  \left(\widehat \M^{\calR(L, c)}_{D_n, \epsilon}\right)_{k, \ell}
  -
  \left(\M^{\calR(L, c)}_{D_n}\right)_{k, \ell}
  \;\right|
  \label{eq:mgerr:goal1}
\end{align}
our goal will be to derive a sufficiently small uniform upper bound on 
\begin{align}
  \left|\;
  \left(\widehat \M^{\calR(L, c)}_{D_n, \epsilon}\right)_{k, \ell}
  \;-\;
  \left(\M^{\calR(L, c)}_{D_n}\right)_{k, \ell}
  \;\right|
  \label{eq:mgerr:goal2}
\end{align}
that holds for any $k\ell \le B(n)$ with probability at least $1-p$. 
Taking a union bound over all possible $k, \ell$ pairs 
(of which there at are most $B(n)^2$) means that the uniform bound
on the absolute difference holds simultaneously
for all $k\ell \le B(n)$ with probability at least $1 - B(n)^2 p$. 

To derive an upper bound on \eqref{eq:mgerr:goal2}, recall that
for a fixed (wlog) $k \le \ell$, the grid $\Gamma := \Gamma_{c, \ell}$
is the master range-equipartition grid for $\calR(L, c, k, \ell)$.
Then
\begin{align}
  \left|\;
  \left(\widehat \M^{\calR(L, c)}_{D_n, \epsilon}\right)_{k, \ell}
  \;-\;
  \left(\M^{\calR(L, c)}_{D_n}\right)_{k, \ell}
  \;\right|
  &\;=\;
  \left|\;
  \max_{G \in \calR(L, c, k, \ell)}
  I^\star\left(
  \widehat \P^{\epsilon}_{D, \Gamma, G}
  \right)
  \;-\;
  \max_{G \in \calR(L, c, k, \ell)}
  I^\star\left(
  \P_{D, \Gamma, G}
  \right)
    \;\right| \nonumber \\
  &\;\le\;
    \max_{G \in \calR(L, c, k, \ell)}
    \left|\;
    I\left(
    \widehat \P^{\epsilon}_{D, \Gamma, G}
    \right)
    \;-\;
    I\left(
    \P_{D, \Gamma, G}
    \right)
    \;\right|
  \label{eq:mgerr:goal3}
\end{align}
where the last line is due to the triangle inequality and since
$\log_2 k \ge 1$.

Now for suitable $B(n)$, and for a fixed $k\le\ell$ such that $k \ell \le B(n)$,
we claim the following probabilistic bound on \eqref{eq:mgerr:goal3}.

\begin{lem}
  \label{lem:mgerr:fixedkl}
  Fix any finite $c > 0$, $\epsilon > 0$,
  $\alpha \in (0, 0.5)$, and $B(n) \le O(n^\alpha)$.
  For sufficiently large $n$ and any $k \ell \le B(n)$, let
  $\Gamma$ denote the master range-equipartition for $\calR(L, c, k, \ell)$.
  Then there exists some $u > 0$ such that, 
  for any dataset $D$ of size $n$ and $G \in \calR(L, c, k, \ell)$,
  \begin{equation*}
    \left|\;
      I\left(
        \widehat \P^{\epsilon}_{D, \Gamma, G}
      \right)
      \;-\;
      I\left(
        \P_{D, \Gamma, G}
      \right)
      \;\right|
    \;\le\;
    O\left(
      \frac{c}{\epsilon n^u}
    \right)
  \end{equation*}
  with probability at least $1 - O(n^{-3})$.
\end{lem}

If we again grant this lemma true for now, the proof of Theorem~\ref{thm:micrgeom-err}
follows:

\begin{proof}[Proof (of Theorem~\ref{thm:micrgeom-err})]
  By Lemma~\ref{lem:mgerr:fixedkl}, for any
  $k\ell \le B(n) = O(n^\alpha)$,
  expression \eqref{eq:mgerr:goal3} is bounded from above by
  $O(c/(\epsilon n^u))$ for some constant $u > 0$ (that depends on $k, \ell$)
  with probability at least $1 - O(n^{-3})$. 
  Let $a > 0$ be the minimum of all such constants over all $k\ell \le B(n)$.
  Then for all $k \ell \le B(n)$,
  \begin{equation*}
    \left|\;
      \left(\widehat \M^{\calR(L, c)}_{D_n, \epsilon}\right)_{k, \ell}
      \;-\;
      \left(\M^{\calR(L, c)}_{D_n}\right)_{k, \ell}
      \;\right|
    \;\le\;
    O\left(
      \frac{c}{\epsilon n^a}
    \right)
  \end{equation*}
  with probability at least $1 - O(B(n)^2/n^{3}) \ge 1 - O(n^{-2})$.
  The error probability holds simultaneously over all $k\ell \le B(n)$
  and follows by taking a union bound over all such entries, of which 
  there at most $B(n)^2 \le O(n^{2\alpha}) \le O(n)$ since $\alpha < 0.5$.
  In particular, this means the bound holds for the maximizing
  entry of expression \eqref{eq:mgerr:goal1}, which completes
  the proof.
\end{proof}

It remains to prove Lemma~\ref{lem:mgerr:fixedkl}, which requires the 
most technical machinery. 

\subsubsection{Proof of Lemma~\ref{lem:mgerr:fixedkl}}

We will use a tool from \cite{reshef16-mice}, which relates
the statistical distance between two discrete distributions to
their difference in mutual information.
For two discrete distributions $\Pi$ and $\Psi$
over $[k] \times [\ell]$, let $D_{TV}(\Pi, \Psi)$ denote
their statistical (total variation) distance, and recall that 
$
  D_{TV}(\Pi, \Psi)
  \;=\;
  \frac{1}{2} \cdot
  \left\|\Pi - \Psi \right\|_1 .
$
We have the following lemma of Reshef et al.:

\begin{lem}[\cite{reshef16-mice} Proposition 40, Appendix B]
  \label{lem:prop40}
  Let $\Pi$ and $\Psi$ be discrete distributions over
  $[k] \times [\ell]$ for $k, \ell \ge 2$. 
  For any $0 < \delta \le 1/4$, if $D_{TV}(\Pi, \Psi) \le \delta$, then
  $
    \left|\;
      I(\Pi) - I(\Psi)
      \;\right|
    \;\le\;
    O\left(\delta \log_2 ((\min\{k, \ell\})/\delta)\right) .
    $
\end{lem}

For a fixed dataset $D$ of size $n$, finite $c > 0$, a fixed $k, \ell$,
and a fixed $G \in \calR(L, c, k, \ell)$ with master
range-equipartition $\Gamma$,
we will bound the difference in mutual information between
$\widehat \P^\epsilon_{D, \Gamma, G}$ and $\P_{D, \Gamma G}$
by providing an upper bound on
$D_{TV}(\widehat \P^\epsilon_{D, \Gamma, G}, \P_{D, \Gamma, G})$.
But given that every $G \in \calR(L, c, k, \ell)$ is a
subpartition of $\Gamma$, it follows by the triangle inequality that
\begin{equation}
  D_{TV}(\widehat \P^\epsilon_{D, \Gamma, G}, \P_{D, \Gamma, G})
  \le
  D_{TV}(\widehat \P^\epsilon_{D, \Gamma}, \P_{D, \Gamma}) \;,
  \label{eq:mgerr:dtvbound}
\end{equation}
and thus our goal will be to provide a probabilistic
upper bound on 
$D_{TV}(\widehat \P^\epsilon_{D, \Gamma}, \P_{D, \Gamma})$.

Assume wlog that $k \le \ell$, which means that the master
range-equipartition grid $\Gamma$ is of size $c \ell \times \ell$. 
Fix $\epsilon > 0$, and let $\hatp(i, j)$ and $p(i,j)$ denote the
$(i,j)$ entries of $\hatP^\eps_{D, \Gamma}$ and $\P_{D, \Gamma}$.
By definition of the $\micrgeom$ mechanism from Section~\ref{sec:micrgeom},
recall that for all $(i, j) \in [c\ell] \times [\ell]$, we define
\begin{align*}
  \hata(i, j)
  &\;\sim\;
  \truncgeom(\eps/2, n, p(i, j)\cdot n) \\
  \hatn 
  &\;=\;
  \sum\nolimits_{i, j} \hata(i, j)\\
  \hatp(i ,j)
  &\;=\;
  \frac{\hata(i,j)}{\hatn}\;.
\end{align*}
For all $(i, j)$ let $\Delta_{i, j}$ be a random variable defined by
\begin{equation}
  \Delta_{i, j}
  \;=\;
  \hata(i, j) - p(i, j) \cdot n \;,
  \label{eq:micrgerr:deltaij}
\end{equation}
which can be thought of as the change in number of datapoints
in cell $(i, j)$ after applying the $\truncgeom$ mechanism.
For all $(i, j)$,  the value $\Delta_{i, j}$ is distributed like
a doubly-geometric random variable centered at $0$ with parameter $e^{(-\epsilon/2)}$,
but with truncation at $n-(p(i, j)\cdot n)$ and $-(p(i, j)\cdot n)$. 

So we can rewrite each $\hatp(i, j)$ as
\begin{equation}
  \hatp(i, j)
  \;=\;
  p(i,j) \cdot \frac{n}{\hatn} + \frac{\Delta_{i, j}}{\hatn} \;,
\end{equation}
and moreover, we can write
\begin{equation}
  \hatn = n + \sum_{i, j} \Delta_{i, j} \;.
  \label{eq:micrgerr:nhat}
\end{equation}

It follows that
\begin{align}
  D_{TV}(\widehat \P^\epsilon_{D, \Gamma}, \P_{D, \Gamma})
  &\;=\;
    \frac{1}{2}
    \sum_{i, j}
    \left|\;
    \hatp(i, j) - p(i, j)
    \;\right| \nonumber \\
  &\;=\;
    \frac{1}{2}
    \sum_{i, j}
    \left|\;
    p(i,j) \cdot \frac{n}{\hatn} + \frac{\Delta_{i, j}}{\hatn}
    \;-\;
    p(i, j)
    \;\right| \;. \nonumber
\end{align}
By the triangle inequality, and increasing the $1/2$ term to $1$, we have
\begin{align}
  D_{TV}(\widehat \P^\epsilon_{D, \Gamma}, \P_{D, \Gamma})  
  &\;\le\;
    \sum_{i, j}
    \left|\;
    p(i,j) \left(\frac{n}{\hatn} - 1\right)  
    \;\right|
    \;+\;
    \sum_{i, j}
    \left|\;
    \frac{\Delta_{i, j}}{\hatn}
    \;\right| \nonumber \\
  &\;=\;
    \left|\;
    \frac{n}{\hatn} - 1
    \;\right|
    \cdot \sum_{i, j} p(i,j)
    \;+\;
    \sum_{i, j}
    \left|\;
    \frac{\Delta_{i, j}}{\hatn}
    \;\right| \nonumber \\
  &\;=\;
    \left|\;
    \frac{n}{\hatn} - 1
    \;\right|
    \;+\;
    \sum_{i, j}
    \left|\;
    \frac{\Delta_{i, j}}{\hatn}
    \;\right| \;. \label{eq:micrgerr:goal1} 
\end{align}
where the final equality holds because
$\sum_{i, j} p(i, j) = 1$.

The following lemma gives a uniform upper bound on $|\Delta_{i,j}|$
that will translate into bounds on both terms of \eqref{eq:micrgerr:goal1}.
Note that the lemma is stated and proven with respect to the
fixed parameters (namely $\epsilon$) already stated in this section.

\begin{lem}
  \label{lem:micrg-err:delta-tail}
  For any $(i, j) \in [c\ell] \times [\ell]$ and
  for any number $x > 0$, 
  \begin{equation*}
    \Pr\left(
      |\Delta_{i, j}| > x
    \right)
    \;\le\;
    \exp(-(\eps/2)\cdot x) \;.
  \end{equation*}
\end{lem}

\begin{proof}
  Recall $\Delta_{i, j}$ is a doubly-geometric random variable
  with parameter $e^{(-\eps/2)}$
  centered at 0 and truncated at $n-(p(i, j) \cdot n)$ and $-(p(i, j) \cdot n)$. 
  Let $\Delta$ denote a \textit{non-truncated} doubly-geometric random 
  variable with parameter $e^{(-\eps/2)}$ centered at $0$.
  It is easy to see that for any number $x > 0$,
  \begin{equation*}
    \Pr\left(
      |\Delta_{i, j}| > x
    \right)
    \;\le\;
    \Pr\left(
      |\Delta| > x
    \right) \;.
  \end{equation*}
  Then using the CDF of doubly-geometric random variables, we have
  \begin{equation*}
    \Pr\left(
      |\Delta| > x
    \right)
    \;=\;
    1 -
    \left(
      \frac{1-\exp(-\epsilon/2)}{1 + \exp(-\epsilon/2)}
    \right)
    \cdot
    \left(
      \sum_{-x}^x\;
      \exp(-(\eps/2) \cdot |x|)
    \right)
  \end{equation*}
  The doubly-geometric distribution with parameter $e^{(-\eps/2)}$ is a discrete
  approximation of a continuous Laplace distribution
  with parameter $2/\epsilon$. It is known that for any $x > 0$
  \begin{equation*}
    \Pr\left(
      |\lap(2/\epsilon)| > x
    \right)
    \;=\;
    \exp(-(\eps/2)\cdot x) \;,
  \end{equation*}
  and that 
  \begin{equation*}
    1 -
    \left(
      \frac{1-\exp(-\epsilon/2)}{1 + \exp(-\epsilon/2)}
    \right)
    \cdot
    \left(
      \sum_{-x}^x\;
      \exp(-(\eps/2) \cdot |x|)
    \right)
    \;\le\;
    \exp(-(\eps/2)\cdot x) \;.
  \end{equation*}
  Thus the tail of every $\Delta_{i, j}$ r.v. with parameter $\exp(-(\eps/2))$
  is dominated by the tail of the non-truncated $\Delta$ r.v. with
  the same parameter,
  which is dominated by the tail of a continuous Laplace r.v.
  with parameter $2/\epsilon$, completing the proof.
\end{proof}

Using this tail bound (which observe applies uniformly to all
$(i,j)$), we have the following corollary.

\begin{cor}
  \label{cor:deltaij-bound}
  For every $(i, j)$ and for any $d > 0$,
  \begin{equation*}
    \Pr\left(
      |\Delta_{i,j}|
      > d \ln n 
    \right)
    \;\le\;
    \exp(- (\eps/2) d \ln n)
    \;=\;
    n^{-((\eps/2)d)} \;.
  \end{equation*}
\end{cor}

By a union bound over all $c\ell^2$ entries $(i, j)$, it follows that
for all $d > 0$, 
\begin{equation}
  \sum_{i, j}\;
  |\Delta_{i, j}|
  \;\le\;
  c\ell^2 \cdot (d \ln n)
  \label{eq:sum-delta-bound}
\end{equation}
with probability all but $(c\ell^2)/(n^{-((\eps/2)d)})$.
Because $\hatn = n + \sum_{i, j} \Delta_{i, j}$
(from expression \eqref{eq:micrgerr:nhat}), then with this
same probability:
\begin{align}
  \hatn
  &\le
    n + \sum_{i, j} |\Delta_{i,j }|
    \;\le\; n + c\ell^2 d \ln n \label{eq:hatn-ub}\\
  \text{and}\;\;\;
  \hatn
  &\ge n - \sum_{i, j} |\Delta_{i, j}|
    \;\ge\; n - c\ell^2 d\ln n \;. \label{eq:hatn-lb}
\end{align}
Using these bounds on $\hat n$ and $\sum_{i, j} \Delta_{i,j}$,
we state and prove the following two bounds on the terms
from expression~\eqref{eq:micrgerr:goal1}.

\begin{lem}
  \label{lem:micrgerr:term1-bound}
  For sufficiently large $n$, and assuming $B(n) = O(n^\alpha)$ for
  $\alpha \in (0, 0.5)$, there exists a constant $u_1 > 0$ such that
  \begin{equation*}
    \left|\;
      \frac{n}{\hatn} - 1
      \;\right|
    \;\le\;
    O\left(
      \frac{c}{\epsilon n^{u_1}}
    \right)
  \end{equation*}
  with probability at least $1 - O(n^{-3})$. 
\end{lem}

\begin{proof}
  First, suppose that $(n/\hatn) \ge 1$.
  Then $|(n/\hatn) - 1|$ is maximized when $\hatn$ is small.
  Using the lower bound on $\hatn$ from \eqref{eq:hatn-lb}, it follows
  that
  \begin{align*}
    \left|\;
      \frac{n}{\hatn} - 1
      \;\right|
    \;=\;
    \left|\;
      \frac{n - \hatn}{\hatn}
    \;\right|
    \;\le\;
    \frac{c\ell^2 d \ln n}{n - c\ell^2 d \ln n}
  \end{align*}
  with probability all but $n^{-((\eps/2)d)}$.
  By the assumption that $\alpha < 0.5$, it follows
  that $\ell^2 \le (B(n))^2 \le O(n^{2\alpha}) = o(n)$.
  Setting $d := 8 / \epsilon$, it follows that
  $c\ell^2 d \ln n \le O((c/\eps) n^{1+t_1})$ for
  some constant $t > 0$, and that
  $n - c\ell^2 d \ln n = \Omega(n)$, which means that
  \begin{align*}
    \left|\;
    \frac{n}{\hatn} - 1
    \;\right|
    \;\le\;
    \frac{c\ell^2 d \ln n}{n - c\ell^2 d \ln n}
    \;\le\;
    O\left(
    \frac{c}{\epsilon n^{t}}
    \right)
  \end{align*}
  with probability all but 
  $(c\ell^2)/n^{-(\eps/2)d}
  = O(c n^{2\alpha} / n^{-4})
  = O(n^{-3})$. 

  On the other hand, suppose $(n/\hatn) < 1$.
  Then $|1 - (n/\hatn)|$ is maximized when $\hatn$ is large, and
  using the upper bound on $\hatn$ from \eqref{eq:hatn-ub},
  it follows that
  \begin{align*}
    \left|\;
    1  - 
    \frac{n}{\hatn}
    \;\right|
    \;=\;
    \left|\;
      \frac{\hatn - n}{\hatn}
    \;\right|
    \;\le\;
    \frac{c\ell^2 d \ln n}{n + c\ell^2 d \ln n} 
  \end{align*}
  with probability all but $n^{-((\eps/2)d)}$.
  By similar arguments as in the first case and again setting
  $d := 8/\eps$, it then follows that for some constant $v > 0$,
  \begin{equation*}
    \left|\;
    1  - 
    \frac{n}{\hatn}
    \;\right|
    \;\le\;
    \frac{c\ell^2 d \ln n}{n + c\ell^2 d \ln n}
    \;\le\;
    O\left(
      \frac{c}{\epsilon n^v}
    \right)
  \end{equation*}
  with probability all but $n^{-3}$.

  Setting $u_1 := \min\{t, v\}$ then concludes the proof. 
\end{proof}

Lemma~\ref{lem:micrgerr:term1-bound} gives a bound on
the first term from expression \ref{eq:micrgerr:goal1}.
We now prove a bound on the second term.

\begin{lem}
  \label{lem:micrgerr:term2-bound}
  For sufficiently large $n$, and assuming $B(n) = O(n^\alpha)$
  for $\alpha \in (0, 0.5)$, there exists a constant $u_2 > 0$
  such that
  \begin{equation*}
    \sum_{i, j}\;
    \left|
      \frac{\Delta_{i, j}}{\hatn}
    \right|
    \;\le\;
    O\left(
      \frac{c}{\epsilon n^{u_2}}
    \right)
  \end{equation*}
  with probability all but $n^{-3}$. 
\end{lem}

\begin{proof}
  As in the proof of Lemma~\ref{lem:micrgerr:term1-bound},
  set $d := 8/\eps$.
  Again using the fact that $\ell^2 \le B(n)^2 \le O(n^{2\alpha}) = o(n)$
  by the assumption that $\alpha < 0.5$, observe from \eqref{eq:hatn-lb} that
  $\hat n > 0$ with probability all but $O(n^{-3})$ for sufficiently large $n$.
  So with this same probability we can write
  \begin{equation*}
    \sum_{i, j}\;
    \left|
      \frac{\Delta_{i, j}}{\hatn}
    \right|
    \;=\;
    \frac{\sum_{i, j}\; |\Delta_{i, j}|}{\hat n}
    \;\le\;
    \frac{c\ell^2 (8/\eps) \ln n}{n - c\ell^2 (8/\eps) \ln n} \;.
  \end{equation*}
  Here, since $\ell^2\ln n \le n^{2\alpha}\ln n$, it follows that
  the denominator of the final expression is $\Omega(n)$,
  and the numerator is $O((c/\eps)n^{1 + u_2})$ for some $u_2 > 0$.
  Thus with probability all but $O(n^{-3})$, it follows that
  \begin{equation*}
    \sum_{i, j}\;
    \left|
      \frac{\Delta_{i, j}}{\hatn}
    \right|
    \;\le\;
    \frac{c\ell^2 (8/\eps) \ln n}{n - c\ell^2 (8/\eps) \ln n}
    \;\le\;
    O\left(
      \frac{c}{\epsilon n^{u_2}}
    \right) \;. \qedhere
  \end{equation*} 
\end{proof}

Together, Lemmas~\ref{lem:micrgerr:term1-bound}
and~\ref{lem:micrgerr:term2-bound} give the following
probabilistic bound on the total variation
distance between $\hatP^\eps_{D, \Gamma}$ and
$\P_{D, \Gamma}$ from expression \eqref{eq:micrgerr:goal1},
and thus a proof of Lemma~\ref{lem:mgerr:fixedkl}.

\begin{proof}[Proof (of Lemma ~\ref{lem:mgerr:fixedkl})]
  By expressions \eqref{eq:mgerr:dtvbound}
  and \eqref{eq:micrgerr:goal1}, and using
  Lemmas~\ref{lem:micrgerr:term1-bound}
  and~\ref{lem:micrgerr:term2-bound}, it follows that
  for sufficiently large $n$, there exist constants $u_1, u_2 > 0$
  such that for any $G \in \calR(L, c, k, \ell)$, 
  \begin{equation*}
    D_{TV}(\widehat \P^\epsilon_{D, \Gamma, G}, \P_{D, \Gamma, G})
    \;\le\;
    D_{TV}(\widehat \P^\epsilon_{D, \Gamma}, \P_{D, \Gamma})
    \;\le\;
    O\left(
      \frac{c}{\epsilon n^{u_1}}
    \right)
    \;+\;
    O\left(
      \frac{c}{\epsilon n^{u_2}}
    \right)
    \;\le\;
    O\left(
      \frac{c}{\epsilon n^{u_3}}
    \right)
  \end{equation*}
  with probability at least $1 - O(n^{-3})$ for sufficiently large $n$,
  where $u_3 := \min\{u_1, u_2\} > 0$. 
 
  Then by Lemma~\ref{lem:prop40}, it follows that 
  \begin{equation*}
    \left|\;
      I\left(
        \widehat \P^{\epsilon}_{D, \Gamma, G}
      \right)
      \;-\;
      I\left(
        \P_{D, \Gamma, G}
      \right)
      \;\right|
    \;\le\;
    O\left(
      \frac{c}{\epsilon n^{u_3}}
      \cdot 
      \log_2 (n^{(\alpha+u_3)})
    \right)
    \;\le\;
    O\left(
      \frac{c}{\epsilon n^{u}}
    \right)
  \end{equation*}
  with probability at least $1 - O(n^{-3})$ 
  for some $u_3 > u > 0$ 
  when $n$ is sufficiently large.
\end{proof}


%
%

\section{Experimental Results Appendix}
\label{sec:app:exps}

In this section, we provide more details on
the methodologies used for our experimental
results from Section~\ref{sec:exps}.

\subsection{Implementation Details}
\label{sec:app:real-exps:imp}


As mentioned in Section~\ref{sec:prelims},
the MICe statistic in Definition~\ref{def:mice}
(and the analogous definitions of MICr and its private variants)
varies slightly from the definition of MICe
from \citet{reshef16-mice}, specifically when defining
the maximization space of individual entries in the
equicharacteristic matrix\footnote{%
  \citet{reshef16-mice} refer to this as
  the ``clumped'' variant of MICe, but to avoid confusion
  we refer to this as just MICe.
  }.
In particular, for $k \le \ell$, when $\ell > \sqrt{B}$,
\citet{reshef16-mice} set the master row partition
used in the optimization as a mass equipartition of
size $c(B/\ell)$.
In contrast, Definition~\ref{def:mice} sets the
master row partition as a mass equipartition of size $c\cdot\ell$,
even when $\ell > \sqrt{B}$ (note that the definitions are identical
in the case when $\ell \le \sqrt{B}$).

The advantage of using these smaller master partitions is computational:
the MICe variant of \citet{reshef16-mice} can be computed
in $O(c^2 B^{2.5})$ time (Appendix H.1, \citet{reshef16-mice}),
in contrast with the $O(c^2 B^4)$ running time from
Theorem~\ref{thm:mice-runtime} using Definition~\ref{def:mice}.
On the other hand, for $B := B(n) = O(n^\alpha)$
where $\alpha \in (0, 0.5)$, the consistency of this variant
can only be shown to hold when the output of the statistic is the maximum
$(k, \ell)$ entry of the equicharacteristic matrix
where $k \cdot \ell \le B$ \textit{and} $k$ and $\ell$
are individually at most $\sqrt{B}$.
Note however that although consistency for this variant
is only shown to hold when imposing this additional constraint
on $k$ and $\ell$, the statistic is still nonetheless
both defined and can be computed (i.e., as in the implementation of
\citet{albanese}) as a maximization over \textit{all}
equicharacteristic matrix entries where $k\cdot \ell \le B$.

The MICr statistic from Definition~\ref{def:micr}
(and its private variants) can be adjusted in an analogous way
to reduce the size of the master row range equipartitions when
$\sqrt{B} < \ell \le B/2$. Again, the benefit is faster computation
at the expense of a slightly different consistency guarantee
(analogous to that of the MICe variant described above).

Primarily for computational considerations, our experiments in this work
use implementations of MICe and MICr (and their private variants)
that follow the original definition of \citet{reshef16-mice}
(i.e., use smaller master partition sizes for $\ell > \sqrt{B}$).
While these variants hold slightly different consistency properties
than their counterparts defined in Sections~\ref{sec:prelims},
\ref{sec:mice-lap}, and \ref{sec:micr}, in practice the outputs
between pairs of corresponding variants are similar.
Note that an end-user with a fixed computational budget
could use either set of definitions and tune the $B$ and $c$
parameters accordingly to meet their budget.

Moreover, we remark that the $\eps$-DP guarantees of the private
MICe and MICr mechanisms introduced in this work apply to
either regime of master grid size settings.


In our experiments, to compute $\mice$ we used the
MINEPY library implementation from \citet{albanese},
and to compute $\micr$ and its private variants,
we developed an implementation available at
\url{https://github.com/jlazarsfeld/dp-mic}.

\subsection{Synthetic Data Experiments}
\label{sec:app:synth-exps}

\subsubsection{Functional Relationships Used to Generate Distributions}

As described in Section~\ref{sec:exps}, we
evaluated our private mechanisms on distributions
based on a set of 21 functional
relationships originally defined by \citet{reshef2011}.
We list these functions here:

\begin{itemize}
\item
  F1: $f(x) = 0.2 \cdot \sin(12x - 6) + 1.1(x-1) + 1$
\item
  F2: $f(x) = 0.15 \cdot \sin(11x \cdot \pi) + (x + 0.05)$
\item
  F3: $f(x) = 0.1 \cdot \sin(48x) + 2(x - 0.05)$
\item
  F4: $f(x) = 0.2 \cdot \sin(48x) + 2(x - 0.05)$
\item
  F5: $f(x) = 0.4 \cdot \cos(7x \cdot \pi) + 0.5$
\item
  F6: $f(x) = 0.4 \cdot \cos(14x \cdot \pi) + 0.5$
\item
  F7: $f(x) = 10 \cdot {(x - 0.6)}^3 + 2 \cdot {x}^2 + (1.5 - 3x)$
\item
  F8: $f(x) = 4 \cdot (10 \cdot {(x - 0.6)}^3 + 2 \cdot {x}^2 + (1.5 - 3x)) - 1.4 $
\item
  F9: if $x \leq 0.99$ then $f(x) = \frac{x}{99}$, else $f(x) = 99x - 98$
\item
  F10: $f(x) = {2}^x - 1$
\item
  F11: $f(x) = {8}^{(x - 0.3)} - 1$
\item
  F12: $f(x) = x$
\item
  F13: $f(x) = 4 \cdot {(x - 0.5)}^2 + 0.1$
\item
  F14: $f(x) = 0.4 \cdot \sin(9x \cdot \pi) + 0.5$
\item
  F15: $f(x) = 0.4 \cdot \sin(8x \cdot \pi) + 0.5$
\item
  F16: $f(x) = 0.4 \cdot \sin(16x \cdot \pi) + 0.5$
\item
  F17: if $x < 0.491$ then $f(x) = 0.05$, else if $x > 0.509$ then $f(x) = 0.95$, else $f(x) = 50 \cdot (x - 0.5) + 0.5$
\item
  F18: $f(x) = 0.4 \cdot \cos(5x \cdot \pi \cdot (1 + x)) + 0.5$
\item
  F19: $f(x) = 0.4 \cdot \sin(6x \cdot \pi \cdot (1 + x)) + 0.5$
\item
  F20: if $x \leq 0.0528$ then $f(x) = 18x$, else if $x \geq 0.1$ then $f(x) = -1 \cdot \frac{x}{9} + \frac{1}{9}$, else $f(x) = -18x + 1.9$
\item
  F21: if $x \leq 0.0051$ then $f(x) = 190x$, else if $x \geq 0.01$ then $f(x) = -1 \cdot \frac{x}{99} + \frac{1}{99}$, else $f(x) = -198x + 1.99$
\end{itemize}

\newpage
For each function, we generated 9 joint distributions by placing $k=100$ ``generating'' points
on the function graph at evenly spaced intervals.
We considered a bivariate Gaussian distribution centered at each point with
$\rho=0$ and identical variances. The magnitude of the variance was set
according to the desired $R^2$ value of the resulting noisy functional
relationship using the method described in \citet{reshef2018empirical}.

\subsubsection{Approximately Computing MIC*}

For a joint distribution $\Pi = (X, Y)$, we used the method described in
Section 3.5 of \citet{reshef16-mice} to approximately compute $\micstar(\Pi)$.
This involves maximizing the mutual information
of $I(X|_Q, Y|_P)$ for increasingly large discrete partitions $P$
of size $k$ chosen from a dense master mass equipartition $\Gamma$,
and where $Q$ is a dense, fixed-sized master mass equipartition .
For this, we set the size of $\Gamma$ to 260 and the size of $Q$ to 360,
and we found the computation insensitive to increases in
master grid sizes beyond this point.

\subsubsection{Tuning Parameters for MICr-Lap and MICr-Geom}
\label{sec:app:synth:params}

To determine optimal $B$ and $c$ parameter settings for
$\micrlap$ and $\micrgeom$ at a particular value of $n$ and $\eps$,
we defined the following objective function WSUM using
the set of 189 distributions in $\calQ$.

First, we sorted $\cal{Q}$ in increasing order of $\micstar$
value (using the method described in the previous subsection).
For the $i$'th distribution $\Pi_i$ in sorted order, we defined a weight
$w_i$ by taking the length of the interval between
the midpoint of $\micstar(\Pi_{i-1})$ and $\micstar(\Pi_{i})$
and the midpoint of $\micstar(\Pi_{i})$ and $\micstar(\Pi_{i+1})$.
For the distributions with smallest and largest $\micstar$ values,
we used $0$ and $1$ as left and right interval endpoints respectively.
Then for a fixed mechanism, $B$, $c$, $n$, and $\eps$, we took
the weighted sum of the mechanism's average unsigned error on
each distribution wrt $\micstar$ using the $w_i$'s as weights.

By minimizing this function wrt the $B$ and $c$ parameters
for each mechanism and $(n, \eps, B, c)$ combination,
our goal was to determine parameter settings that ensured
equal levels of error across the entire spectrum of
low-correlation (low $\micstar$) and high-correlation (high $\micstar$)
distributions.

\begin{table}[h!]
  \centering
  \small
  \begin{tabular}{r|cccc}
    & \multicolumn{2}{c}{$\micrgeom$} & \multicolumn{2}{c}{$\micrlap$} \\
    & $\eps = 1.0$ & $\eps = 0.1$ & $\eps = 1.0$ & $\eps = 0.1$  \\
    \hline
    $n=25$   & (2, 12)   & (2, 6)    & (5, 8)    & (5, 6)     \\
    $n=250$  & (1, 40)   & (2, 10)   & (5, 40)   & (5, 40)    \\
    $n=500$  & (1, 40)   & (2, 20)   & (5, 60)   & (5, 80)    \\
    $n=1000$ & (1, 60)   & (2, 40)   & (5, 80)   & (5, 100)   \\
    $n=5000$ & (1, 150)  & (1, 40)   & (5, 150)  & (5, 125)   \\
    $n=10000$ & (1, 150)  & (1, 80)   & (5, 150)  & (5, 150)
\end{tabular}
\caption{Each entry is the optimal $(c, B)$ settings
  for the corresponding mechanism and ($n$, $\eps$)
  that was found by minimizing a weighted sum of the
  mechanism's absolute error (wrt $\micstar$)
  over all $\Pi \in \calQ$.}
\label{tab:opt-BC}
\end{table}

Table~\ref{tab:opt-BC} shows these optimized parameters
for the $\micrlap$ and $\micrgeom$ mechanisms.
For both mechanisms and both $\epsilon$ values,
the optimal $B$ values are generally increasing with $n$,
which aligns with the intuition that both mechanisms' outputs
converge toward $\micstar$ with larger $n$.
Notice also that for $\eps$=0.1 and small $n$,
the optimal $c$ value of $\micrgeom$ is 2.
Although the error of $\micrgeom$ wrt to $\micr$
decreases with $c$, we expect
the error of $\micr$ wrt  $\micstar$
to decrease with \textit{larger} $c$ and $B$.
For small $n$ and $\eps$, this leads to an optimization
tradeoff that is more pronounced.

\subsubsection{Bias/Variance Analysis for $\eps=0.1$}

\begin{figure}[h!]
  \begin{center}
    \includegraphics[width=0.5\linewidth]{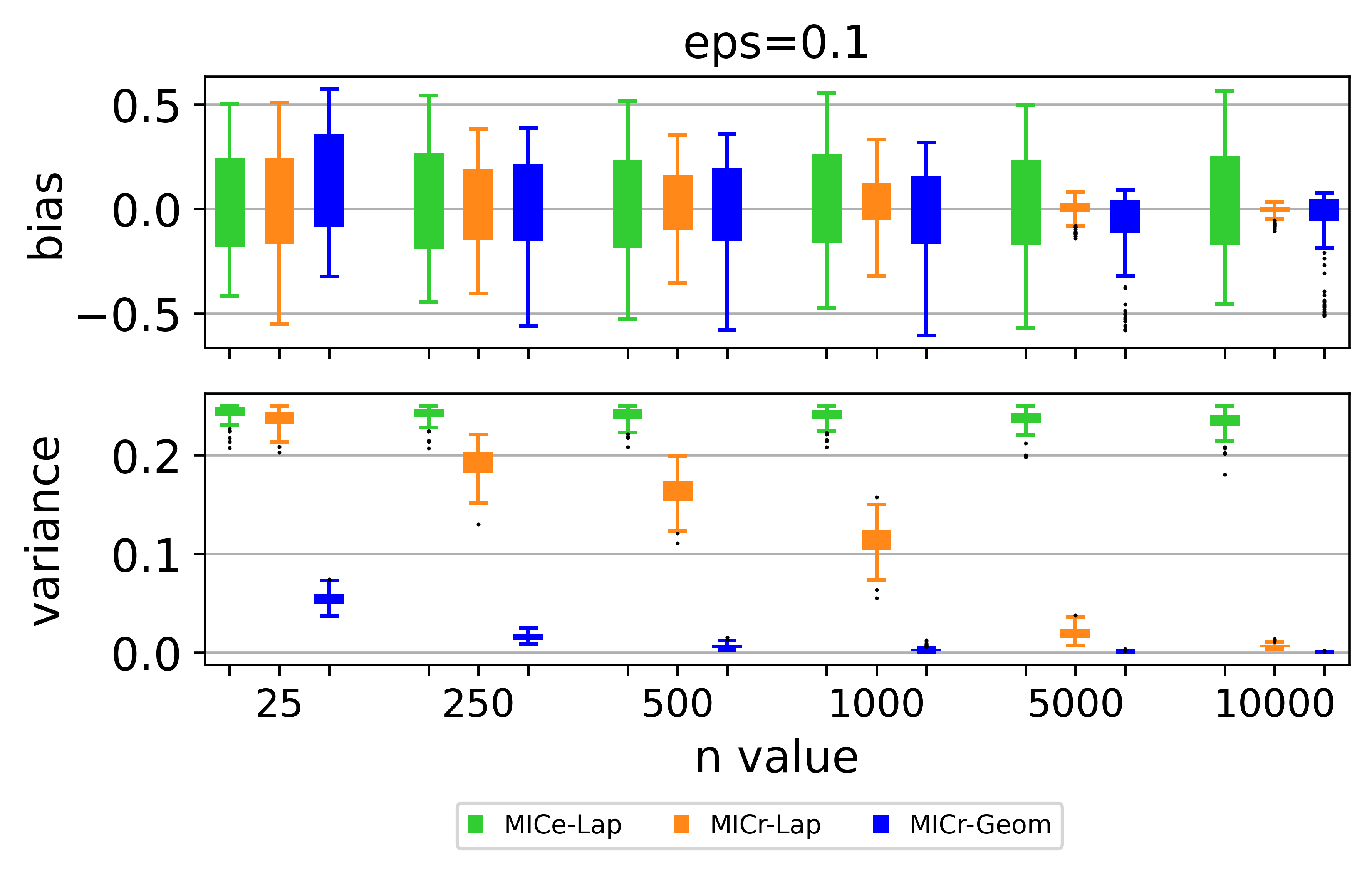}
  \end{center}
  \vspace*{-0.5em}
  \caption{
      Boxplots of bias (top) and variance (bottom)
      of each private mechanism (over 50 iterations)
      over all distributions in $\mathcal{Q}$ for $\eps$=0.1
      and varying $n$.
    }
  \label{fig:synthetic-bv:eps10}
\end{figure}

Figure~\ref{fig:synthetic-bv:eps10} shows the boxplots
of the bias and variance of each private mechanism
over the set of all distributions in $\mathcal{Q}$ as $n$ varies
and $\eps=0.1$. These are the analogous plots for
Figure~\ref{fig:synthetic-bv} in Section~\ref{sec:exps}
for $\eps=1$.
Again notice that as $n$ grows, the IQR of bias
tends to decrease for the $\micrlap$ and $\micrgeom$ mechanisms,
but this is less pronounced for $\micrgeom$ compared to
the $\eps=1$ case. Here, the bias/variance tradeoff
between $\micrlap$ (lower bias, higher variance)
and $\micrgeom$ (higher bias, lower variance) is much more
apparent, especially at smllaller values of $n$.

\subsection{Real Data Experiments}
\label{sec:app:real-exps}

Table~\ref{tab:real-bv:eps10} shows the median bias
and variance for each mechanism over all datasets in each collection
for $\eps=0.1$ (analogous to Table~\ref{tab:real-bv} from
Section~\ref{sec:exps}).

\begin{table}[h!]
  \centering
  \small
  \begin{tabular}{r | rrr}
  & $\micelap$
  & $\micrlap$
  & \micrgeom \\
  \hline
  Spellman23
  & 0.20 \;(0.24)
  & 0.19 \;(0.24)
  & 0.31 \;(0.05)  \\
  Baseball
  & 0.36 \;(0.24)
  & 0.25 \;(0.18)
  & 0.33 \;(0.01) \\
  Spellman4381
  & 0.41 \;(0.24)
  & 0.03 \;(0.01)
  & 0.14 \;(8e-4)
\end{tabular}
\caption{The median bias (average signed error wrt $\mice$ over 100 runs)
  and median variance (over 100 iterations) of each private mechanism
  across all datasets of each collection for $\eps$=0.1.
}
\label{tab:real-bv:eps10}
\end{table}

\begin{figure}[htb!]
  \centering
  \begin{minipage}{.5\textwidth}
    \centering
    \includegraphics[width=0.9\linewidth]{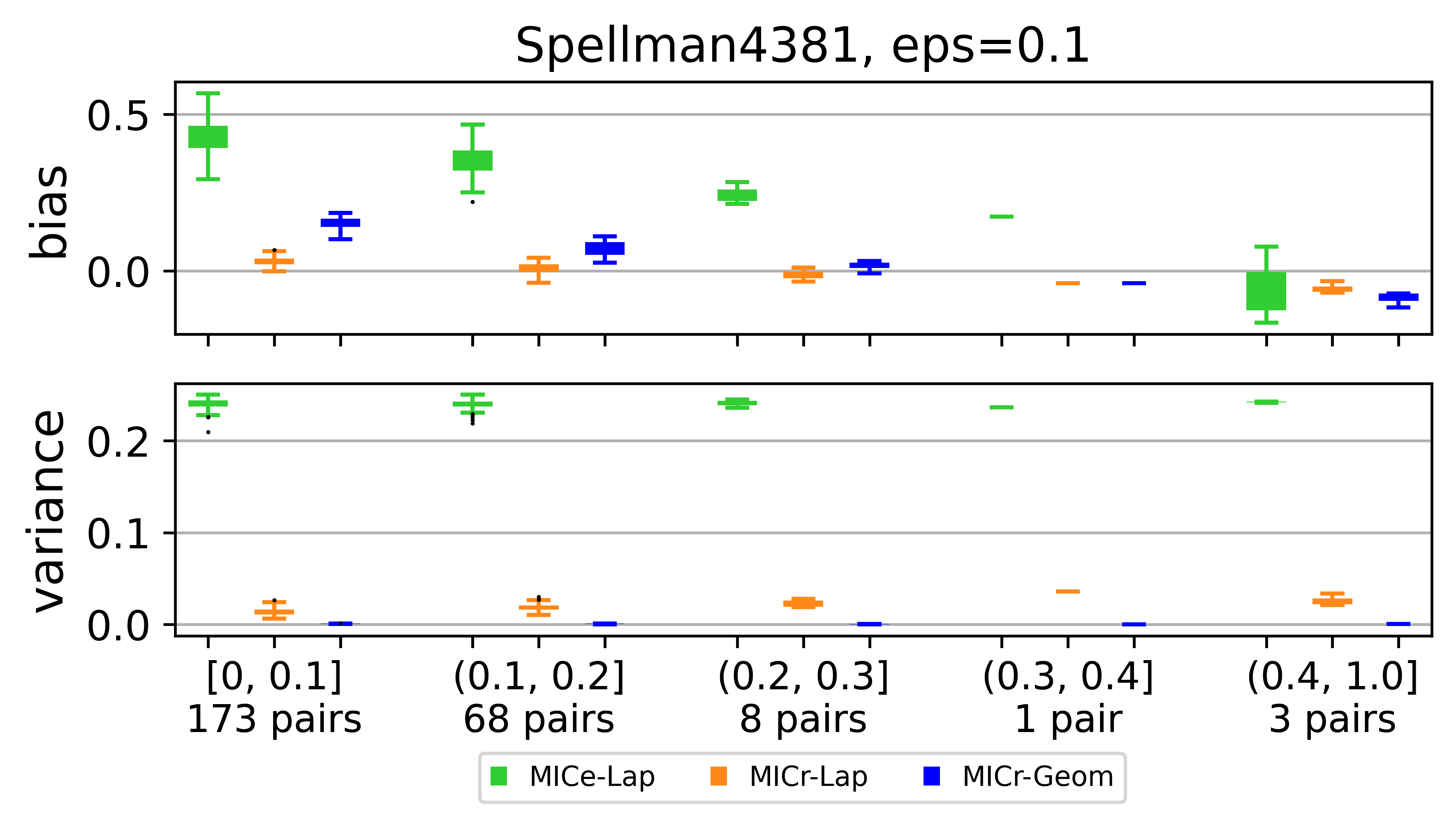}
  \end{minipage}%
  \begin{minipage}{.5\textwidth}
    \centering
    \includegraphics[width=0.9\linewidth]{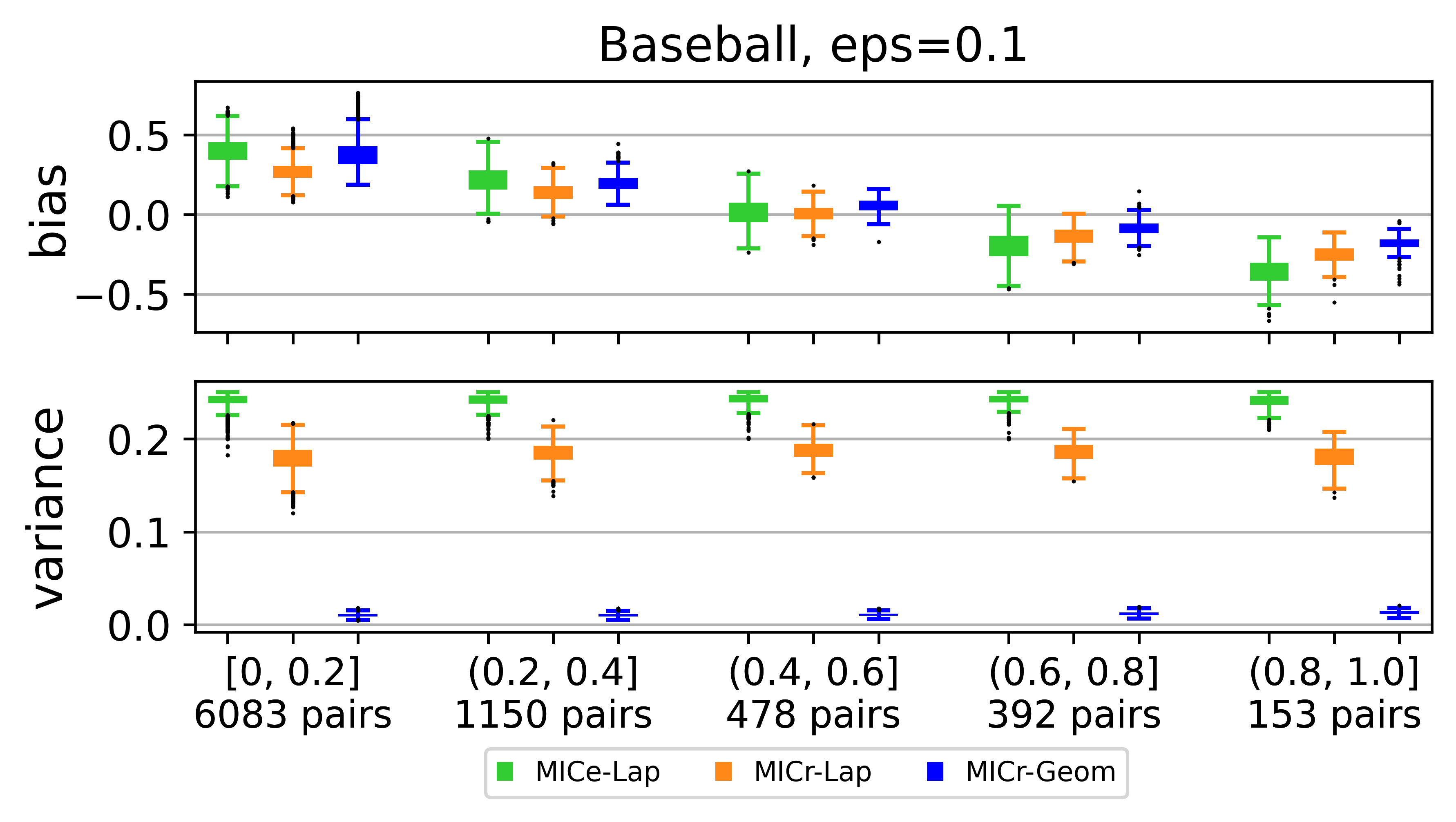}
  \end{minipage}
  \caption{%
      Bias and variance boxplots for each mechanism
      over datasets (pairs) in the Spellman4381 collection (left)
      and Baseball collection (right) binned by
      non-private $\mice$ score for $\eps$=0.1.
    }
  \label{fig:real-bw:eps10}
\end{figure}

Figure~\ref{fig:real-bw:eps10} shows the bias and
variance of each mechanism for the Spellman4381 and Baseball
collections at $\eps=0.1$ in the binned setting
(analogous to Figure~\ref{fig:real-bw} in Section~\ref{sec:exps}).
Again notice the bias/variance tradeoff between the $\micrlap$
and $\micrgeom$ mechanisms, which is especially apparent
at this level of $\eps$ in the smaller ($n=337$) Baseball
datasets.


\end{document}